\documentclass[12pt]{article}
\usepackage{a4wide}

\usepackage[dvipdfmx,hiresbb]{graphicx} 
\usepackage{mediabb}                    
\pdfoutput=1

\usepackage{color}

\newcommand{\al}[1]{\begin{align}#1\end{align}}

\newcommand{\paren}[1]{\left(#1\right)}

\newcommand{\sqbr}[1]{\left[#1\right]}
\newcommand{\br}[1]{\left\{#1\right\}}
\newcommand{\vev}[1]{\left\langle#1\right\rangle}

\newcommand{\GeV}{\ensuremath{\,\text{GeV} }}
\newcommand{\meV}{\ensuremath{\,\text{meV} }}
\newcommand{\eV}{\ensuremath{\,\text{eV} }}
\newcommand{\nn}{\nonumber\\}

\newcommand{\p}{\partial}

\usepackage{amsmath,amssymb}
\usepackage{epsf}

\begin{document}
\title{\vbox{
\baselineskip 14pt
\hfill \hbox{\normalsize 
MAD-TH-17-04
}
} \vskip 1cm
\bf \Large 
Weak Gravity Conjecture, Multiple Point Principle and the Standard Model Landscape
\vskip 0.5cm
}
\author{
Yuta~Hamada$^{1,2}$\thanks{E-mail: \tt yhamada@wisc.edu} ~and
Gary~Shiu$^{1}$\thanks{E-mail: \tt shiu@physics.wisc.edu}
\bigskip\\
$^1$\it\normalsize
Department of Physics, University of Wisconsin-Madison, Madison, WI 53706, USA\\
$^2$\it\normalsize
KEK Theory Center, IPNS, KEK, Tsukuba, Ibaraki 305-0801, Japan\\
}
\date{\today}


\maketitle

\abstract{\noindent \normalsize
The requirement for an ultraviolet completable theory to be well-behaved upon compactification  has been suggested as a guiding principle for distinguishing the landscape from the swampland.
Motivated by the weak gravity conjecture and the multiple point principle, we investigate the vacuum structure of the standard model compactified on  $S^1$ and $T^2$.
The measured value of the Higgs mass implies, in addition to the electroweak vacuum, the existence of a new vacuum where the Higgs field value is around the Planck scale.
We explore two- and three-dimensional critical points of the moduli potential arising from compactifications of the electroweak vacuum as well as this high scale vacuum, in the presence of Majorana/Dirac neutrinos and/or axions.
We point out potential sources of instability for these lower dimensional critical points in the standard model landscape. 
We also point out that a high scale $AdS_4$ vacuum of the Standard Model, if exists, would be at odd with the conjecture that all non-supersymmetric $AdS$ vacua are unstable.
We argue that, if we require a degeneracy between three- and four-dimensional vacua as suggested by the multiple point principle, the neutrinos are predicted to be Dirac, with the mass of the lightest neutrino  $\approx \mathcal{O} (1\text{--}10)$ meV, which may be tested by future CMB, large scale structure and $21$cm line observations.
}

\newpage
\tableofcontents
\newpage
\normalsize
\section{Introduction}
String theory is one of the most promising candidates for a consistent quantum theory of gravity.
While there is no free parameter in string theory, there appears to be an enormous large number of vacua, usually dubbed as the string theory landscape.
A natural question is whether the theory is so rich that {\it any} low energy effective theory can be realized in the string landscape?
At present, the space of low energy theories that can(not) be realized in string theory is not entirely known. 
The set of classically consistent effective field theories which turn out to be inconsistent when coupled to quantum gravity is referred to as the swampland \cite{Vafa:2005ui}.
Identifying the boundary between the landscape and the even vaster swampland has become an active research area in recent years.

Among the vast number of seemingly viable low energy effective theories, particularly interesting are those that reproduce the standard model (SM) spectrum at energies below the electroweak scale. If string theory is the ultraviolet completion of the SM, it is certainly important to examine the region of the string landscape where the SM vacuum resides. Understanding how our SM vacuum arises from compactifications of string theory may give us insights to the principle behind how our vacuum is selected. But equally interesting are 
vacua that arise from compactifying the SM down to {\it lower} dimensions, as they show that the rich structure of a landscape is not unique to ultraviolet complete theories of quantum gravity, but is already manifest in well understood theories such as the SM. It was in this spirit that the vacuum structure of the SM 
upon compactification on $S^1$ and $T^2$  
was investigated in Refs.~\cite{ArkaniHamed:2007gg,Arnold:2010qz,Arnold:2011cz,Fornal:2011tw}.


In this paper, we improve on these earlier works in several fronts.
First of all, in light of the discovery of the Higgs boson~\cite{Aad:2012tfa,Chatrchyan:2012xdj}, we can now provide a more accurate analysis up to the electroweak scale while in lack of the LHC data, previous works 
only focussed on the contributions from physics at the $\meV$ scale.
Moreover, the measured value of the Higgs mass implies the existence of a new vacuum where the Higgs field value is around the Planck scale (see e.g. Ref.~\cite{Degrassi:2012ry, Hamada:2012bp,Buttazzo:2013uya}).
Thus, in addition to mapping out the landscape of the standard model upon compactifying the electroweak vacuum, we also analyzed
the landscape arising from this high scale vacuum.
On a technical level, we also generalized these earlier studies to include the most general boundary conditions for the SM fields in the compact space, and with general fluxes supported on the internal cycles. These generalizations allow us to find many more lower-dimensional vacua in the SM landscape.
We also performed a careful analysis of the perturbative stability of the candidate vacua in two dimensions.
Our results can thus be taken as a starting point for future systematic studies of the SM landscape. 
As we shall see, some of the salient features of the SM landscape can be exhibited in a simpler setting. 
To this effect, we have examined the vacuum structure of the compactified $U(1)$ gauge theory with matter. We will first present our results for the $U(1)$ case as a warmup before we discuss our findings for the full-fledged SM landscape.


There are several motivations to study compactifications of the SM, along the lines we developed in this paper.
First of all, our analysis may lend insights to the weak gravity conjecture~\cite{ArkaniHamed:2006dz}, see also Refs.~\cite{Cheung:2014vva,delaFuente:2014aca,Brown:2015iha,Hebecker:2015rya,Bachlechner:2015qja,Brown:2015lia,Junghans:2015hba,Heidenreich:2015wga,Kooner:2015rza,Harlow:2015lma,Ibanez:2015fcv,Montero:2016tif,Cottrell:2016bty,Hebecker:2017wsu,Palti:2017elp} for some recent studies.
A proposed criterion for a low energy theory to be ultraviolet complete with quantum gravity is that it should be consistently behaved upon compactification, and requiring this consistency leads to non-trivial constraints on the low energy effective theory~\cite{Brown:2015iha,Heidenreich:2015nta,Heidenreich:2016aqi,Montero:2017yja}. 
This criterion implicitly assumes that the size of the compactification can be chosen freely. While this assumption may hold if supersymmetry is preserved in the lower dimensional theory\footnote{Even in this case, non-perturbative corrections can generate a potential for the moduli.}, it may not hold for non-supersymmetric theories.
Therefore, it is important to explore the conditions under which lower dimensional vacua exist.
Furthermore, it has recently been conjectured that non-supersymmetric $AdS$ vacua are unstable~\cite{Ooguri:2016pdq,Freivogel:2016qwc}.
This conjecture is based on the picture that $AdS$ vacua can be identified as the near horizon limit of the extremal black hole.\footnote{This point of view was emphasized in Ref.~\cite{Danielsson:2016mtx}.} 
Ref.~\cite{Ooguri:2016pdq} further pointed out that their argument rules out minimal Majorana neutrino masses for the SM if they give rise to stable non-supersymmetric $AdS_3$ vacua, and thus suggested a novel connection between the weak gravity conjecture and neutrino physics.
Key to this line of arguments is an understanding of the vacuum structure and the possible sources of instabilities.
Our work therefore sets the stage for a systematic investigation of this picture at the quantum level.
We found that the moduli potential in some cases develops a runaway behavior 
in the small compactification scale region ($\lesssim$ GeV$^{-1}$).
Therefore, even though there exist two- and three-dimensional AdS critical points of the SM landscape, we argue that 
these candidate vacua may be subject to quantum tunneling instabilities.

Another motivation for investigating the SM landscape is to explore the implications of the multiple point criticality principle~\cite{Froggatt:1995rt,Bennett:1996hx}, see also App.~D of Ref.~\cite{Hamada:2015ria} for a review and Refs.~\cite{Hamada:2014ofa,Hamada:2014xra,Hamada:2015dja} for its possible interpretations. 
This principle requires that the parameter of the theory to be tuned so that there are multiple vacua that are degenerate in energy.
Based on this principle, Froggatt and Nielsen~\cite{Froggatt:1995rt,Nielsen:2012pu} predicted the mass of the Higgs boson in 1995.
In the present work, we show that the mass of the lightest neutrino may also be predicted using this principle, by requiring the $3$-dimensional vacuum is close to the flat vacuum.
Thus, while it is unclear whether the aforementioned 2- and 3-dimensional AdS vacua in the SM landscape are stable, we still find an
intriguing constraint between the neutrino mass and the observed cosmological constant based on a rather general principle that had some success in a different particle physics context.

Finally, our investigation of the SM landscape provides a starting point to discuss the possibility of dynamical compactification of the SM~\cite{Carroll:2009dn}, which may determine the final fate of our universe. 

This paper is organized as follows.
In Sec.~\ref{Sec:4d SM vacua}, we review the $4$ dimensional SM Higgs vacua.
In Sec.~\ref{Sec:SM S1 vacua} and Sec.~\ref{Sec:SM T2 vacua}, we present our results for $S^1$ and $T^2$ compactifications of the SM.
We summarize our findings in Sec.~\ref{Sec:summary}.
Some detailed calculations are relegated to the appendices.
For convenience, we summarize the models which will be analyzed in this paper in Table~\ref{Table:Introduction}.

\begin{table}
  \begin{center}
  \scalebox{0.9}[0.9]{  
    \begin{tabular}{|c||c|c|} \hline
		& model 			                          & Reference, Section 						\\ \hline
		& $U(1)$, neutral  		            	 & Sec.~\ref{Sec:S1 U(1) neutral} 				\\ 
		& $U(1)$, charged   				 & Sec.~\ref{Sec:S1 U(1) charged} 				\\ \cline{2-3}
		& SM, $\nu_M$ 				 & Ref.~\cite{ArkaniHamed:2007gg},  	Sec.~\ref{Sec:S1 SM}  			\\ 	
$S^1$	& SM, $\nu_D$         			  	& Ref.~\cite{ArkaniHamed:2007gg},  	Sec.~\ref{Sec:S1 SM}  			\\
		& SM, $\nu_M$, high scale	  	& Sec.~\ref{Sec:S1 SM}						\\	
		& SM, $\nu_D$, high scale  		& Sec.~\ref{Sec:S1 SM}						\\ \cline{2-3}						
		& axion  						& Sec.~\ref{Sec:S1 flux vacua}					\\ \hline
		& $U(1)$, neutral 				& Sec.~\ref{Sec:T2 U(1) charged}				\\
		& $U(1)$, charged 				& Sec.~\ref{Sec:T2 neutral}					\\ \cline{2-3}
$T^2$	& SM, $\nu_M$ 				& Ref.~\cite{Arnold:2010qz}, Sec.~\ref{Sec:T2 SM}	\\
		& SM, $\nu_D$, NH  				& Ref.~\cite{Arnold:2010qz}, Sec.~\ref{Sec:T2 SM}	\\
		& SM, $\nu_D$, IH  				& Ref.~\cite{Arnold:2010qz}, Sec.~\ref{Sec:T2 SM}	\\ \cline{2-3}	
		& axion 						& Sec.~\ref{Sec:T2 flux vacua}					 \\ \hline
\end{tabular}
}
\end{center}
\caption{
The models which will be analyzed in this paper. Related earlier works are also shown.
}
\label{Table:Introduction}
\end{table}

\subsection*{Note added}
While this work was being written, Ref.~\cite{Ibanez:2017kvh} appeared
where the constraints on the neutrino mass and the cosmological constant from the weak gravity conjecture were considered.

\section{The SM vacua in four dimension}\label{Sec:4d SM vacua}
In this section, we review the SM vacua in four dimensions with the current experimental values of the SM parameters, see e.g. Ref.~\cite{Hamada:2012bp,Buttazzo:2013uya,Hamada:2014wna} for details.
The Higgs potential is written as
\al{\label{Eq:Higgs potential}
V_H=-m^2|H|^2+\lambda\paren{|H|^2}^2,
}
and our electroweak vacuum corresponds to
\al{
\vev{|H|^2}={m^2\over2\lambda}.
}

At a high scale compared with the electroweak one, we can neglect the quadratic term in the potential, and obtain
\al{
V_H=\lambda_\text{eff}(\mu){h^4\over4}+c_6{h^6\over M_P^2}+...,
}
where $\lambda_\text{eff}$ is the effective quartic coupling which includes the quantum corrections to the Higgs potential, $M_P$ is the reduced Planck scale,
$h=\sqrt{2}\vev{H}$ is the physical Higgs field, and $\mu$ is the renormalization scale.
Usually, $\mu=h$ is taken in order to optimize the log term in the quantum correction.
The Planck suppressed term represents the effect of gravity.

\begin{figure}
\begin{center}
\hfill
\includegraphics[width=.49\textwidth]{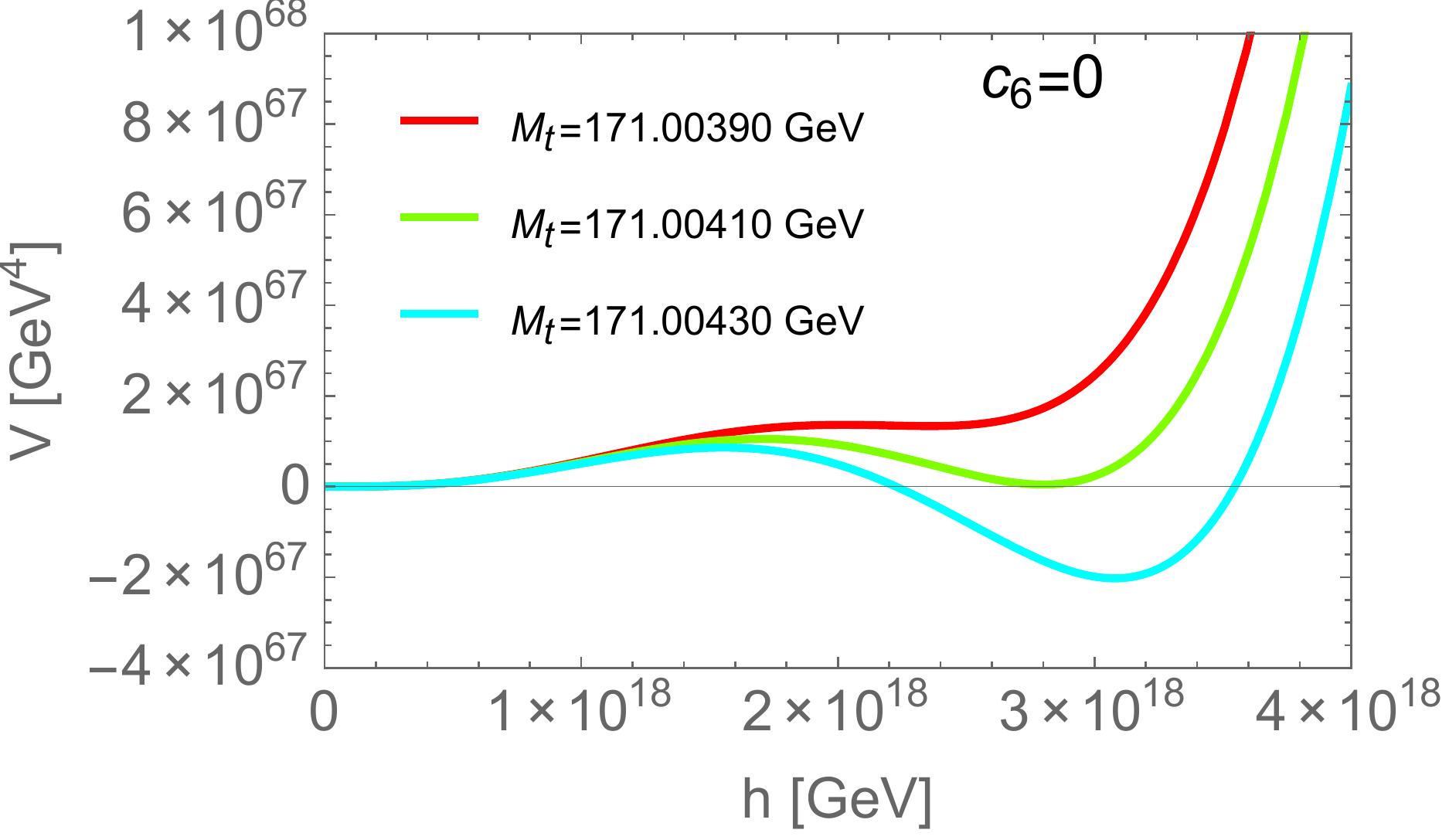}
\hfill
\includegraphics[width=.49\textwidth]{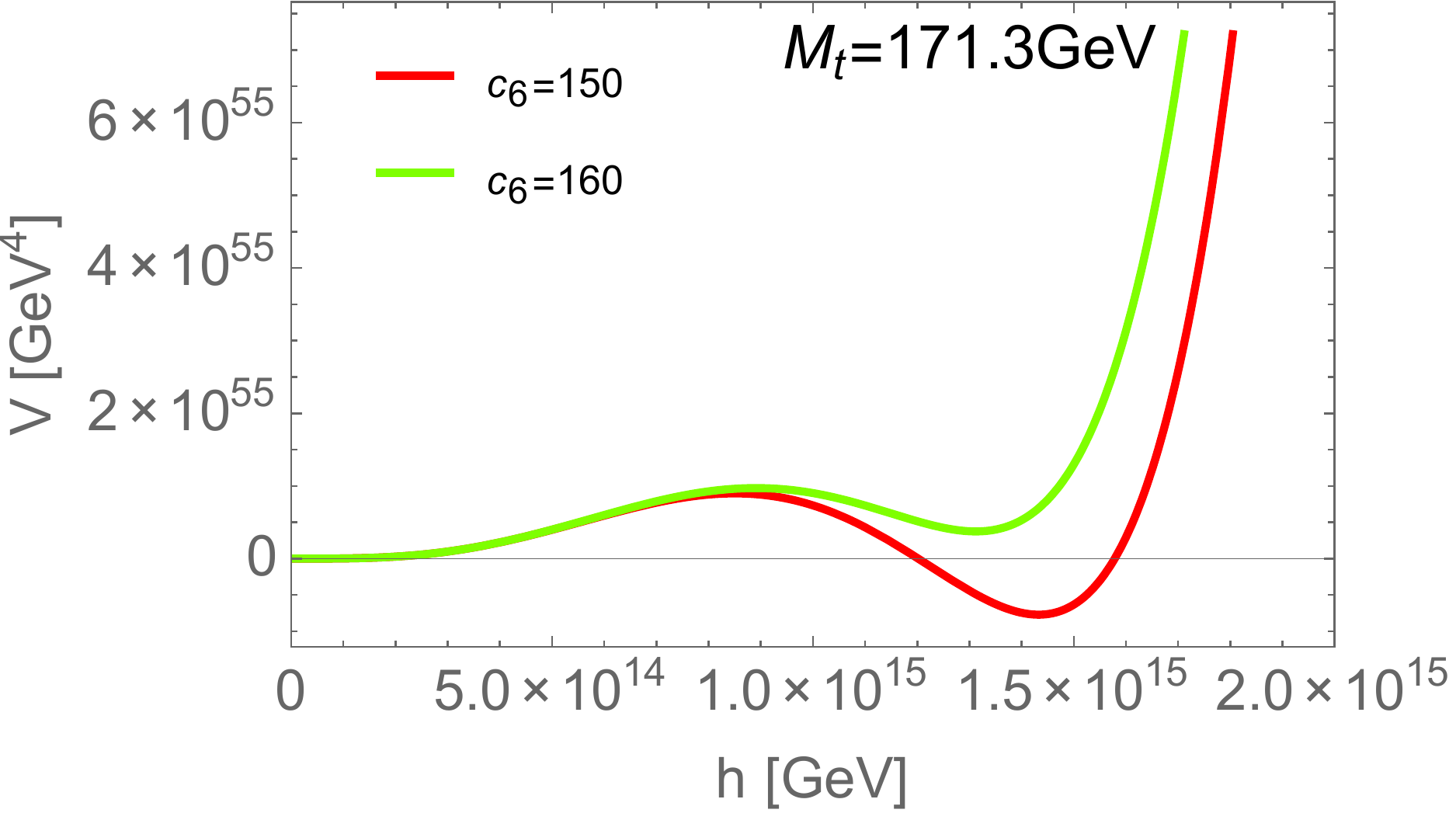}
\hfill\mbox{}
\end{center}
\caption{The $4$-dimensional Higgs potential as a function of  the physical Higgs field $h$, Eq.~\eqref{Eq:Higgs potential}.
In the left panel, we put $c_6=0$. The potential has AdS, flat or dS vacua depending on the value of the top mass.
In the right panel, the $c_6$ term is included while $M_t$ is fixed. We again have AdS, flat or dS minima corresponding to the value of $c_6$.
}
\label{Fig:Higgs potential}
\end{figure}

Interestingly, the current values of the SM parameter indicates the existence of a new vacuum at the high scale.\footnote{
Here we assume that the SM plus Einstein gravity is valid up to the high scale. 
Any new physics beyond the SM may change the structure of the second minimum, or eliminate it altogether. 
} 
In Fig.~\ref{Fig:Higgs potential}, we plot the Higgs potential as a function of $h$.
Depending on the mass of the top quark and the value of $c_6$, the cosmological constant of the high scale vacuum can be positive, zero or negative.

In the following, we consider compactification of the SM where the Higgs field takes either the electroweak scale or the vacuum value at the high scale. 

Before going to the analysis of the compactification, we would like to comment on the relation between the 4D SM vacua and the conjecture that all non-supersymmetric $AdS$ vacua are unstable~\cite{Ooguri:2016pdq,Freivogel:2016qwc}.
If the high scale vacuum of the SM has a negative cosmological constant and is stable, it would seem to be at odd with the weak gravity conjecture.
 This may indicate an interesting connection between the Higgs potential and the weak gravity conjecture, which we plan to investigate in future work.

\section{The SM vacua from $S^1$ compactification}\label{Sec:SM S1 vacua}
In this section, we consider the compactification of the SM on $S^1$.
We calculate the one-loop effective potential, and investigate the vacuum structure.
\subsection{Effective action}
Let us consider the situation where the SM is compactified on $S^1$.
First, the four dimensional action is
\al{
S=\int d^4x \sqrt{-g}\paren{{1\over2}M_P^2R-\Lambda_4-V_{S^1}^\text{all}-{1\over4}F_{\mu\nu}F^{\mu\nu}+...},
}
where $L$ is the radion field of $S^1$, $F_{\mu\nu}$ is the field strength of the $U(1)$ field, and $\Lambda_4$ is the cosmological constant of the four dimensional theory.
We adapt the mostly positive metric convention.
In our universe, we have $\Lambda_4\simeq3.25\times10\meV^4$.
If we consider the high scale vacuum in four dimensions, $\Lambda_4$ can take other values.
We also add $V_{S^1}^\text{all}$, the one-loop Casimir energy, for later convenience. 
The remaining terms include the Higgs boson, fermions and the $SU(3)\times SU(2)$ gauge fields.

Since the radius of $S^1$ is denoted by $L$, the volume of the compactified space is $2\pi L$, and so the momentum is quantized as $2\pi n/L$.
The metric of this $S^1$ compactification is
\al{
ds^2=\paren{g_{ij}+L^2 A_iA_j}dx_idx_j+2L^2 A_idx_idx_3+L^2\paren{dx_3}^2,
}
where $x_3$ is the compactified dimension, $0\leq x_3\leq2\pi$, $A_i$ is the graviphoton, and $i, j=0, 1, 2$.
Then, we have the following decomposition:
\al{
\det\paren{-g_{\mu\nu}}=L^2\det\paren{-g_{ij}},\quad R=R^{(3)}-2{1\over L}\nabla^2 L-{1\over4}L^2 F_{\mu\nu}F^{\mu\nu},
}
where $\mu, \nu=0,1,2,3$, $R^{(3)}$ is the Ricci scalar constructed from $g_{ij}$.
The dimensional reduction yields
\al{
S&=\int d^3x \sqrt{-g^{(3)}}(2\pi L)\sqbr{{1\over2}M_P^2\br{R^{(3)}-2{1\over L}\nabla^2 L-{1\over4}L^2 F_{\mu\nu}F^{\mu\nu}}-\Lambda_4-V_{S^1}^\text{all}}
\nn
&=\int d^3x \sqrt{-g^{(3)}}(2\pi L)\sqbr{{1\over2}M_P^2\br{R^{(3)}-{1\over4}L^2 F_{\mu\nu}F^{\mu\nu}}-\Lambda_4-V_{S^1}^\text{all}},
}
where the total derivative is omitted in the last equality.
Performing the redefinition of the metric $g_{ij}=(2\pi L/L_0)^{-2}g^E_{ij}$, we obtain\footnote{$L_0$ is introduced in order to keep $g_{ij}$ dimensionless.}
\al{
\sqrt{-g^{(3)}}&=(2\pi L/L_0)^{-3}\sqrt{-g^{E(3)}},\quad
R^{(3)}=(2\pi L/L_0)^2
\br{
R^{E(3)}+4\nabla^2\ln(2\pi L)-2{g^{Eij}\partial_i L \partial_j L\over L^2}
},\nn
g^{ij}\partial_i L \partial_j L&=(2\pi L/L_0)^2g^{Eij}\partial_i L \partial_j L,\quad
g^{\mu\rho}g^{\nu\sigma}F_{\mu\nu}F_{\rho\sigma}=(2\pi L/L_0)^4g^{E\mu\rho}g^{E\nu\sigma}F_{\mu\nu}F_{\rho\sigma}.
}

Note that the formula for $D$-dimensional Weyl transformation is
\al{
\tilde{R}=e^{-2\omega}\br{R-2(D-1)\nabla^2\omega-(D-2)(D-1)\partial_\mu\omega\partial^\mu\omega},
}
where $\tilde{R}$ and $R$ are constructed by $\tilde{G}_{\mu\nu}=e^{2\omega}G_{\mu\nu}$ and $G_{\mu\nu}$, respectively.

The resultant action is
\al{
S
&=\int_{x_{3d,E}} \sqbr{{1\over2}L_0M_P^2\br{R^{E(3)}-2{g^{Eij}\partial_i L \partial_j L\over L^2}-\paren{2\pi L\over L_0}^2{1\over4}L^2 F_{\mu\nu}F^{\mu\nu}}-{L_0^3\Lambda_4\over(2\pi L)^2}-{L_0^3V_{S^1}^\text{all}\over(2\pi L)^2}}\nn
&=\int_{x_{3d,E}}\sqbr{{1\over2}L_0M_P^2R^{E(3)}-L_0M_P^2{g^{Eij}\partial_i L \partial_j L\over L^2}-{L_0M_P^2\over8}\paren{2\pi L\over L_0}^2L^2 F_{\mu\nu}F^{\mu\nu}-{L_0^3\Lambda_4\over(2\pi L)^2}-{L_0^3V_{S^1}^\text{all}\over(2\pi L)^2}},
}
where $\int_{x_{3d,E}}:=\int d^3x \sqrt{-g^{E(3)}}$.
Furthermore, by performing $A_i\to {1\over \sqrt{2}\pi M_P L_0}B_i$ and denoting the field strength for $B_i$ by $B_{ij}$, we arrive at
\al{\label{Eq:3D Einstein frame}
S
&=\int_{x_{3d},E} (L_0)\sqbr{{1\over2}M_P^2R^{E(3)}-M_P^2{g^{Eij}\partial_i L \partial_j L\over L^2}-{1\over4}\paren{L\over L_0}^4 B_{ij}B^{ij}-{\Lambda_4 L_0^2\over(2\pi L)^2}-{V_{S^1}^\text{all} L_0^2\over(2\pi L)^2}},
}
which agrees with Ref.~\cite{ArkaniHamed:2007gg}.

Let us calculate the one-loop correction to the effective potential.
The procedure is the same as that of thermal effective potential, see Apps.~\ref{App:curved space} and~\ref{App:T2 calculation} for the details.
As a result, we obtain\footnote{At the next order in perturbation theory, we may need to include the effect of the ring (or daisy) diagram, which will be presented elsewhere.}~\cite{ArkaniHamed:2007gg}
\al{\label{Eq:S1 result}
{V_{S^1}^\text{all}\over(2\pi L)^2}&=\sum_{\text{particle}}(-1)^{2s_p}n_p {V_{S^1}^{(1)}\paren{L,M_p,q_pA_\phi+{1-z_p\over2}}\over(2\pi L)^2},\nn
V_{S^1}^{(1)}(L,M,\theta)&=-{M^4\over2\pi^2}\sum_{n=1}^\infty {\cos(2\pi n\theta)\over(2\pi L M n)^2}K_2(2\pi L M n),\nn
V_{S^1}^{(1)}(L,0,0)&=-{1\over360 L^4}{1\over(2\pi)^2},
\quad
V_{S^1}^{(1)}(L,0,1/2)={7\over2880 L^4}{1\over(2\pi)^2},
}
as the one-loop Casimir energy.
Here $M_p$ is the mass of the particle, $s_p$ is the spin of the particle, $n_p$ is the number of degrees of freedom of the particle, $A_\phi$ is the Wilson line modulus and $z_a$ is the boundary condition of the particle which we discuss below. $z_a=0$ and $1$ correspond to anti-periodic and periodic boundary conditions, respectively.
Now we can see that
\al{
{L_0^2\over(2\pi L)^2}\paren{\Lambda_4+{V_{S^1}^\text{all}}}
}
is the Einstein frame effective potential in $3$ dimensions.

Note that the canonically normalized field $\chi$ is related to $L$ by the relation
\al{
L=e^{-{\chi\over M_P\sqrt{L_0}}}.
}

\subsection{boundary condition}
In order to define the theory on a compactified spacetime, we have to specify the boundary condition of each field as well as the action.
The restriction is the requirement of the single valuedness of the action, from which one can see that the gauge boson should be periodic because the covariant derivative term is linear in the gauge field.
Similarly, the graviton should be periodic because the Einstein Hilbert term behaves as
\al{
\int d^4x\sqrt{-g}R\to\int d^4x\sqrt{-g}R e^{i\alpha},
}
under $g_{\mu\nu}\to e^{i\alpha} g_{\mu\nu}$.

On the other hand, fermions can have non-trivial boundary condition (spin structure):
\al{
\psi_\text{lepton}(x_3+2\pi L)&= 
\begin{cases}
			\pm\psi_\text{lepton}(x_3)	&	
				\text{for Majorana neutrino,}\\ \\
			e^{iQ_L}\psi_\text{lepton}(x_3) &	
				\text{for Dirac neutrino.}		
		\end{cases}\nn
\psi_\text{baryon}(x_3+2\pi L)&= e^{iQ_B}\psi_\text{baryon}(x_3).		
}
These correspond to the symmetries of the classical action, $U(1)_L$ and $U(1)_B$, respectively. 
In terms of Eq.~\eqref{Eq:S1 result}, the fermion behaves as
\al{
\psi(x_3+2\pi L)=e^{2\pi\paren{q_pA_\phi+{1-z_p\over2}}}\psi(x_3).
}

\subsection{$U(1)$ gauge theory on $S^1$}
\subsubsection{with charged matter}\label{Sec:S1 U(1) charged}
Before we get into the complicated structure of the SM, it is instructive as a warmup exercise to first analyze the vacuum structure of a $U(1)$ gauge theory.
The field content includes a charged Dirac fermion as well as a $U(1)$ gauge field.
The one-loop potential is given by
\al{\label{Eq:charged S1 potential}
V_{S^1}^\text{charged}=
{L_0^2\over(2\pi L)^2}\br{\Lambda_4-{1\over180L^4(2\pi)^4}-4V_{S^1}^{(1)}\paren{L,M_e, q_e A+{1-z_e\over2}}}
}
where $M_e, q_e$ are the mass and charge of the fermion, $z_e$ is the boundary condition of the fermion and $A$ is the $U(1)$ Wilson line.
The second and third terms correspond to the photon and charged matter contributions, respectively.

We recall that $L$ is not the canonically normalized field.
However, the extrema of the potential in term of $L$ corresponds to extrema in terms of canonically normalized field $\chi$ because $\p_\chi V\propto \p_L V$.
In this sense, the potential in terms of $L$ is useful.
Moreover, the curvature of the potential is obtained by
\al{
{\p^2 V\over\p\chi^2}=
\paren{\p L\over \p \chi}^2 {\p^2 V\over \p L^2}+{\p L\over\p\chi}\paren{{\p\over\p L}{\p L\over \p\chi}}{\p V\over\p L} 
={1\over M_P^2 L_0}\paren{L^2{\p^2 V\over\p L^2}+L{\p V\over\p L}}.
}
Therefore, at the extreme $\p_L V= \p_\chi V=0$, the positive curvature condition $\p_\chi^2V>0$ is equivalent to the condition $\p_L^2 V>0$.

\begin{figure}
\begin{center}
\hfill
\includegraphics[width=.55\textwidth]{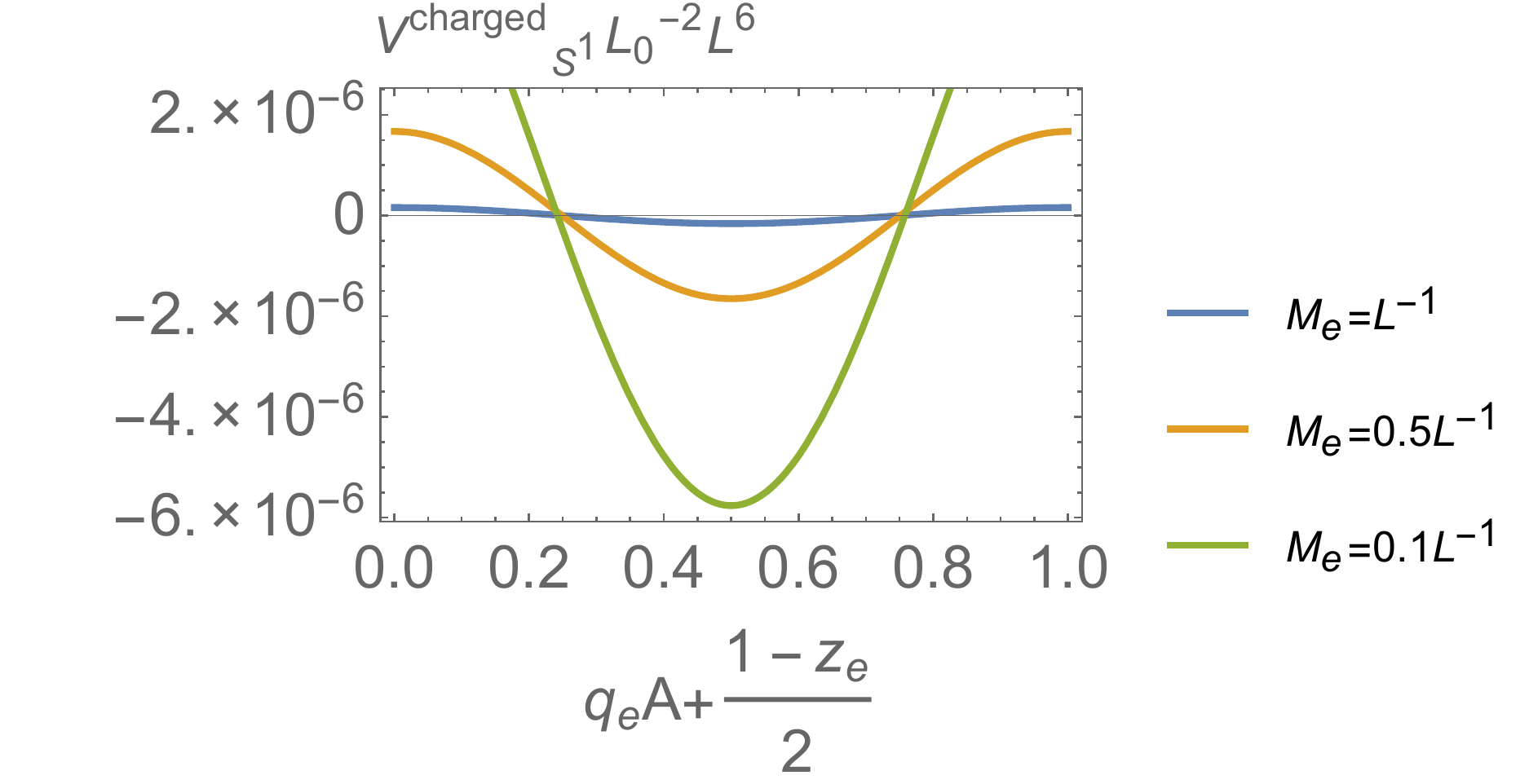}
\hfill
\includegraphics[width=.4\textwidth]{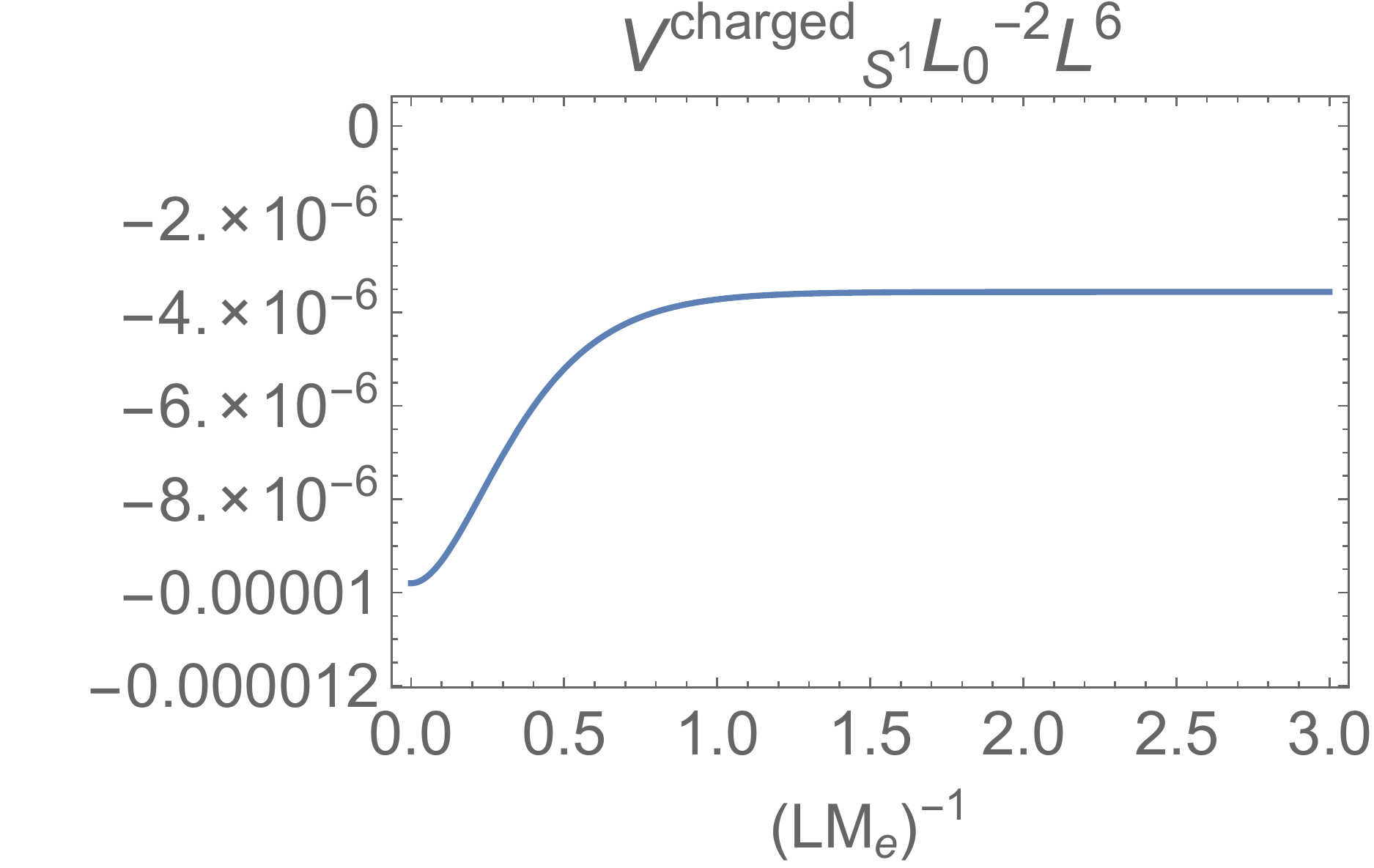}
\hfill\mbox{}
\end{center}
\caption{
Left: The potential of the $U(1)$ gauge theory with a charged Dirac fermion, Eq.~\eqref{Eq:charged S1 potential}, is plotted as a function of the Wilson line. The potential takes minimum at $q_e+(1-z_e)/2=1/2$.
Here we take $\Lambda_4=0$.
For the illustration, the vertical axis is not the potential itself, but the potential multiplied by $L_0^{-2}L^6$.
Right: The potential as a function of $L$, the radius of $S^1$. The value of the Wilson line is set to be at the minimum of the potential.
}
\label{Fig:QED_S1_charged_potential}
\end{figure}

In the left panel of Fig.~\ref{Fig:QED_S1_charged_potential}, we can numerically see that, when $q_e A+{1-z_e\over2}=1/2$, the potential $V$ takes its minimum with respect to $A$, and  $-4V_{S^1}^{(1)}$ takes negative value at the minimum.
Setting $q_e A+{1-z_e\over2}=1/2$, the potential for the $L$ field is plotted in the right panel of Fig.~\ref{Fig:QED_S1_charged_potential}.
No local minimum appears in the potential.\footnote{
We note that, in the figures, the potential is multiplied by $L^6$ for illustration. We need to be careful when we see the conditions $\p_L V=0$ and $\p_L^2 V>0$ from the figures.
}
This conclusion is valid if we add a four dimensional cosmological constant.

Therefore, there are no vacua in $S^1$ compactification of QED.
One may think that the Wilson line field need not be fixed at the minimum because tachyons are allowed if the three dimensional space is $AdS_3$.
As discussed in App.~\ref{App:vacuum condition S1}, this does not help. 
While the typical mass scale of Willson line is determined by compactification scale $L^{-1}$, the scale of Ricci curvature is $L^{-4}/M_P^2$.
Hence, as long as the compactification scale is below the Planck scale, the stability condition is effectively the same as that in flat spacetime.\footnote{
This is not obvious for $T^2$ compactifications, which we will see in Sec.~\ref{Sec:SM T2 vacua}.
}

\subsubsection{with neutral matter}\label{Sec:S1 U(1) neutral}
\begin{figure}
\begin{center}
\hfill
\includegraphics[width=.45\textwidth]{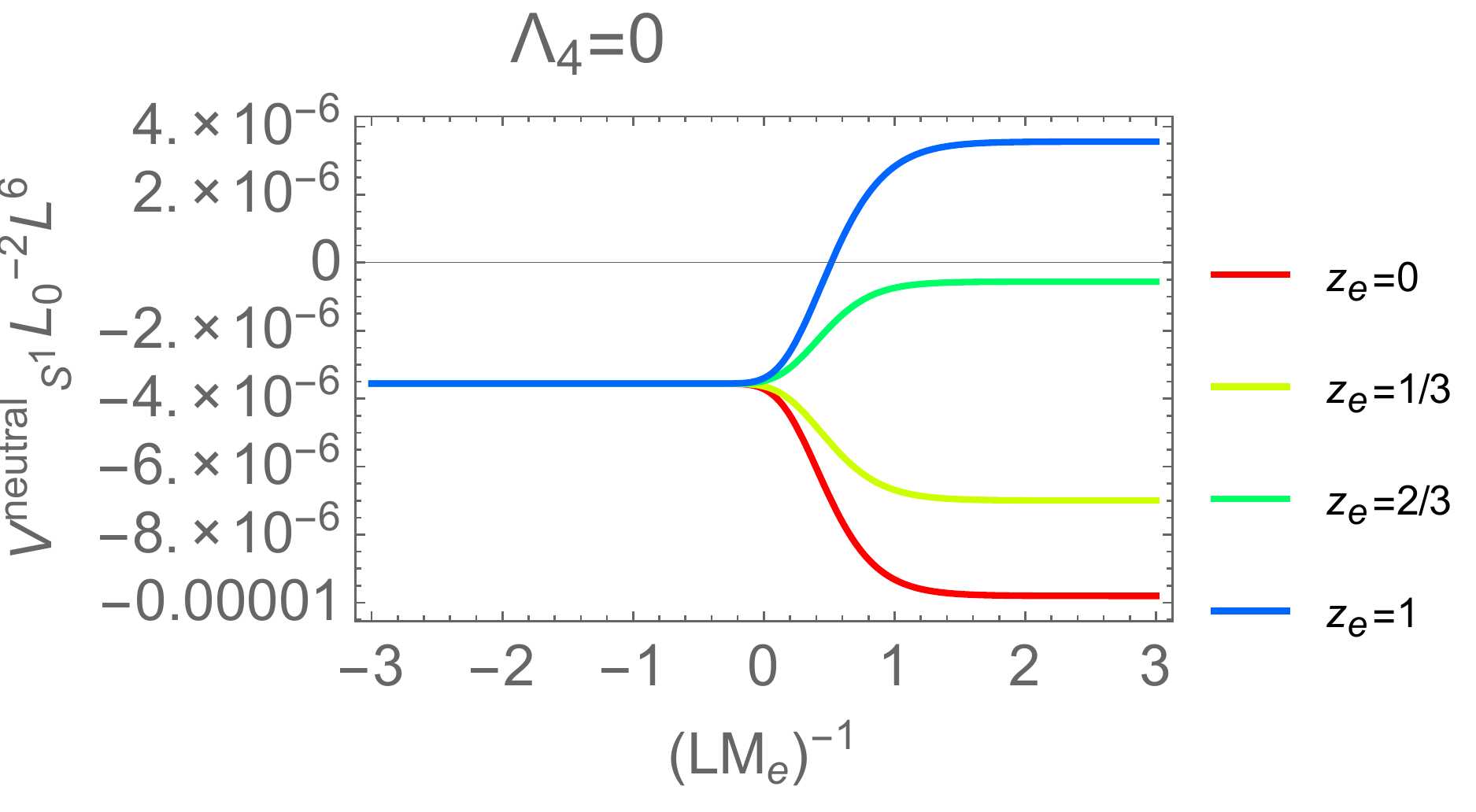}
\hfill
\includegraphics[width=.49\textwidth]{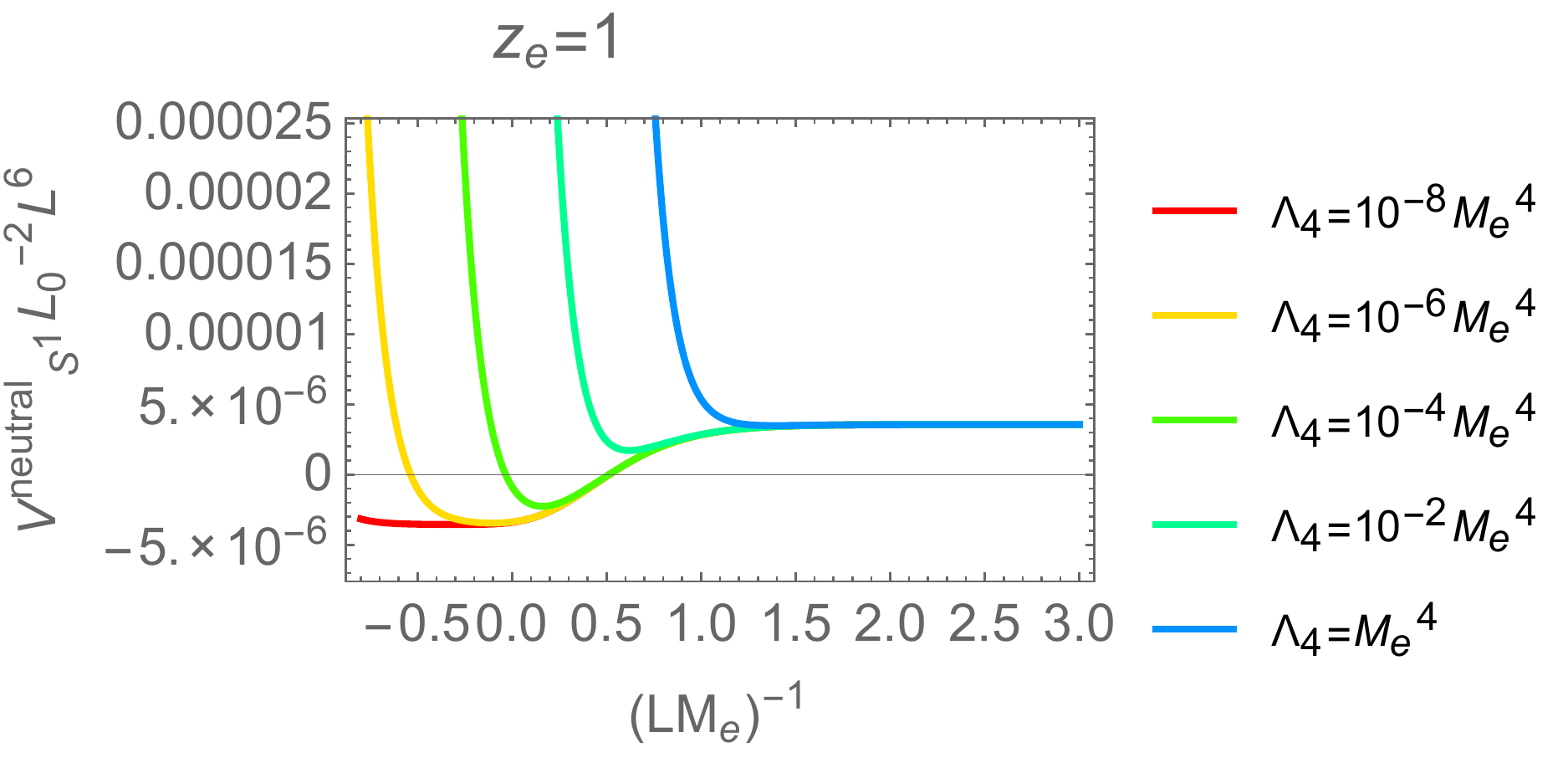}
\hfill\mbox{}
\end{center}
\caption{
The potential of compactified $U(1)$ gauge theory with neutral matter.
In the left figure, $\Lambda_4$ is set to be zero.
In the right figure, periodic boundary condition, $z_e=1$, is taken.
}
\label{Fig:QED_S1_neutral_potential}
\end{figure}

In contrast, compactified vacua can appear if the matter field is neutral under $U(1)$, where the potential is given by
\al{
V_{S^1}^\text{neutral}=
{L_0^2\over(2\pi L)^2}
\br{
\Lambda_4-{1\over180L^4(2\pi)^4}-4V_{S^1}^{(1)}\paren{L,M_e, {1-z_e\over2}}
}.
}
We can plot the potential as a function of $L$ for various value of $z_e$, which is shown in the left panel of Fig.~\ref{Fig:QED_S1_neutral_potential}. Here $\Lambda_4=0$ is taken.
We can see that, if the boundary condition is close to the periodic one, a stable vacuum appears.
In the right panel, we plot the potential for various $\Lambda_4$ with a fixed $z_e=1$.
If the value of  $\Lambda_4$ is small, the minimum corresponds to $AdS_3$. For the larger value of $\Lambda_4$, the vacuum becomes $M_3$ or $dS_3$ . This is shown in the right panel of Fig.~\ref{Fig:QED_S1_neutral_potential}.
The lower dimensional $AdS_3$, $M_3$ or $dS_3$ vacua are obtained for $\Lambda_4\lesssim10^{-2.8}M_e^4$, $\Lambda_4\simeq10^{-2.8}M_e^4$ and $10^{-2.8}M_e^4\lesssim\Lambda_4\lesssim 10^{-2.6}M_e^4$, respectively.

\subsection{SM on $S^1$}\label{Sec:S1 SM}
\begin{table}
  \begin{center}
    \begin{tabular}{|c||c|c|c|} \hline
particle & mass & $(-1)^{2s_p} n_p$ & $q_{U(1)_\text{EM}}$ \\ \hline \hline
graviton  & $0$  & $2$ &$0$ \\
photon & $0$  & $2$ &$0$ \\
$\nu$ & $\lesssim$\,0.1eV  & $-6$ or $-12$ &$0$ \\
$e$ &  $0.511$MeV & $-4$ &$-1$ \\
$\mu$ &  $100$MeV & $-4$ & $-1$\\
$\pi$& $140$MeV & $3$ & $(1,0,-1)$\\
$K$& $500$MeV & $4$ & $(1,0,-1)$\\
$\eta_8$& $550$MeV & $1$ & $0$\\
\hline
    \end{tabular}
  \end{center}
\caption{
The particle contents contributing to the Casimir energy below the GeV scale.
}
\label{Table:particle charge1}
\end{table}

\begin{table}
  \begin{center}
    \begin{tabular}{|c||c|c|c|c|c|} \hline
particle & mass & $(-1)^{2s_p} n_p$ & $q_{U(1)_\text{EM}}$ &  $q_{SU(3)1}$ & $q_{SU(3)2}$\\ \hline \hline
graviton  & $0$  & $2$ &$0$&$0$&$0$ \\
photon & $0$  & $2$ &$0$&$0$&$0$ \\
gluon & $0$  & $2$ &$0$&$(2,1,1,0,0,-1,-1,-2)$&$(1,1,0,0,0,0,-1,-1)$ \\
$\nu$ & $\lesssim$\,0.1eV  & $-6$ or $-12$ &$0$ &$0$&$0$\\
$e$ &  $0.511$MeV & $-4$ &$-1$&$0$&$0$ \\
$\mu$ &  $100$MeV & $-4$ &$-1$&$0$&$0$ \\
$u$ & $300$MeV & $-12$ &$2/3$&$(1,0,-1)$&$(1,1,-2)$ \\
$d$ & $300$MeV & $-12$ &$-1/3$&$(1,0,-1)$&$(1,1,-2)$ \\
$s$ & $300$MeV & $-12$ &$-1/3$&$(1,0,-1)$&$(1,1,-2)$ \\
$c$ & $300$MeV & $-12$ &$2/3$&$(1,0,-1)$&$(1,1,-2)$  \\
$\tau$& $1$GeV & $-4$ &$-1$&$0$&$0$ \\
$b$ & $3$GeV &$-12$ &$-1/3$&$(1,0,-1)$&$(1,1,-2)$ \\
$W$ & $80$GeV & $6$ &$(1,0,-1)$& $0$ & $0$\\
$Z$  & $90$GeV & $3$ & $0$ & $0$ & $0$\\
Higgs & $125$GeV & $1$ &$0$&$0$&$0$ \\
$t$ & $173$GeV & $-12$ &$2/3$&$(1,0,-1)$&$(1,1,-2)$ \\
\hline
    \end{tabular}
  \end{center}
\caption{
The particle contents contributing to the Casimir energy above the GeV scale.
}
\label{Table:particle charge2}
\end{table}

Next, let us move on to the vacuum structure of the SM.
The particle contents contributing to the Casimir energy in the SM are shown in  Tables~\ref{Table:particle charge1} and \ref{Table:particle charge2}. 
The potential of the standard model is given by
\al{\label{Eq:S1 SM potential}
V_{S^1}^\text{SM}=
{L_0^2\over(2\pi L)^2}\paren{\Lambda_4+V_{S^1}^\text{all}}.
}

In our calculation of the Casimir energy, the neutrino masses were chosen numerically as $m_2^2-m_1^2=7.53\times10^{-5}\eV^2, |m_3^2-m_2^2|=2.44\times10^{-3}\eV^2$~\cite{Gando:2013nba}.
The lightest neutrino mass, $m_{\nu,\text{lightest}}$, is $m_1$ for the normal hierarchy (NH), and is $m_3$ for the inverted hierarchy (IH).

\begin{figure}
\begin{center}
\hfill
\includegraphics[width=.49\textwidth]{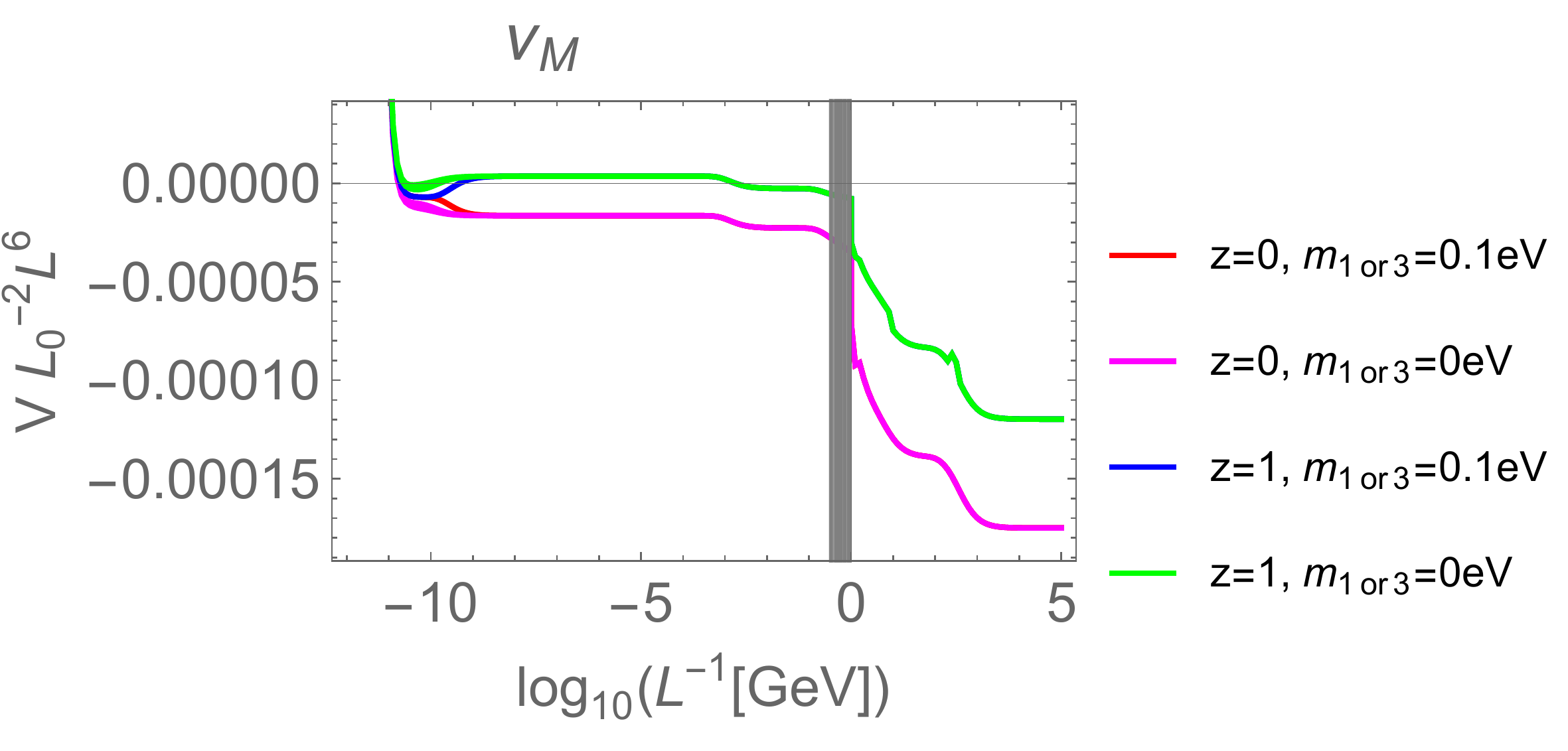}
\hfill
\includegraphics[width=.49\textwidth]{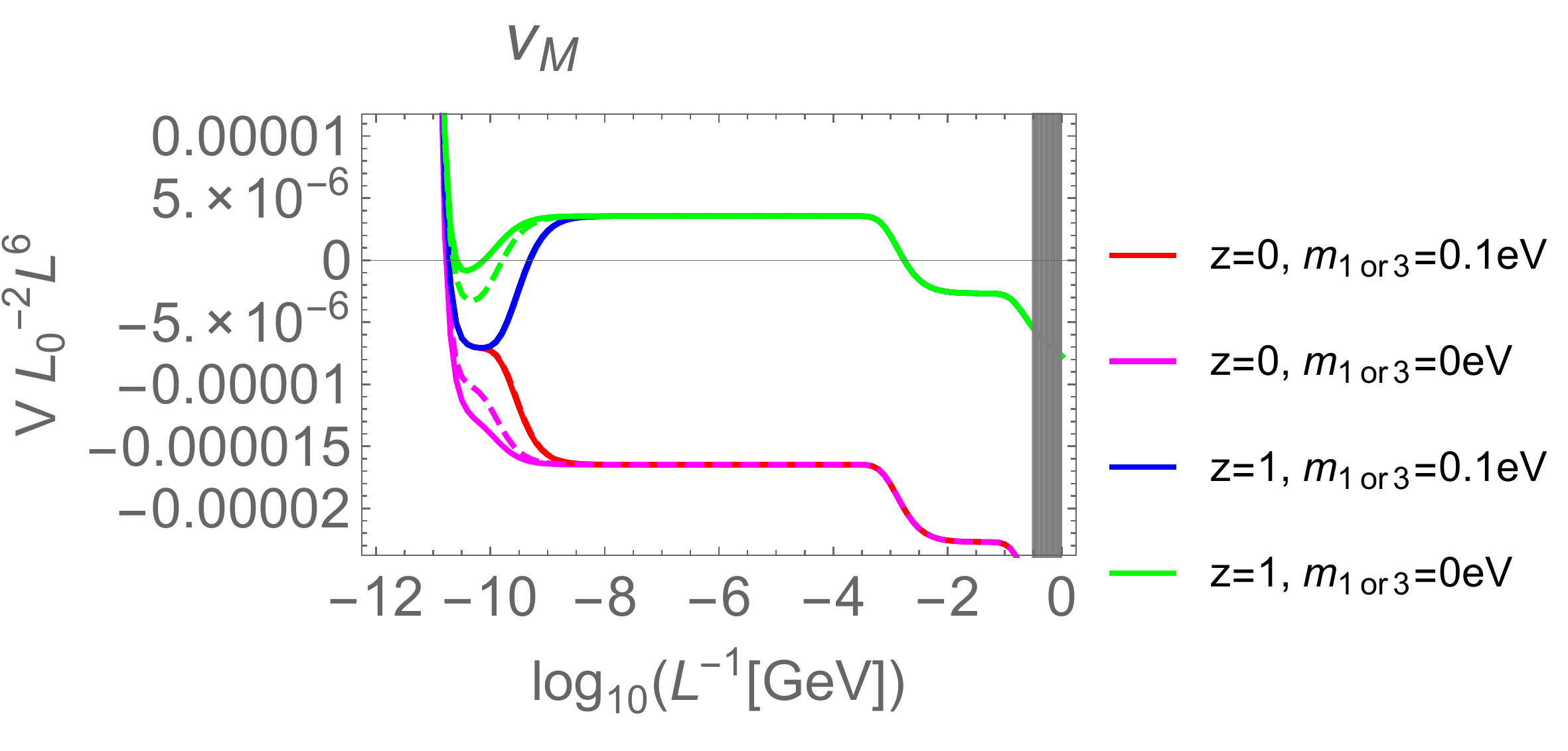}
\hfill\mbox{}
\\
\includegraphics[width=.49\textwidth]{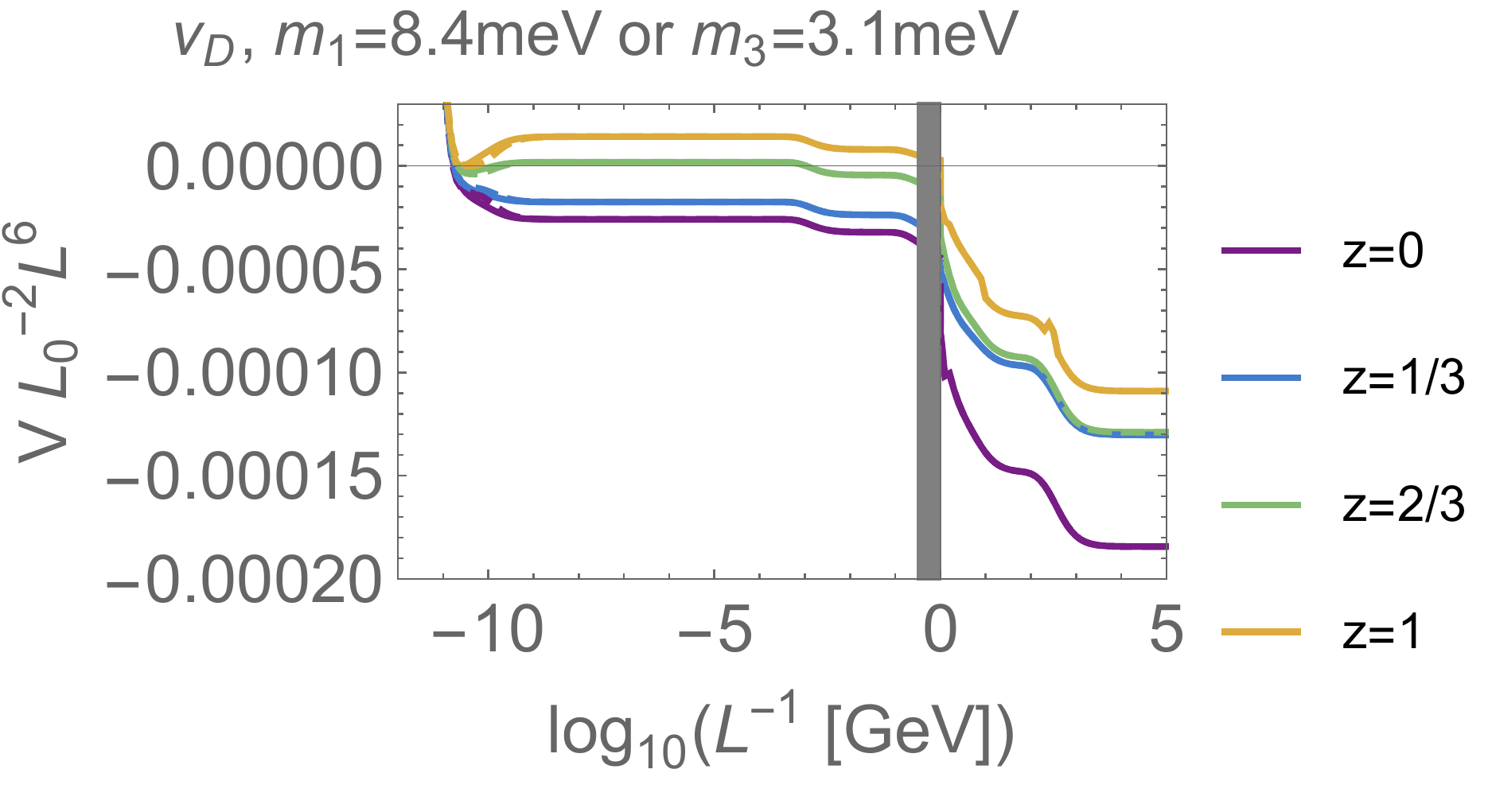}
\hfill
\includegraphics[width=.49\textwidth]{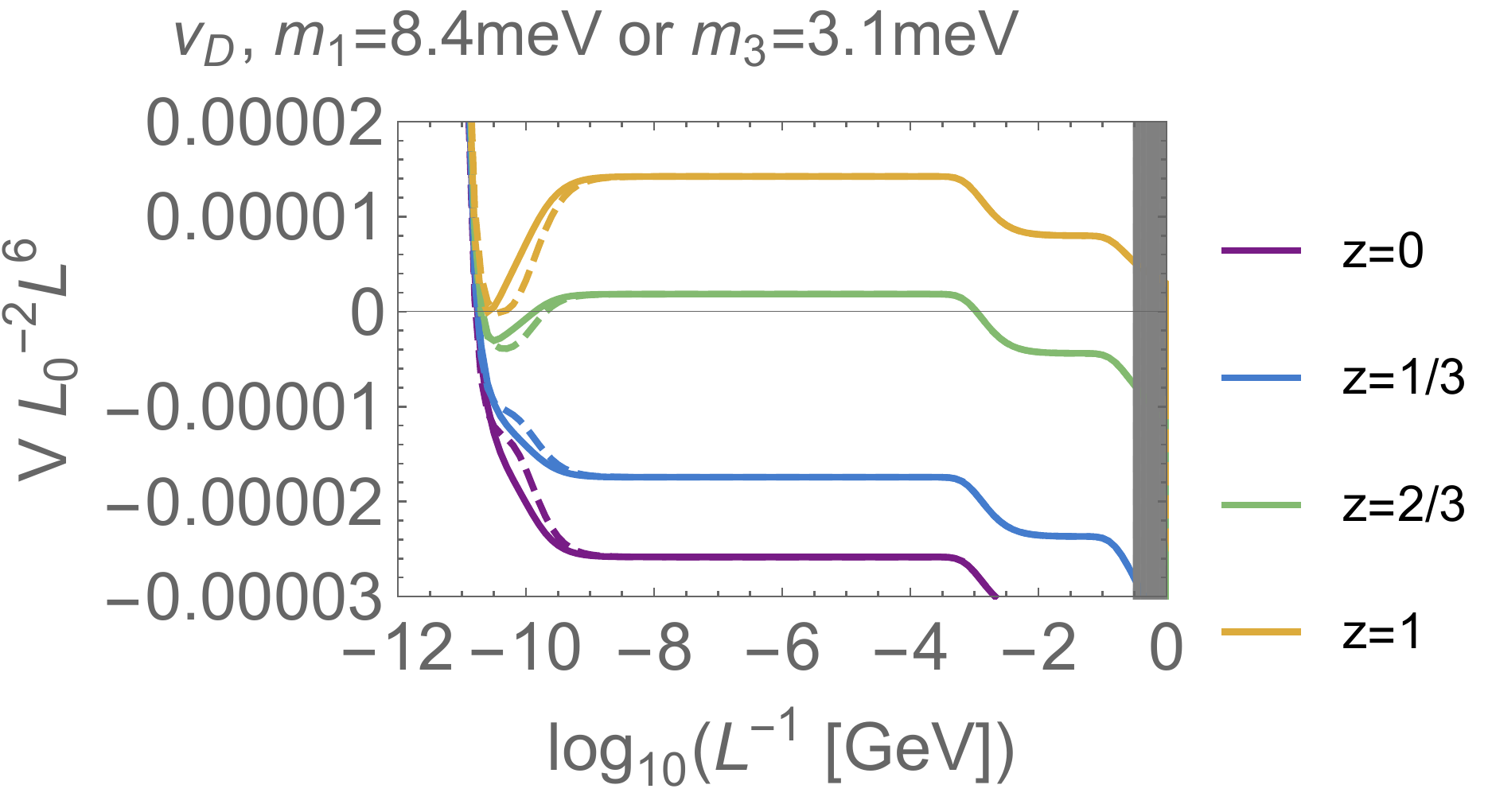}
\hfill\mbox{}
\end{center}
\caption{
$S^1$ compactification of the SM.
The effective potential as a function of the radion $L$. Here the Wilson lines are fixed at the potential minimum.
``$\nu_{M(D)}$" represents Majorana (Dirac) neutrino, and $z$ is the boundary condition of fermion $\psi\to -e^{-i\pi z}\psi$.
The shaded region is close to the QCD scale, $0.3\text{--}1$ GeV, around which perturbation theory is not good.
Right figures are enlarged view of the left figures.
The solid and dashed line correspond to the normal and inverted hierarchy, respectively.
We can see that there is vacuum at around the neutrino mass scale if the boundary condition is close to the periodic one.
}
\label{Fig:S1_result}
\end{figure}

We plot the Casimir energy as a function of $L$ in Fig.~\ref{Fig:S1_result}.
Below the QCD scale $\sim0.3$ GeV, we use the particle contents in Table~\ref{Table:particle charge1} while we use Table~\ref{Table:particle charge2} above $1$ GeV.\footnote{For simplicity, we neglect the effect of $SU(2)_L$ Wilson line, $w$, which would not change our result.
Neglecting the effect of $w$ is equivalent to fixing the value of  $w$ to be zero. If we consider the dynamics of the Wilson line moduli, we may find the true minimum which has an even smaller energy, but this only strengthens the runaway behavior, and our qualitative conclusion about the runaway behavior does not change.
}
The vertical axis is the height of the potential normalized by $L_0^2L^{^6}$.
The Wilson line moduli is fixed to be at the minimum of the potential.
The upper and lower figures correspond to Majorana and Dirac neutrinos respectively.
For simplicity, we take the same boundary condition for leptons and baryons.
It would be interesting to consider different boundary conditions.
The right figures are the enlarged view of the left figures.
We note that
the vertical axis of the figures is the potential multiplied by $L_0^{-2}L^6$, so one has to be careful in locating the stationary points from the figures.
For example, at the mass threshold of the electron $\sim 10^{-3}\GeV$, the vertical axis exhibits a step function-like behavior because of this normalization we have chosen, making it seem like there is a stationary point at that mass scale.
However, a stationary point exists only if the sign of the vertical axis changes at around the mass threshold.

\begin{figure}
\begin{center}
\hfill
\includegraphics[width=.49\textwidth]{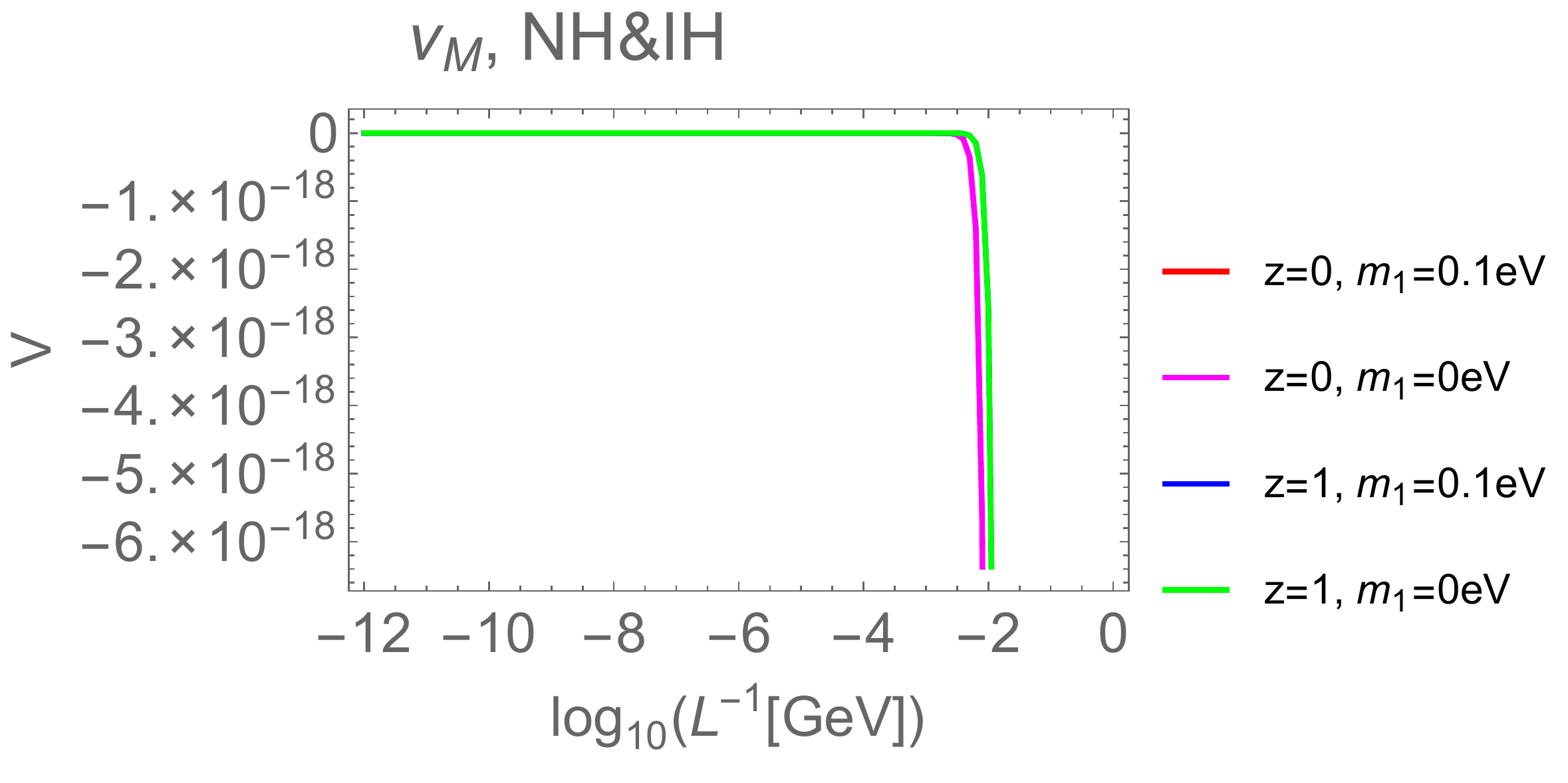}
\hfill
\includegraphics[width=.49\textwidth]{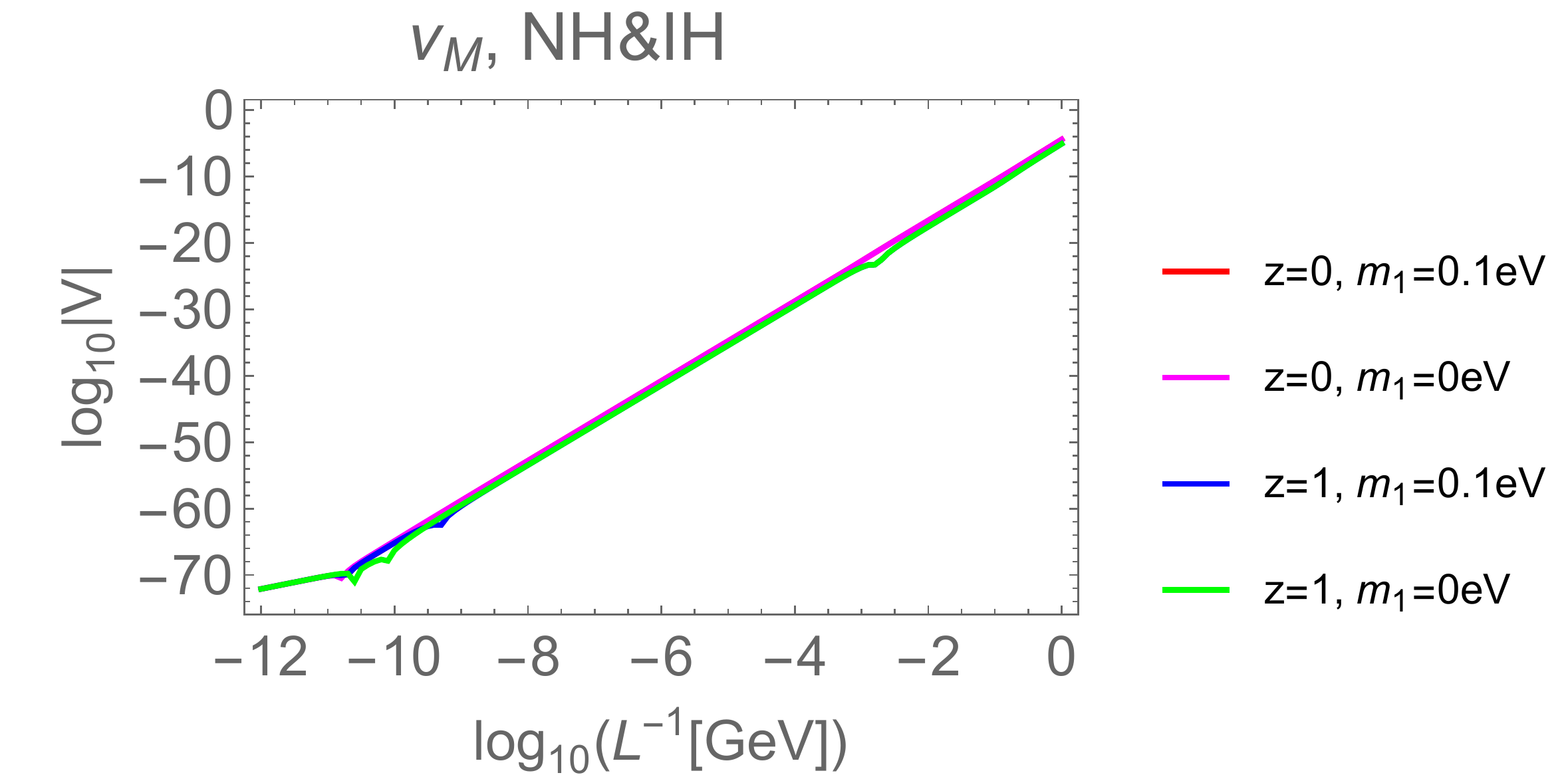}
\hfill\mbox{}
\end{center}
\caption{
The potential is the same as  the upper right panel of Fig.~\ref{Fig:S1_result}, but the vertical axes are $V$(left), and $\log_{10} |V|$(right), respectively. Here the scale $L_0$ is taken to be $1\GeV^{-1}$.
}
\label{Fig:V figures}
\end{figure}

\begin{figure}
\begin{center}
\hfill
\includegraphics[width=.55\textwidth]{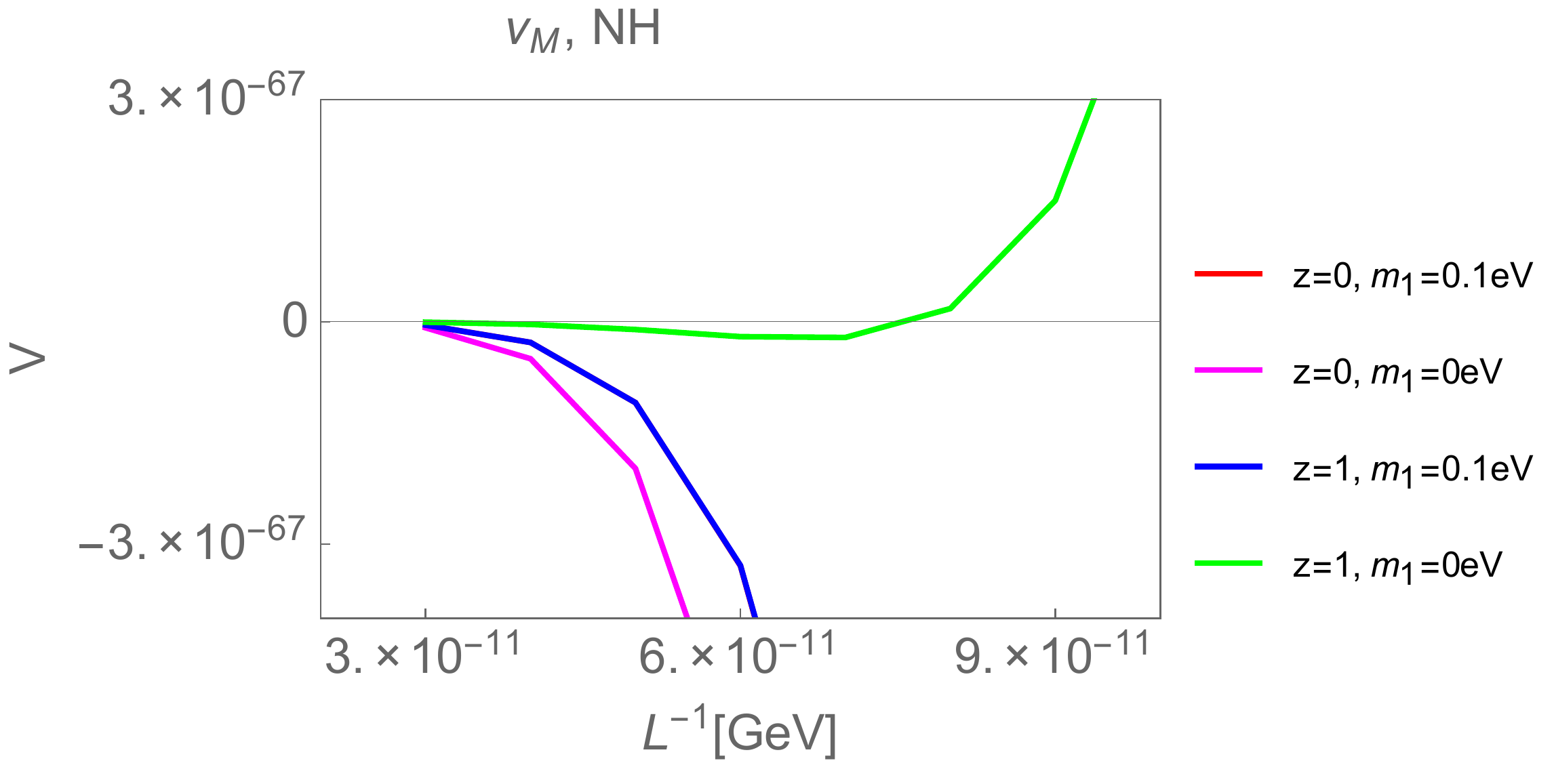}
\hfill
\includegraphics[width=.44\textwidth]{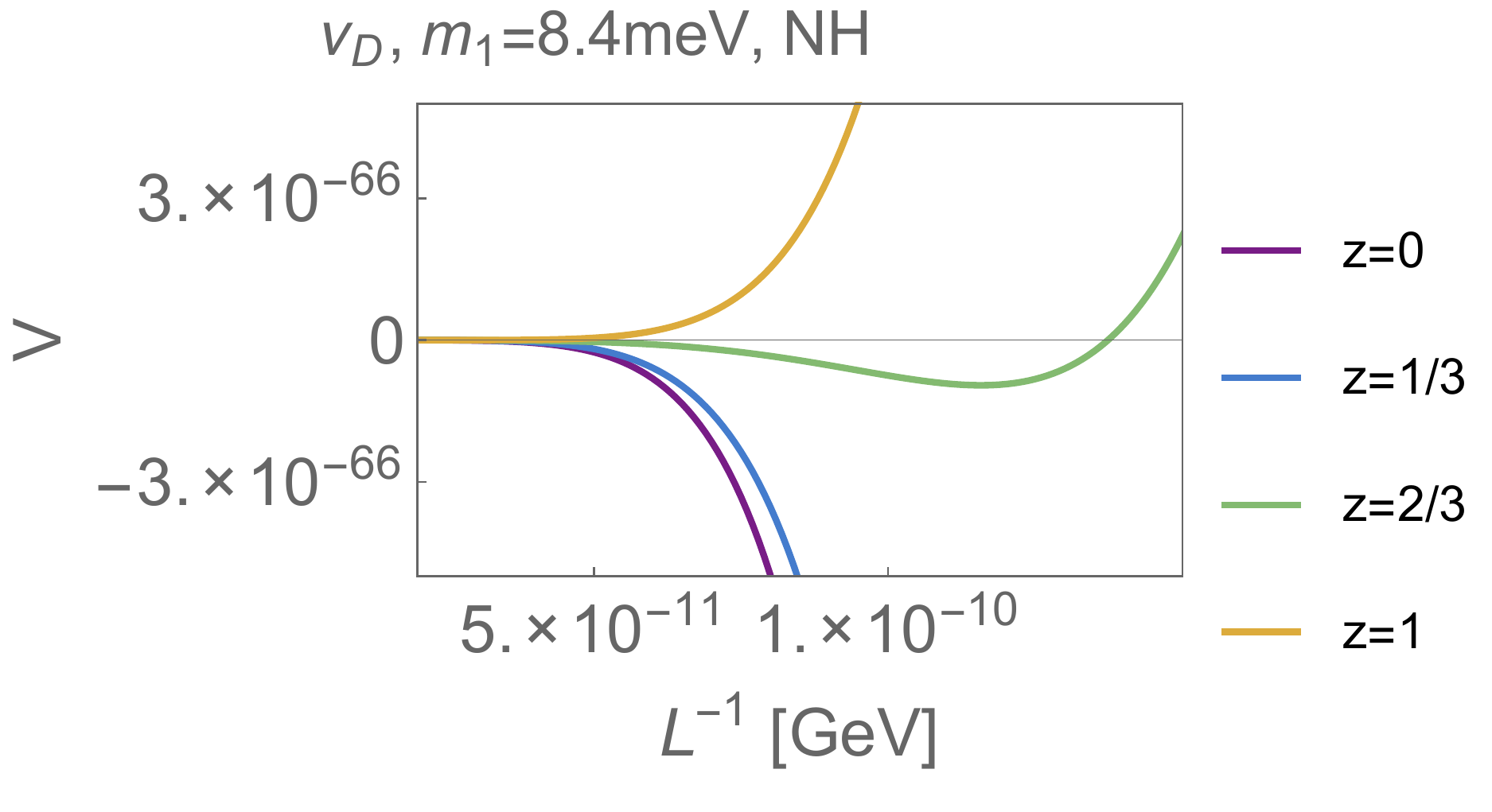}
\hfill\mbox{}
\end{center}
\caption{
The radion potential around the neutrino mass scale for NH. Here the scale $L_0$ is taken to be $1\GeV^{-1}$. The potential is same as that in Fig.~\ref{Fig:S1_result}, but the vertical axis is the potential $V$ itself, and the horizon axis is the small segment of $L^{-1}$ around the neutrino mass scale.
As in Fig.~\ref{Fig:S1_result}, for the Majorana case, the $z=1$ plots correspond to $AdS$ minima. For the Dirac case, $z=1$ and $z=2/3$ correspond to the flat and $AdS$ minima, respectively.
}
\label{Fig:S1 neutrino minima}
\end{figure}

The reader may wonder why we choose $L_0^{-2}L^6 V$ as the vertical axis rather than the potential $V$ itself. The reason is as follows. Since $V$ is very steep, it is unfortunately difficult to find its minima from the figure where $V$ itself is the vertical axis and the horizontal axis $L^{-1}$ covers the wide range of values that we consider. For example, if we try to draw the figure corresponding to the upper right panel in Fig.~\ref{Fig:S1_result} without the $L^6$ normalization, we obtain Fig.~\ref{Fig:V figures}. The left panel is a linear plot of $V$, and the right panel is a log plot of the absolute value of $V$. It is not easy to find the neutrino minimum from these figures. If we concentrate on a small segment of $L^{-1}$ which is close to neutrino minima, then a figure where the vertical axis is $V$
(which we show in Fig.~\ref{Fig:S1 neutrino minima}) is more illustrative of the features of the potential.
On the other hand, if one wants to see the full behavior of the potential for a wide range of $L^{-1}$, the figure with the $L_0^{-2}L^6$ normalization is more appropriate.

We also note how one can infer the existence of the neutrino minimum from Fig.~\ref{Fig:S1_result}. In Fig.~\ref{Fig:S1_result}, we have plotted the flat and the AdS neutrino minima. For the flat case, the minima of $L_0^{-2}L^6V$ is same as that of $V$ itself. For the $AdS$ minima, the point is that the sign of $L_0^{-2}L^6V$ is the same as that of $V$ itself. Then, if the sign of $L_0^{-2}L^6V$ changes as plus$\to$ minus$\to$ plus as we increase $L^{-1}$, then $V$ should follow the same sign change, and hence there should be an $AdS$ minimum. In this way, the existence of the $AdS$  minima is common for $L_0^{-2}L^6V$ and $V$ itself although the precise value of $L^{-1}$ corresponding to the minima is different. To summarize, the change of the sign of the vertical axis signals the existence of a stationary point.

We can see that, if the boundary condition is close to the periodic one, the potential has a minimum at around the neutrino mass scale, and this vacuum is likely to unstable under tunneling to the runaway vacuum at high energy scale because the potential behaves as $V\propto-L^{-6}$  at high scale, and the runaway vacuum has a smaller energy than the neutrino vacuum, see the left panel of Fig.~\ref{Fig:V figures}.\footnote{Whether this runaway behavior continues to smaller $L$ or becomes an extremum depends on the UV completion of the SM.
It would also be interesting to investigate the robustness of the runaway behavior by considering the contributions of new particles in various extensions of the SM.
}
We leave the construction of the concrete bounce solution describing the tunneling to a future work.
On the contrary, if it is found that the $AdS_3$ vacuum is stable\footnote{See, e.g. Sec.~4.2 of Ref.~\cite{Banks:2010tj} for the claim that tunneling to and from AdS space cannot occur.
Another possibility is that the bubble size is larger than the AdS length so the decay does not happen.
}, we can constrain the mass of the neutrino, and exclude the Majorana neutrino alone the lines of Refs.~\cite{Ooguri:2016pdq,Freivogel:2016qwc}.
Note that, since this vacuum requires a non-trivial spin structure of the fermion, it does not decay by the Witten's bubble of nothing~\cite{Witten:1981gj}.\footnote{
Even if the fermion has a non-trivial spin structure, the Witten bubble of nothing can happen if the fermion couples with the Wilson line and the boundary condition becomes anti-periodic by the background Wilson line value~\cite{Blanco-Pillado:2016xvf}. However, this subtlety does not change the arguments that follow as the neutrinos are uncharged under the Wilson line.
}
For Majorana neutrino, we show the results for $m_{\nu,\text{lightest}}=0$ and $0.1\eV$, where $m_{\nu,\text{lightest}}$ is the mass of the lightest neutrino. Both of them leads to an $AdS_3$ vacuum.
On the other hand, $m_{\nu,\text{1}}=8.4$ or $m_{\nu,\text{3}}=3.1$ meV is taken for the Dirac neutrino case, which give a flat $3$-dimensional vacuum with periodic neutrinos. The vacuum becomes $dS_3 (AdS_3)$ for smaller (larger) $m_{\nu,\text{lightest}}$.
Explicitly, $AdS_3$ is obtained for $8.4(3.1)\meV\lesssim m_{\nu,1(3)}$ and $dS_3$ is obtained for $7.3(2.5)\meV\lesssim m_{\nu,1(3)}\lesssim8.4(3.1)\meV$ for NH (IH). 
This result is independent of whether the hierarchy is normal or inverted.

In the analysis above, we take the Wilson line to be at the global minimum of the potential.
Here we examine the possibility of local minima of the Wilson line potential.
For a massless particle, we approximately have
\al{
V_{S^1,M=0}\simeq -(-1)^{2s_p}n_p{1\over16\pi^6 L^4}\cos\paren{2\pi\theta},
}
see around  Eq.~\eqref{Eq:S1 massless} for the derivation.
As for the quarks, leptons and gluons, it is obtained that
{\footnotesize
\al{
V_{S^1,M=0}^{\text{SM}}&\simeq
{L_0^2\over 16\pi^8 L^6}\bigg[
3\cos\br{2\pi \paren{-A_e+{1-z_L\over2}}}
+
3\cos\br{2\pi\paren{{2\over3}A_e+A_{g1}+A_{g2}+{1-z_B\over2}}}
\nn
&+
3\cos\br{2\pi\paren{{2\over3}A_e+A_{g2}+{1-z_B\over2}}}
+
3\cos\br{2\pi\paren{{2\over3}A_e-A_{g1}-2A_{g2}+{1-z_B\over2}}}
\nn
&+
3\cos\br{2\pi\paren{-{1\over3}A_e+A_{g2}+{1-z_B\over2}}}
+
3\cos\br{2\pi\paren{-{1\over3}A_e-A_{g1}-2A_{g2}+{1-z_B\over2}}}
\nn&
+
3\cos\br{2\pi\paren{-{1\over3}A_e-A_{g1}-2A_{g2}+{1-z_B\over2}}}
\nn&
-
\sqbr{\cos\br{2\pi\paren{2A_{g1}+A_{g2}}}+\cos\br{2\pi\paren{A_{g1}+A_{g2}}}+\cos\paren{2\pi A_{g1}}}
\bigg]
+...}}
$\!\!$where $...$ represents functions which do not depend on the Wilson line moduli.
If a local minimum  with positive value of $V_{S^1,M=0}^\text{SM}$ exists with respect to the Wilson line, it may indicate the existence of a new local minimum in the $S^1$ compactification.
Positivity of the potential at its minimum would be needed because $\p_\chi^2 V$ should be positive in order to obtain a minimum of the potential.\footnote{
Precisely speaking, the coefficient of $L^{-6}$ should change from negative to positive around the mass threshold of  the new particle. However, the positive minimum condition is sufficient for the following discussion.
}

Although we do not exclude this possibility completely, within our numerical analysis, we do not find positive energy minima in $V_{S^1,M=0}^{\text{SM}}$ with respect to the Wilson line.

We also consider the lower dimensional vacuum corresponding to the high scale Higgs vacuum, whose cosmological constant can take  positive, zero, or negative value.
For definiteness, we take $\vev{H}=10^{16}\GeV$, and assume the existence of heavy right handed neutrinos whose masses are smaller than $10^{16}\GeV$ in the case of Majorana neutrino. 
In the high scale vacuum, the SM mass spectrum drastically changes.
The Dirac neutrino mass, $y_\nu \vev{H}$, can become larger than the Majorana mass.
The QCD scale increases, and becomes around $10^6\GeV$. The masses of the quarks and charged leptons are given by
\al{&
m_q=m_{q, EW}\paren{\vev{H}\over \vev{H}_{EW}},
&&
m_\ell=m_{\ell, EW}\paren{\vev{H}\over \vev{H}_{EW}},
}
where $m_{q, EW}$ and $m_{\ell, EW}$ are masses of our electroweak vacuum.
If the neutrino is of the Dirac type, the mass is given by $m_{\nu, EW}\vev{H}/\vev{H}_{EW}$. For the Majorana fermion, the mass matrix and mass eigenvalues are
\al{&
\begin{pmatrix}
0 & y_\nu \vev{H}\\
y_\nu \vev{H} & M_N
\end{pmatrix},
&&
m_\nu=
{1\over2}\paren{M_N\pm\sqrt{M_N^2+4y_\nu^2\vev{H}^2}},
}
where $M_N$ is the Majorana mass of the neutrino.
Note that the neutrino mass in the electroweak vacuum is $m_{\nu, EW}\simeq \vev{H}_{EW}^2y_\nu^2/M_N$.
Therefore, even if we fix $m_{\nu, EW}$, there remains a freedom to choose $M_N$.
In our numerical calculations, we take $M_N=10^{12}\GeV$ as a canonical value.

\begin{figure}
\begin{center}
\hfill
\includegraphics[width=.49\textwidth]{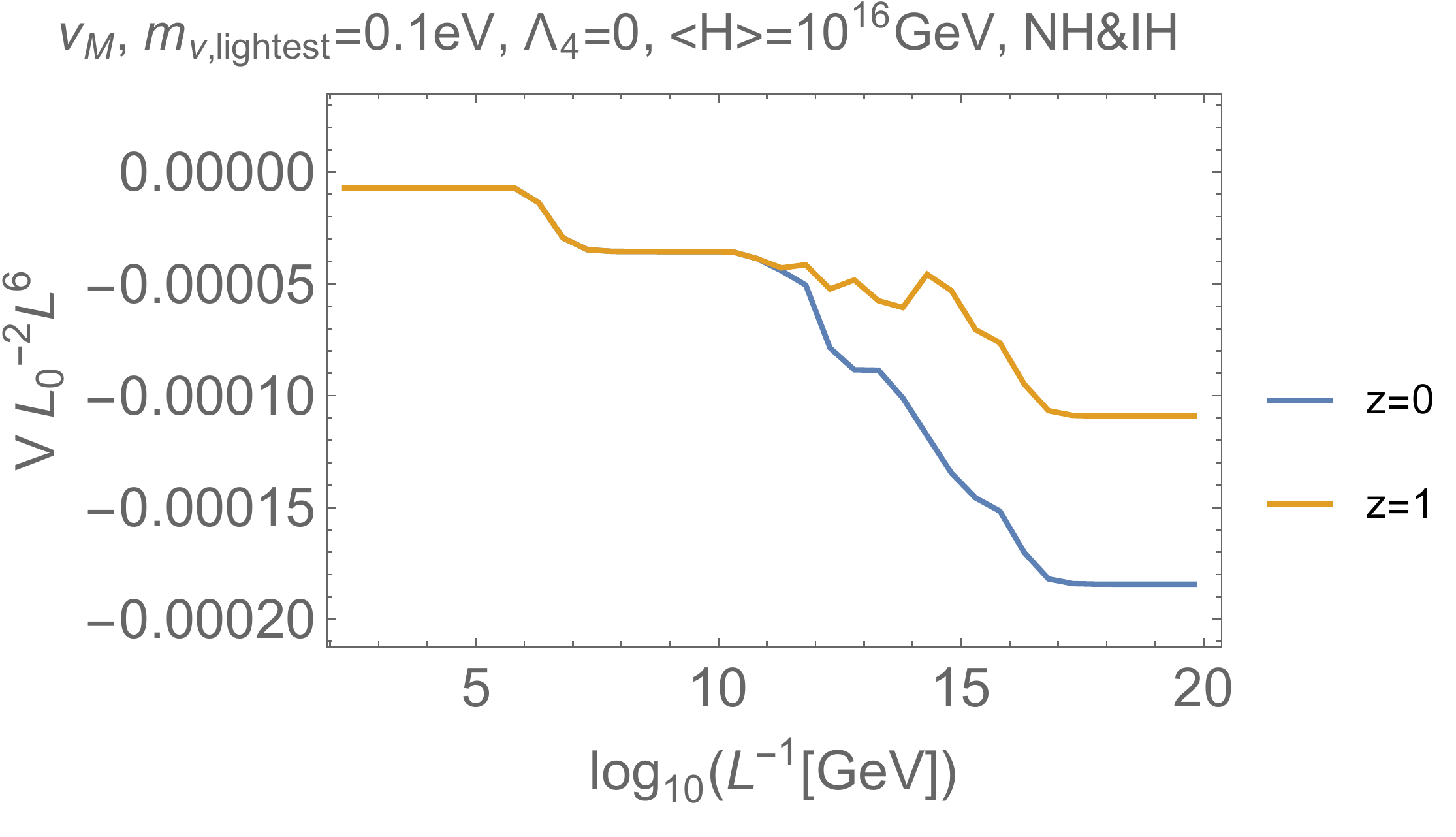}
\hfill
\includegraphics[width=.49\textwidth]{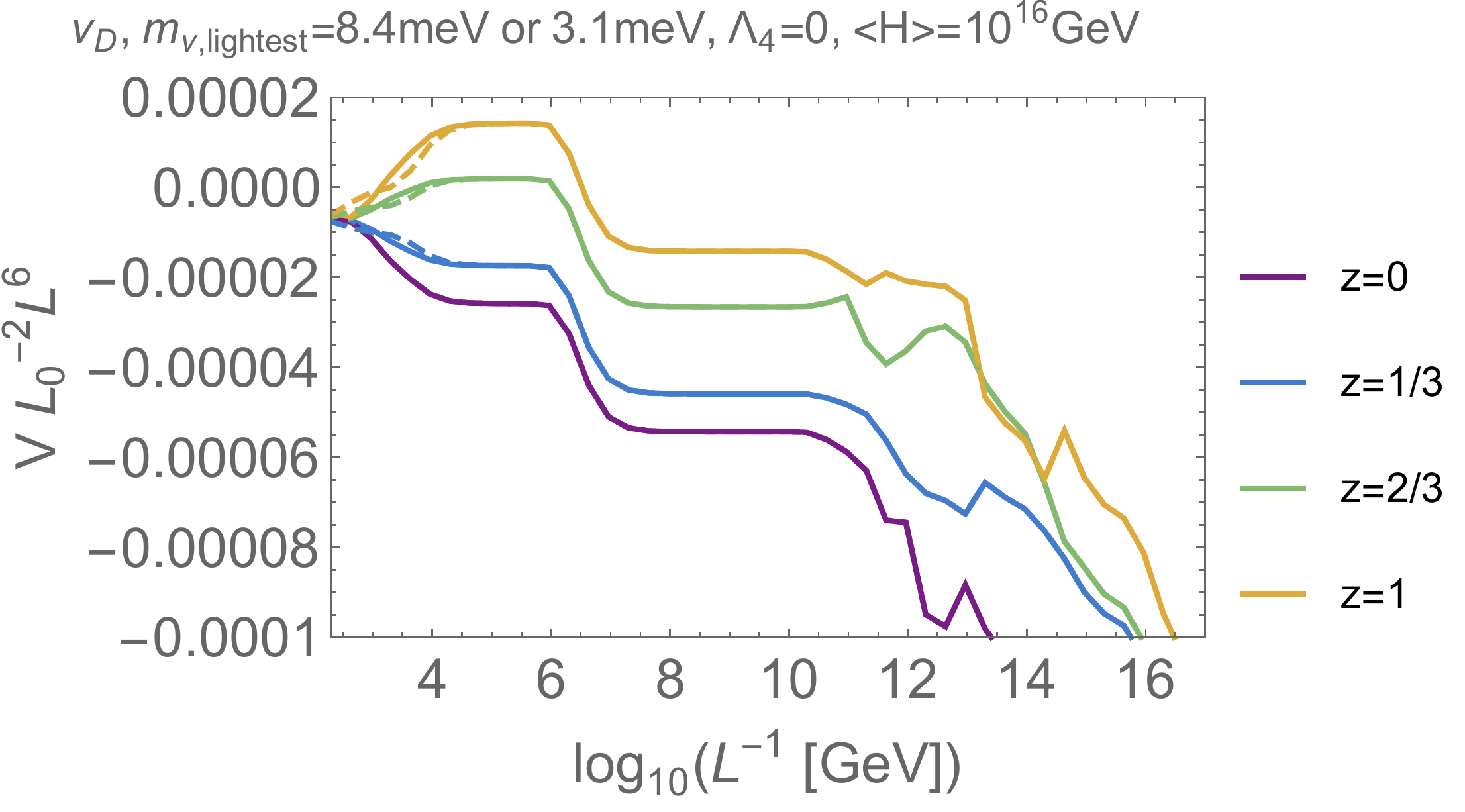}
\hfill\mbox{}
\\
\hfill
\includegraphics[width=.49\textwidth]{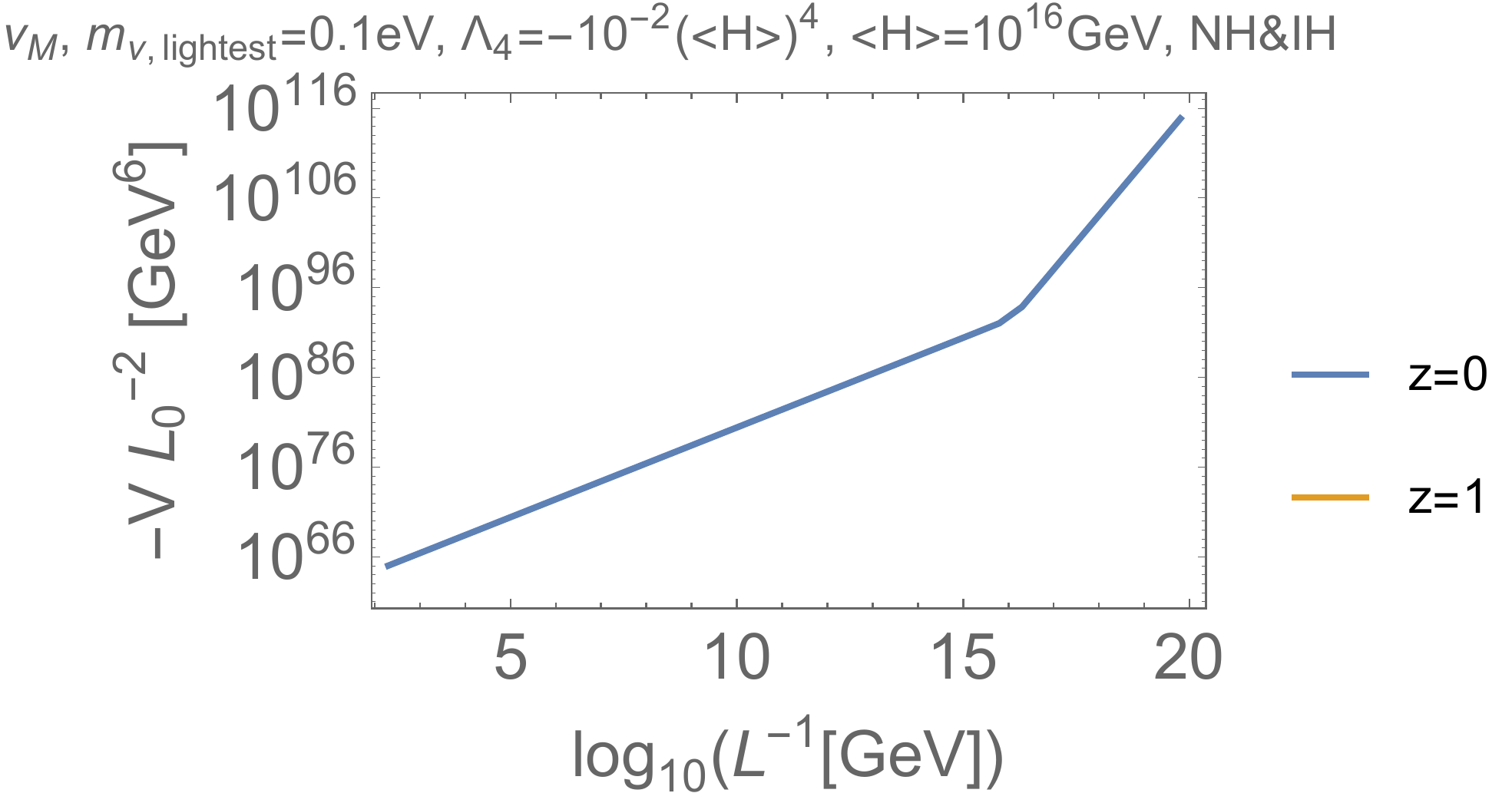}
\hfill
\includegraphics[width=.49\textwidth]{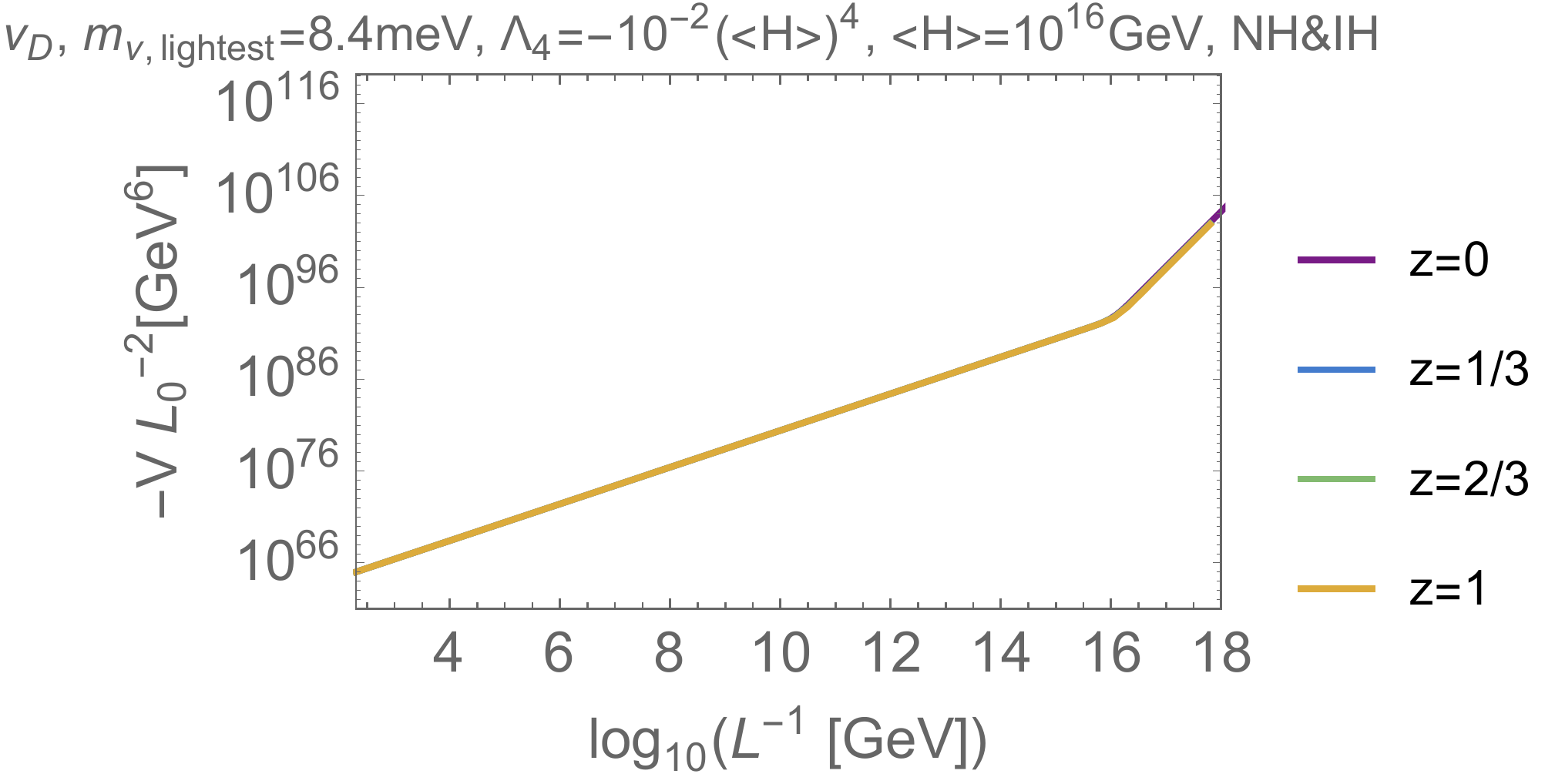}
\hfill\mbox{}
\\
\hfill
\includegraphics[width=.49\textwidth]{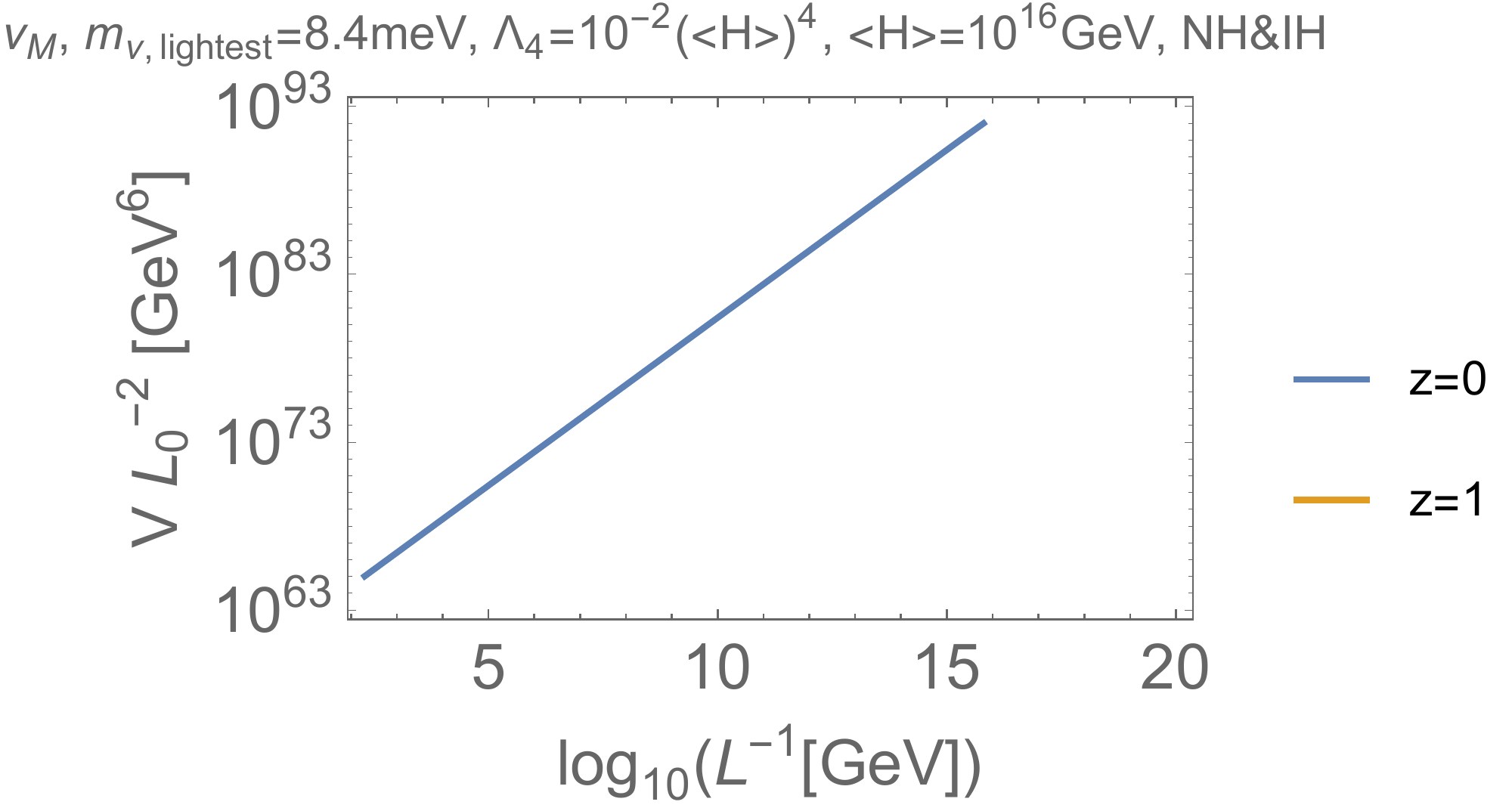}
\hfill
\includegraphics[width=.49\textwidth]{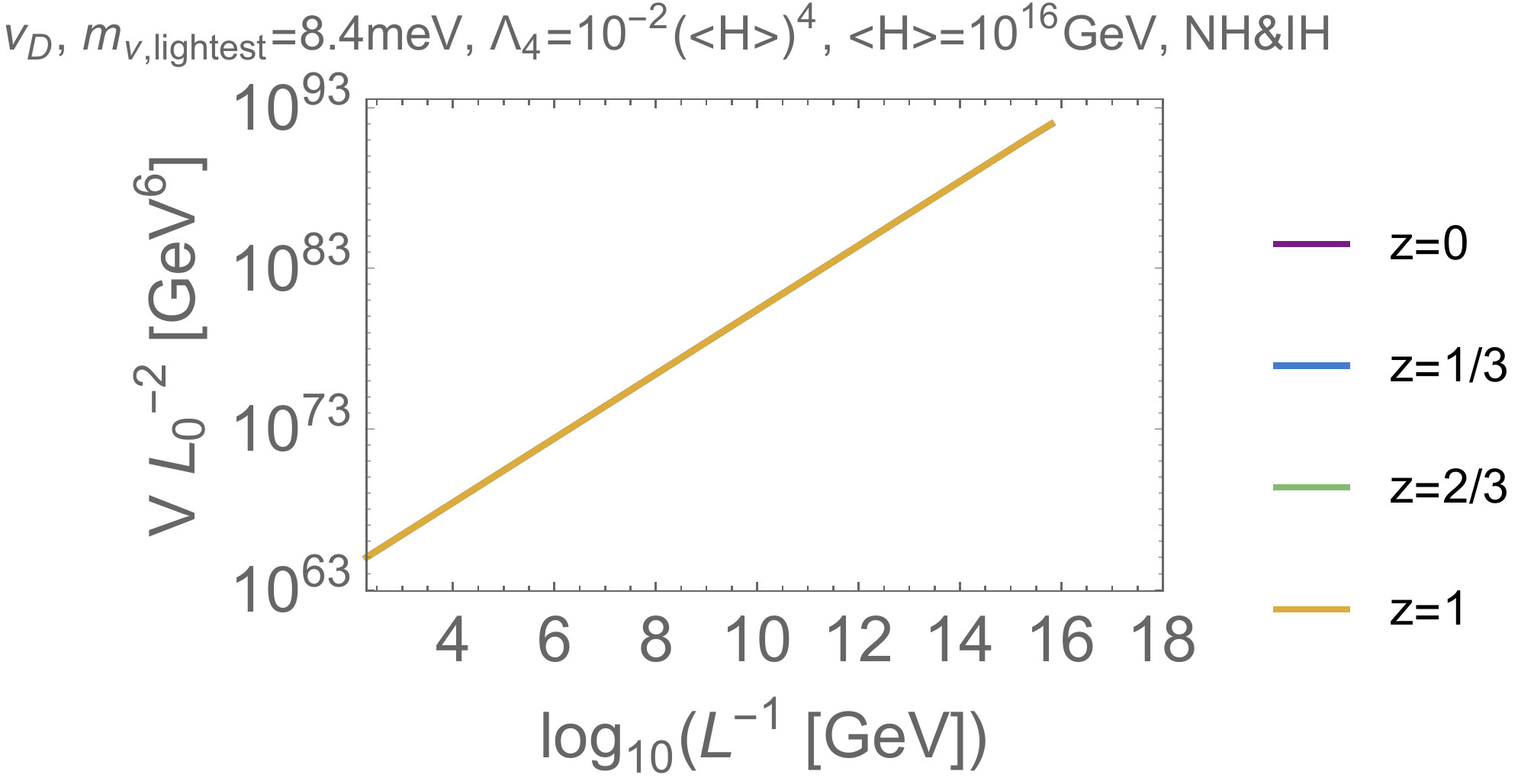}
\hfill\mbox{}
\end{center}
\caption{
Upper: $S^1$ compactification of the SM where $\Lambda_4=0$ and $\vev{H}=10^{16}\GeV$.
For $\nu_D$, the vacuum exists around $L^{-1}\sim10^{-3}\GeV$.
Middle: $S^1$ compactification of the SM where $\Lambda_4=-10^{-2}\vev{H}^4$ and $\vev{H}=10^{16}$ GeV.
For $L^{-1}\lesssim10^{16}$ GeV, the main contribution is the cosmological constant while the Casimir energy dominates for $L^{-1}\gtrsim10^{16}$ GeV.
There are no vacua.
Lower: The $S^1$ compactification of the SM where $\Lambda_4=10^{-2}\vev{H}^4$ and $\vev{H}=10^{16}\GeV$.
There are no vacua.
}
\label{Fig:S1_result2}
\end{figure}

\begin{figure}
\begin{center}
\hfill
\includegraphics[width=.44\textwidth]{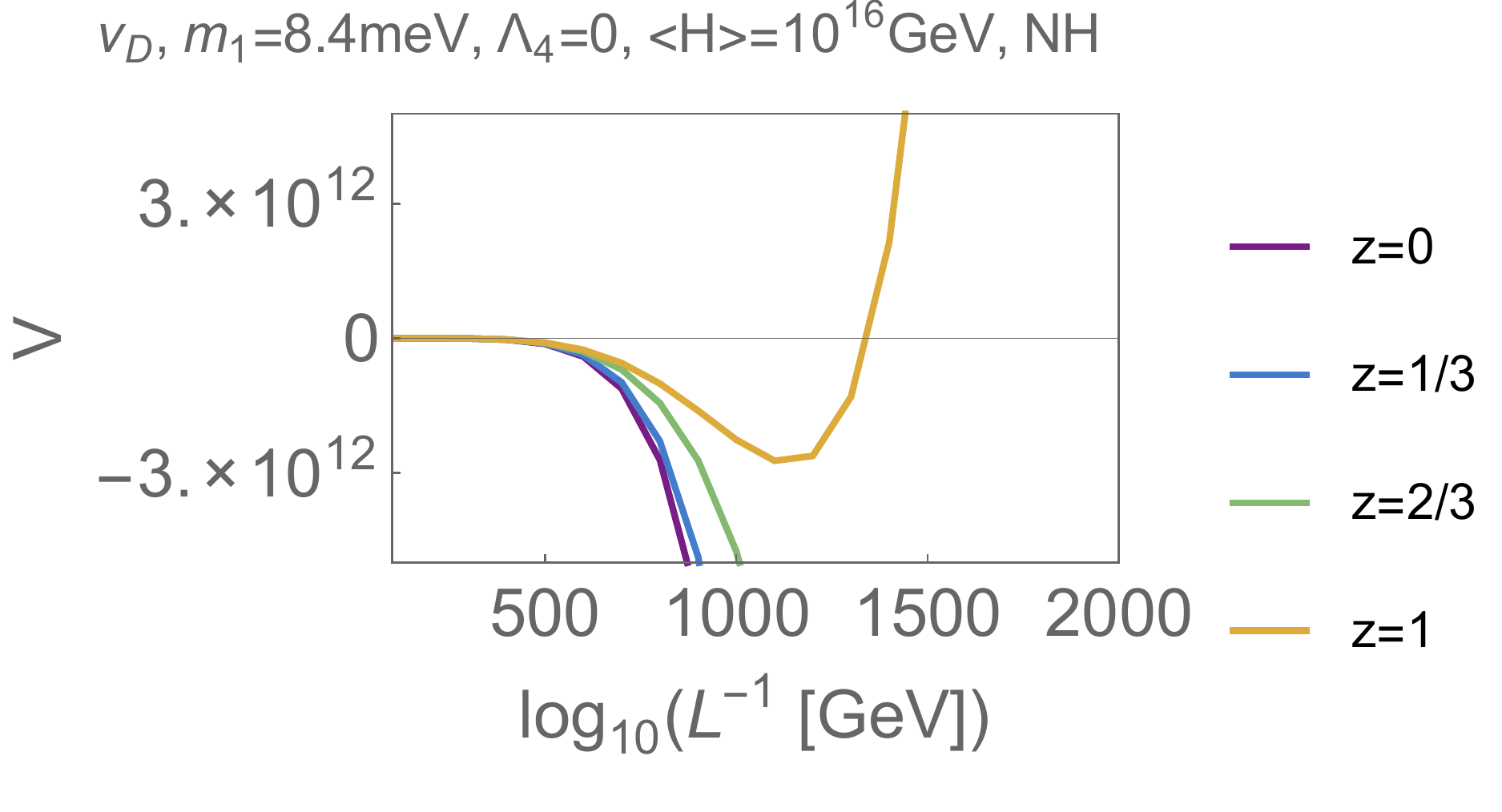}
\hfill\mbox{}
\end{center}
\caption{
The potential of the $S^1$ compactification of the SM where $\Lambda_4=0$ and $\vev{H}=10^{16}\GeV$ around the neutrino mass scale with NH. Here the scale $L_0$ is taken to be $1\GeV^{-1}$. The potential is same as upper right panel of Fig.~\ref{Fig:S1_result2}, but the vertical axis is the potential $V$ itself, and the horizon axis is the small segment of $L^{-1}$ around the neutrino mass scale.
}
\label{Fig:S1 neutrino minima2}
\end{figure}

We summarize the numerical results in Fig.~\ref{Fig:S1_result2}.
It is found that a perturbative stable vacuum only appears for $\Lambda_4=0$ and Dirac neutrino.\footnote{
As in Fig.~\ref{Fig:S1_result}, the vertical axis is $L_0^{-2}L^6V$ in Fig.~\ref{Fig:S1_result2}.
We can guess the existence of  minima in the upper right figure in Fig.~\ref{Fig:S1_result2} in the following way. If $z\gtrsim 2/3$ is satisfied, $L_0^{-2}L^6V$ becomes positive around $L^{-1}\sim10^{4}$GeV, and hence $V$ is also positive there. Moreover, $V$ behaves as $V\propto -L^{-6}$ for smaller $L^{-1}$ where only the gauge boson and graviton contributions are present, see Eq.~\eqref{Eq:S1 result}. Combining the fact that $V$ is negative and a monotonically decreasing function at small $L^{-1}$ and is positive around $L^{-1}\sim10^{4}$GeV, we can see that there should exist $AdS$ minima.
We also plot the figure where the vertical axis is $V$ around the neutrino mass scale in Fig.~\ref{Fig:S1 neutrino minima2}.
}
This can be understood intuitively.
If the neutrino is  of the Majorana type and the neutrino Yukawa coupling is not small\footnote{Here we use $y_\nu\sim0.01$ which is obtained from $M_N=10^{12}\GeV$ as a canonical value.}, 
the neutrino is not the lightest matter in the theory. The electron, up quark and down quark become lighter than the neutrino due to their small Yukawa couplings, $y_e, y_u, y_d \sim10^{-6}$.
Therefore, the lightest particle is the charged one, and the vacuum can not be found as in the compactification of $U(1)$ gauge theory.
This is why the vacuum disappears for Majorana neutrinos.
Even if the neutrinos are of the Dirac type, the neutrino vacuum does not appear if the absolute value of the cosmological constant is large compared with the mass of the neutrino.
In this case, $\Lambda_4$ term dominates the potential~\eqref{Eq:S1 SM potential} up to $L^{-1}\sim (\Lambda_4)^{1/4}$, where the charged particle contribution becomes large.
Therefore, the effect of the neutrino loop is not effective, and the vacuum does not appear.

To summarize, there are no vacua except for the neutrino one, and this neutrino vacuum in $3$ dimensions is likely to be unstable through tunneling to the runaway solution.

We comment on the relation of our results with that in previous works~\cite{ArkaniHamed:2007gg,Ibanez:2017kvh}. In Ref.~\cite{ArkaniHamed:2007gg,Ibanez:2017kvh}, the Wilson line was taken to be zero (or $\pi$).\footnote{In Table 1 in Ref.~\cite{ArkaniHamed:2007gg}, the coefficients of the Casimir energy at each mass threshold were presented for a fixed Wilson line zero or $\pi$.} Since the potential around the neutrino vacuum is very flat\footnote{Because the lightest charged particle (electron) is much heavier than the neutrino, the potential of the Wilson line is exponentially suppressed~\cite{ArkaniHamed:2007gg}.} and the maximum of the potential satisfies the Breitenlohner-Freedman (BF) bound in AdS, vanishing Wilson line is a valid solution. It would be interesting to study if there can be tunneling transitions from those vacua with a zero Wilson line to the runaway found in this paper which has a different value for the Wilson line.

\subsection{Multiple point principle and prediction on the neutrino mass}

Here we briefly review the multiple point principle and apply this principle to the SM landscape. See also Ref.~\cite{Froggatt:1995rt} for the original argument, and App.~D of Ref.~\cite{Hamada:2015ria} for a review of this material.
In the standard argument of statistical mechanics, the fundamental concept is the principle of equal a priori probabilities in the micro-canonical ensemble. The canonical ensemble is derived by dividing a large system into a heat bath and a small system, and applying the micro-canonical ensemble to the whole system. On the other hand, the starting point of  quantum field theory is the path integral which may correspond to the canonical ensemble of statistical mechanics. The natural question is what happens if we start from a micro-canonical type path integral.

With this motivation in mind, Froggatt and Nielsen~\cite{Froggatt:1995rt} started from the micro-canonical type path integral
\al{
\int D\phi \,\delta\paren{\int d^4x |H|^2-I_2} e^{-S},
}
where the delta function is the analogue of micro-canonical ensemble where the energy is fixed. Instead of the energy, the spacetime integral of the Higgs field squared is fixed to be some constant $I_2$. Here $S$ is the action of the SM other than the Higgs mass term. They argued that, if there is a new vacuum around the Planck scale which is degenerate in energy with the electroweak vacuum, then the delta functional constraint can be satisfied by considering the coexisting phase/superposition of the high scale and the electroweak scale vacua.

Here, we further speculate that there is a micro-canonical type constraint
\al{
\delta\paren{\int d^4x L^2-I'_2},
}
in the path integral, and the coexisting phase/superposition of  the two vacua of the radion field realizes the delta functional constraint.
In this respect, it is interesting that the $S^1$ vacuum can be $dS_3$, $M_3$ or $AdS_3$.
If we apply the multiple point principle, it would be natural to require that the three dimensional vacuum to be close to $M_3$, otherwise either the 3 dimensional or the 4 dimensional vacuum is favored  from energetic considerations, and it is difficult to maintain the coexisting phase/superposition.
Then, we can predict that the mass of the lightest neutrino to be $\mathcal{O}(1\text{--}10)\meV$.
The multiple point principle provides an interesting suggestion that the measure of the possibility of the vacuum selection in the string landscape is not equally distributed, but there is some bias. It is important to clarify the phenomenological predictions of the multiple point principle, and compare them with experiment.


\subsection{Flux vacua}\label{Sec:S1 flux vacua}
So far, we have considered a constant background for the Wilson line.
However, in general, we can also consider flux vacua if we add an axion-like particle $a$ to the theory.
Then, the following term is added to the action:
\al{
\Delta S&= \int d^4x \sqrt{-g}\paren{-{1\over2}{f_a^2\paren{\p_\mu a}^2}}
\nn
&\simeq \int \paren{2\pi L} d^3x \sqrt{-g^{(3)}}\paren{-{1\over2}f_a^2g^{ij}\p_i a \p_j a-{f_a^2\over2L^2}\paren{\p_3 a}^2+...}
\nn
&= \int d^3x \sqrt{-g^{E(3)}}\paren{-{L_0^3\over2(2\pi)^2L^4}f_a^2\paren{\p_3 a}^2+...}
}
where $f_a$ is the decay constant of the axion.
The flux vacua is given by $a=w x_3$ where $w$ is the winding number, which gives a positive contribution to the tree-level potential in the Einstein frame,
\al{
\Delta V^{(E)}={L_0^3\over2(2\pi)^2L^4} w^2 f_a^2.
}
The contribution of the flux is stronger than the Casimir energy but weaker than the cosmological constant for large $L$.
Typically, this erases the vacua with $L\gtrsim f_a^{-1}$.
This is reasonable because the flux effect is classical while the Casimir effect is quantum, and the classical term is expected to be dominant at  low energy, i.e., large radius.

\begin{figure}
\begin{center}
\hfill
\includegraphics[width=.6\textwidth]{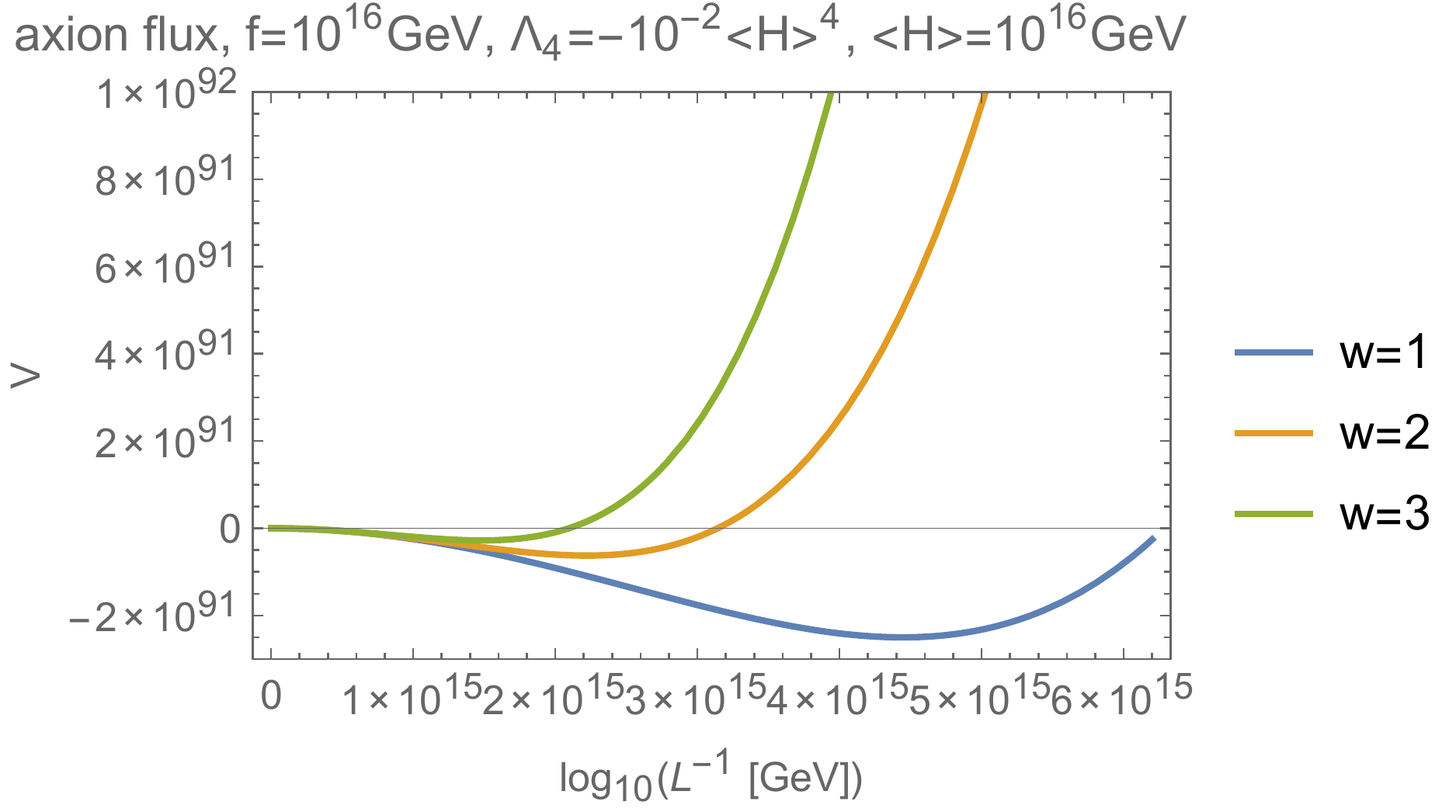}
\hfill\mbox{}
\end{center}
\caption{
The tree level potential of $S^1$ compactification of the SM with the high scale vacuum, Eq.~\eqref{Eq: S^1 axion potential}.
}
\label{Fig:axion_flux}
\end{figure}

If we consider the high scale vacuum with $\Lambda_4<0$, we have many $AdS_3\times S^1$ minima corresponding to $w$.
Indeed, the classical potential in the Einstein frame becomes
\al{\label{Eq: S^1 axion potential}
V^{(E)}={L_0^3\over\paren{2\pi L}^2}\paren{\Lambda_4+{w^2\over2}{f_a^2\over L^2}}.
}
The expression in the parenthesis in the potential is shown in Fig.~\ref{Fig:axion_flux}.
We can see that there are many $AdS$ minima.
This vacuum is stable at least at tree level.
Thus, the SM supplemented by an axion (and nothing else) seems to be at odd with the conjecture ~\cite{Ooguri:2016pdq,Freivogel:2016qwc} if the high scale vacuum has negative cosmological constant.\footnote{
In Sec.~\ref{Sec:4d SM vacua}, we have seen that the SM with the high scale $AdS_4$ vacuum is at odd with the conjecture. Here we point out that this potential conflict with the conjecture can also be found in $S^1$ compactification.
}
It would be interesting to look for the corresponding extremal black hole solutions.
Notice that this potential is similar to that employed by Bousso and Polchinski~\cite{Bousso:2000xa} in illustrating the flux landscape.

\section{The SM vacua from $T^2$ compactification}\label{Sec:SM T2 vacua}
In this section, we consider the vacuum structure of $T^2$ compactification of the SM.
The same issue was discussed in Refs.~\cite{Arnold:2010qz,Arnold:2011cz,Fornal:2011tw}, where only periodic fermion and the potential around the neutrino mass scale was discussed.
In contrast to these earlier works, the generalized formulae and analysis we present here allow for general spin structures of the fermions. As a result, we can carefully consider the vacuum condition for general compactifications on $T^2$.
\subsection{Effective action}
In the $T^2$ compactification, the metric is decomposed as
\al{
ds^2=g_{\alpha\beta}dx^\alpha dx^\beta+\rho\gamma_{ij}dy^i dy^j+B_\alpha^i dx^\alpha dy^i,
}
where $\tau$ is the shape moduli, $\rho$ is the volume moduli of $T^2$, $\alpha, \beta=0,1$, $i, j=2, 3$, $B_\alpha^i$ are graviphotons, and $\gamma_{ij}$ is the metric of the two-torus:
\al{
\gamma_{ij}=
{1\over\tau_2}
\begin{pmatrix}
1 &\tau_1\\
\tau_1 & |\tau|^2
\end{pmatrix}.
}
The Laplacian on $T^2$ is
\al{
\Delta f= {1\over \rho\,\tau_2}\paren{\p_2^2f-\tau_1\p_1\p_2 f+\paren{\tau_1^2+\tau_2^2}\p_1^2 f},
}
and hence the normalized eigenfunction which is periodic on $T^2$ is obviously
\al{&
\psi_{m,n}={1\over2\pi\sqrt{\rho}}\exp\sqbr{im y_2 + i n y_1},
&&
\int d^2y\sqrt{\rho \gamma_{ij}}\psi_{m,n}^*\psi_{m,n}=\delta_{m,n},
&&
0\leq y_{1}, \,y_2\leq2\pi.
}
The corresponding eigenvalue is
\al{
\Delta \paren{\exp\sqbr{im y_2 + i n y_1}}
&=
-{|m-n\tau|^2\over \rho\,\tau_2}\exp\sqbr{im y_2 + i n y_1}.
}
The extension to other boundary conditions is not difficult:
\al{\label{Eq:T2 eigenvalue}
&
\psi(y_1+2\pi)=e^{2\pi i\theta_1}\psi(y_1),\quad\quad \psi(y_2+1)=e^{2\pi i\theta_2}\psi(y_2),
\nn
&
\psi_{m,n}={1\over2\pi\sqrt{\rho}}\exp[i(n+\theta_1)y_1+i(m+\theta_2)y_2],
\quad
\Delta \psi_{m,n}=-{|(m+\theta_2)-(n+\theta_1)\tau|^2\over \rho\,\tau_2} \psi_{m,n}
}

Once we solve the eigenvalue problem, we can calculate the one-loop potential with general boundary conditions by evaluating the one-loop determinant. 
As calculated in App.~\ref{App:T2 calculation}, 
the total Casimir energy after renormalization is
\al{
V^{\text{all}}_{T^2}(\rho,\tau,\theta_1,\theta_2)
=\sum_\text{particle} (-1)^{2s_p}n_p V^{(1)}_{T^2}\paren{\rho,\tau,q_p A_1+{1-z_{1p}\over2},q_p A_2+{1-z_{2p}\over2}},
}
where $s_p$ is the spin, $n_p$ is the number of degrees of freedom, $z_p=0 (1)$ corresponds to anti-periodic (periodic) boundary condition.
If $\theta_1=\theta_2=0$, there is modular invariance in the $\tau$ plane, and the potential has its extrema at $\tau=e^{i\pi/2}$ and $e^{i\pi/3}$.
As in $S^1$ compactification, we have to specify the boundary conditions for two $1$-cycles of $T^2$ in order to define the theory.
The fermions can have non-trivial spin structures corresponding to $U(1)_L$ and $U(1)_B$.
As in $S^1$ compactification, for simplicity, we choose the same boundary condition for the leptons and baryons in the numerical analysis.
Here $V^{(1)}_{T^2}$ is
\al{\label{Eq:T2 result}
&V^{(1)}_{T^2}(\rho, \tau,\theta_1,\theta_2)=
-\sum_{l=1}^\infty{\tau_2 M^2\over8\pi^4 l^2\rho}\cos\paren{2\pi l\theta_1}K_2\paren{2\pi l \sqrt{\rho} M\over\sqrt{\tau_2}}
\nn&
-\frac{1}{64\pi^5 \rho^2\tau_2}\sum_{n=-\infty}^\infty\bigg[2\pi  \sqrt{(n+\theta_1)^2\tau_2^2+M^2 \rho \tau_2}
   \br{\text{Li}_2\paren{e^{\sigma_+}}
   +\text{Li}_2\left(e^{\sigma_-}\right)}
   +\br{\text{Li}_3\left(e^{\sigma_+}\right)
   +\text{Li}_3\left(e^{\sigma_-}\right)}\bigg],
}
where 
\al{&
\sigma_\pm:=2\pi\paren{\pm i\br{-(n+\theta_1)\tau_1+\theta_2}-\sqrt{(n+\theta_1)^2\tau_2^2+M \rho^2\tau_2}}.
}

Now the action including  the Casimir energy is
\al{\label{Eq: T^2 effective action}
S&=\int d^4x \sqrt{-g}\paren{{1\over2}M_P^2R-\Lambda_4-V_{T^2}^\text{all}-{1\over4}F_{\mu\nu}F^{\mu\nu}+...}
\nn&\simeq
\int d^2x\sqrt{-g_{(2)}}
\bigg[
{1\over2}M_P^2\br{\rho R_{(2)}-{\rho\over2\tau_2^2}\br{(\partial_\alpha\tau_1)^2+(\partial_\alpha\tau_2)^2}}-\rho\Lambda_4-\rho V_{T^2}^\text{all}\nn
&-{\rho\over2}F_{01}F^{01}-{\rho\over2}F_{23}F^{23}-{\rho\over2}\br{\paren{\p_\alpha A_2}^2+\paren{\p_\alpha A_3}^2}+...
\bigg].
}

Note that, in addition to the $T^2$ moduli $\tau$ and $\rho$, we have the Wilson line moduli corresponding to the extra dimensional component of the gauge field.
The conditions of a vacuum for these moduli to be stable against  localized perturbations is derived in App.~\ref{App:vacuum condition}.
It should be stressed that, among $\tau, g_{\alpha\beta}$ and $\rho$, only $\tau$ has dynamical degrees of freedom with a kinetic term in the action.

The condition of  vacuum stability is summarized as follows (see App.~\ref{App:vacuum condition} for the derivation).
First, in order to obtain the $2$d spacetime independent solution, it is needed
\al{&
V=0,
&&
\p_{\tau_a,w}V=0.
}
Here $V$ is the full $2$ dimensional potential term, $\p_{\tau_a}$ and $\p_w$ refer to the derivatives with respect to $\tau_a$ and the Wilson line moduli, respectively.
Since the $\rho$ field is not dynamical, it is fixed by the constraint equation $V=0$.
The curvature of $2d$, $R_{(2)}$, is not determined by the height of the potential, but by $R_{(2)}=2\p_\rho V/M_P^2$. 
Therefore, $\p_\rho V>0, \p_\rho V=0$ and $\p_\rho V<0$ correspond to $dS_2$, $M_2$ and $AdS_2$, respectively.
Next, to guarantee the stability of the vacuum against localized perturbations, it is required that 
\al{\label{Eq:stability1}&
\p_{\tau_a}\p_{\tau_b}V\geq 0,\, \p_{w}\p_{w}V\geq 0, \quad\text{for $dS_2$ and $M_2$},
\nn
&
{16\tau_2^2\over\rho M_P^2}\p_{\tau_a}\p_{\tau_b}V\geq R_{(2)},\,\p_w\p_w V\geq R_{(2)}, \quad\text{for $AdS_2$},
}
where $w$ is the dimensionless Wilson line field.
In terms of the field in Eq.~\eqref{Eq: T^2 effective action}, $w$ corresponds to $A_i=w_i/\sqrt{\rho}$.
Notice that, in the case of $AdS_2$ vacua, some amount of tachyonic mass is not in contradiction with the stability condition, known as BF bound~\cite{Breitenlohner:1982bm,Breitenlohner:1982jf}.

\subsection{$U(1)$ gauge theory on $T^2$}
\subsubsection{with charged matter}\label{Sec:T2 U(1) charged}
As in the $S^1$ compactification case, we start as a warmup analyzing the compactification of $U(1)$ gauge theory with matter, before turning to the more complicated structure of the SM landscape.
As we will see, just as in the $S^2$ case, we can not find perturbatively stable solution of $T^2$ compactification.
More explicitly, the potential is given by
\al{
V_{T^2}^\text{charged}=
\Lambda_4+2V_{T^2}^{(1)}(\rho,\tau,0,0)\bigg|_{M=0}-4V_{T^2}^{(1)}\paren{\rho,\tau,q_e A_1+{1-z_{1e}\over2},q_e A_2+{1-z_{2e}\over2}},
}
where the second and the third terms correspond to the photon and electron contributions, respectively.

\begin{figure}
\begin{center}
\hfill
\includegraphics[width=.35\textwidth]{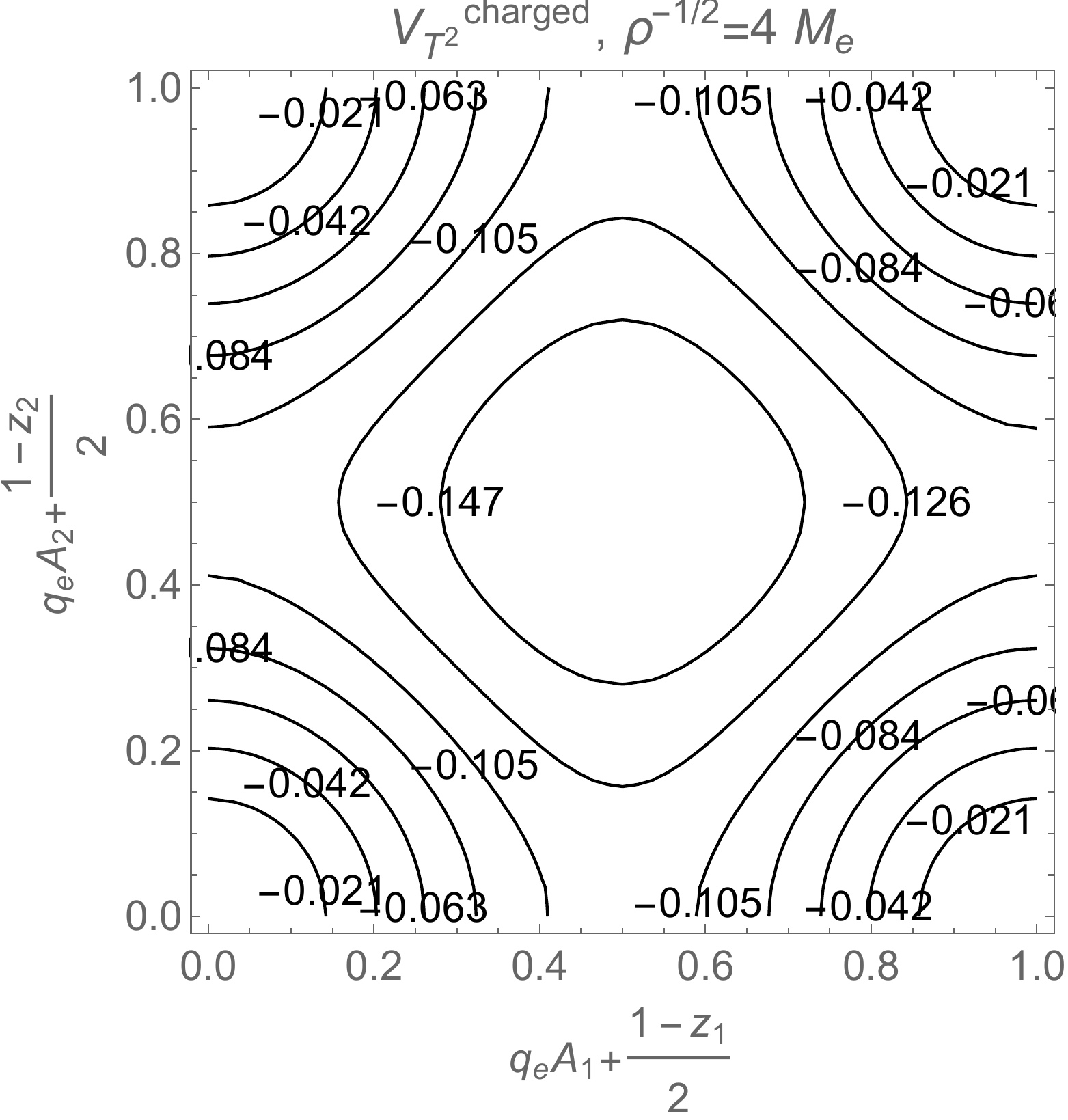}
\hfill
\includegraphics[width=.4\textwidth]{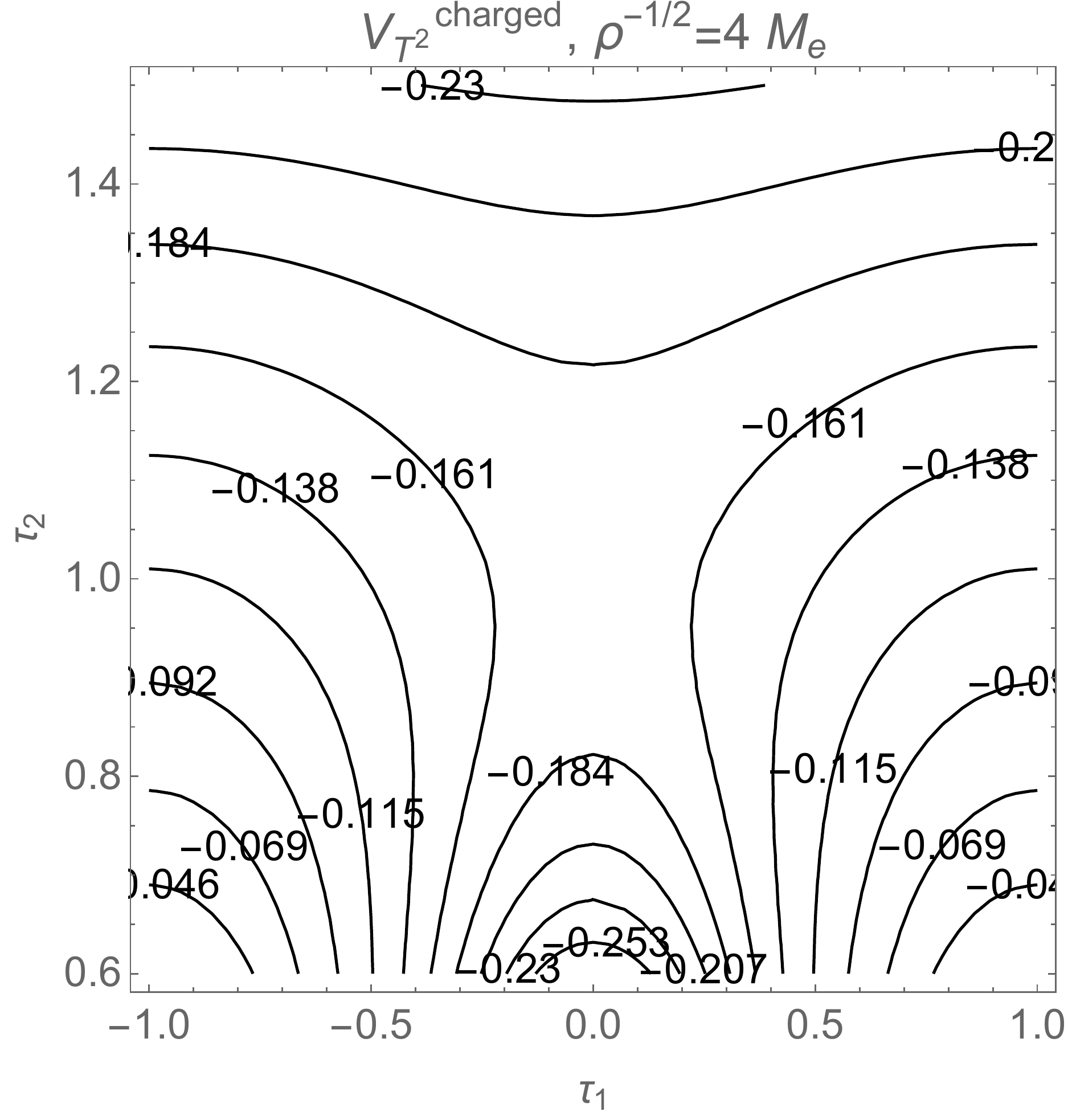}
\hfill\mbox{}
\end{center}
\caption{
$U(1)$ gauge theory with charged matter.
Left: The potential as a function of the Wilson line moduli.
Right: The potential as a function of  the $\tau$ moduli.
}
\label{Fig:QED_S1_potential}
\end{figure}

We plot $V_{T^2}^\text{charged}$ as a function of the Wilson line moduli in the left panel of Fig.~\ref{Fig:QED_S1_potential}, from which we can see that the potential is minimized when the Wilson line is at $q_e A_1+{1-z_{1e}\over2}={1\over2}$ and $q_e A_2+{1-z_{2e}\over2}={1\over2}$.
However,  stabilization of the $\tau$ moduli cannot be achieved in this case.
In the right panel, the potential of the $\tau$ moduli is plotted, from which we can see that the potential is unbounded, and so there is no vacuum in this compactification.

\subsubsection{with neutral matter}\label{Sec:T2 neutral}
\begin{figure}
\begin{center}
\hfill
\includegraphics[width=.6\textwidth]{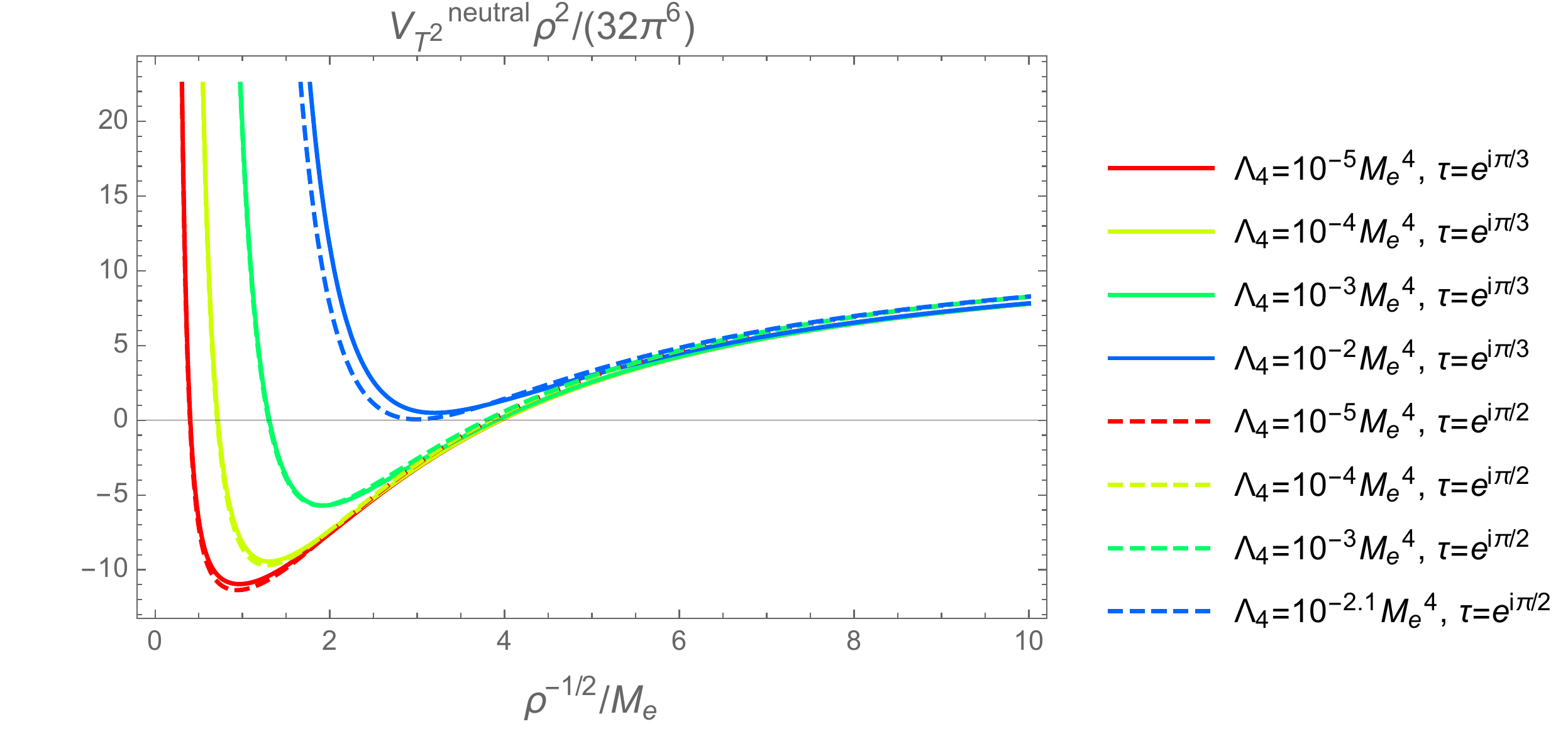}
\hfill\mbox{}
\end{center}
\caption{
$U(1)$ gauge theory with neutral matter.
The potential as a function of $\rho$, the volume of $T^2$.
}
\label{Fig:QED_T2_neutral_potential}
\end{figure}

Next we consider $T^2$ compactification of $U(1)$ gauge theory with neutral matter.
We show that perturbatively stable $dS_2, M_2$ or $AdS_2$ vacua can be obtained.
The potential is
\al{
V_{T^2}^\text{neutral}=
\Lambda_4+2V_{T^2}^{(1)}(\rho,\tau,0,0)\bigg|_{M=0}-4V_{T^2}^{(1)}\paren{\rho,\tau, {1-z_{1e}\over2}, {1-z_{2e}\over2}},
}
First, we consider a neutral Dirac fermion with periodic boundary condition.
In this case, the potential possesses modular invariance, and the fixed points $\tau=e^{i\pi/3}, e^{i\pi/2}$ are extrema of the potential.
Therefore, we fix $\tau=e^{i\pi/3}$ or $e^{i\pi/2}$, and analyze the potential for $\rho$, which is shown in Fig.~\ref{Fig:QED_T2_neutral_potential}.
Depending on the value of $\Lambda_4$, there exists two, one or zero solution(s) of the Hamiltonian constraint $V=0$.
These are candidates for a vacuum. 
Notice that $\p_{\rho^{-1/2}}V<0, \p_{\rho^{-1/2}}V=0$ and $\p_{\rho^{-1/2}}V>0$ correspond to $dS_2$, $M_2$ and $AdS_2$, respectively.

By looking at the figure, we can see that, for $\tau=e^{i\pi/3}$ and $\Lambda_4\lesssim10^{-2}M_e^4$, we have one vacuum candidate for $dS_2$, and one for $AdS_2$. For $\Lambda_4\simeq10^{-2}M_e^4$, we have a vacuum candidate for $M_2$.
Similarly, for $\tau=e^{i\pi/2}$, $dS_2$ and $AdS_2$ vacuum candidates exist for $\Lambda_4\lesssim10^{-2}M_e^4$, and  a $M_2$ vacuum candidate appears for $\Lambda_4\simeq10^{-2}M_e^4$.

\begin{figure}
\begin{center}
\hfill
\includegraphics[width=.49\textwidth]{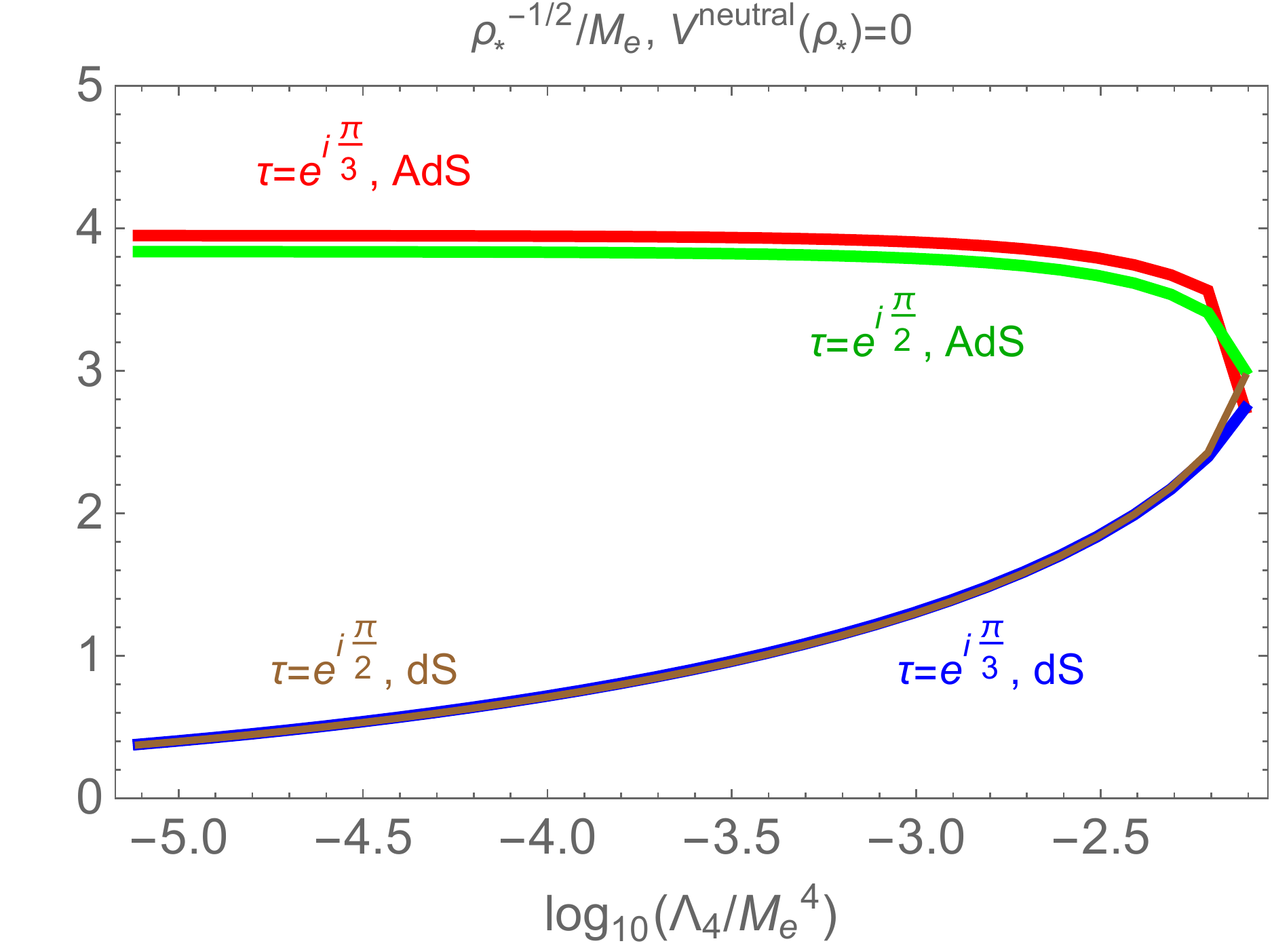}
\hfill
\includegraphics[width=.5\textwidth]{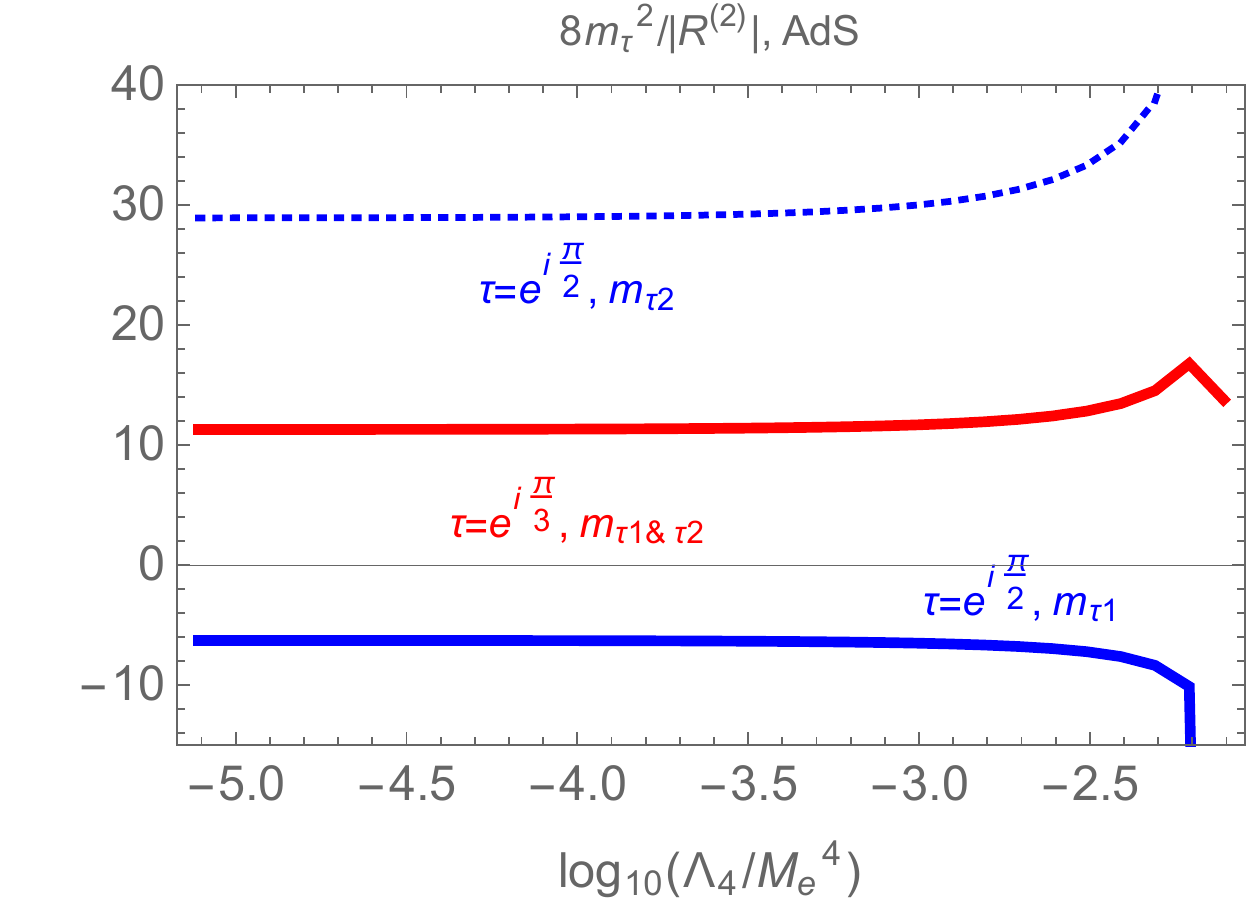}
\hfill\mbox{}
\hfill
\includegraphics[width=.49\textwidth]{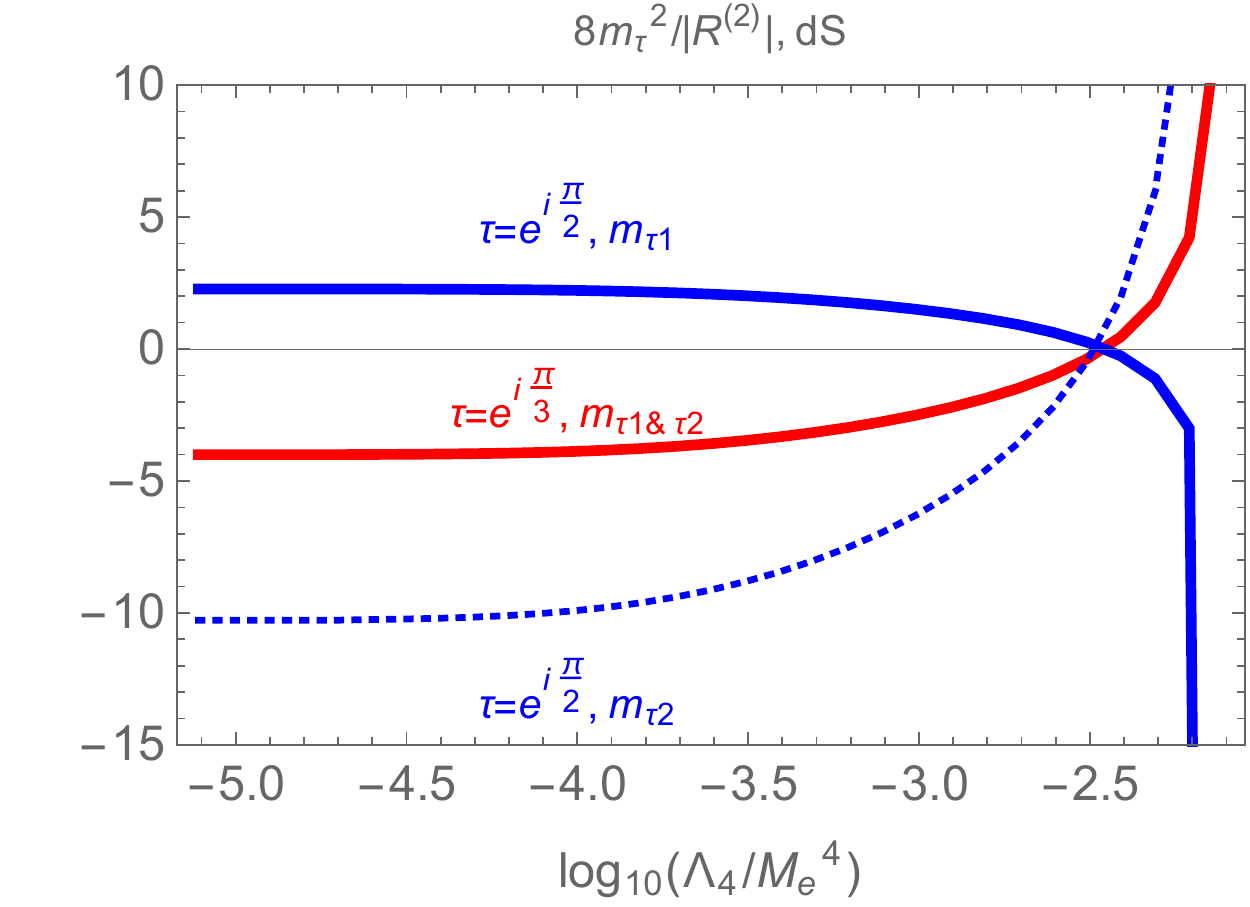}
\hfill
\includegraphics[width=.49\textwidth]{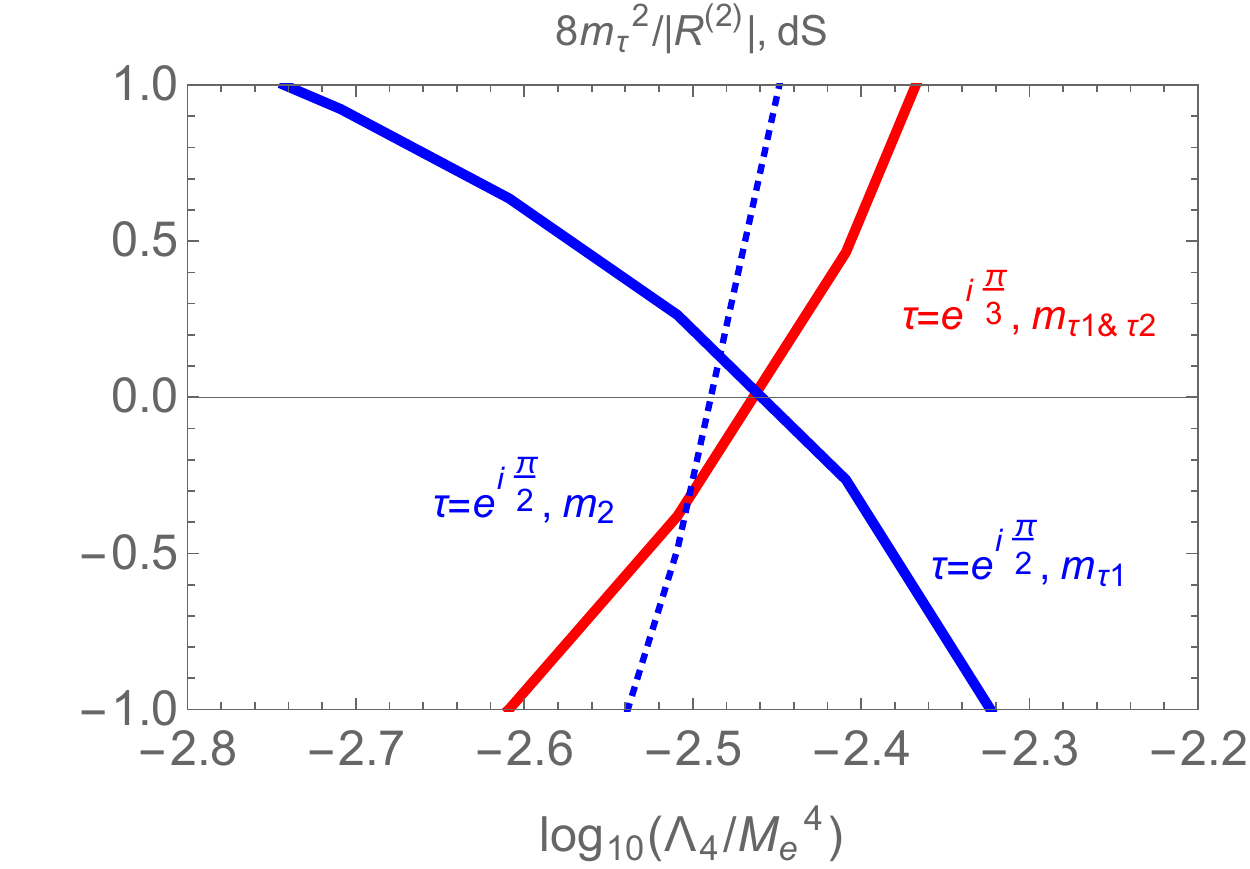}
\hfill\mbox{}
\end{center}
\caption{
$U(1)$ gauge theory with neutral matter.
In the upper left figure, the value of $\rho_*$ which satisfies $V(\rho_*)=0$ is plotted.
The ratio between the mass of the $\tau$ moduli and the curvature of $2$-dimensional spacetime is depicted in the other plots.
The vacuum is perturbatively unstable if this is negative ($dS_2, M_2$) or smaller than $-1$ ($AdS_2$).
}
\label{Fig:QED_T2_neutral_massratio}
\end{figure}

\begin{figure}[t]
\begin{center}
\hfill
\includegraphics[width=.6\textwidth]{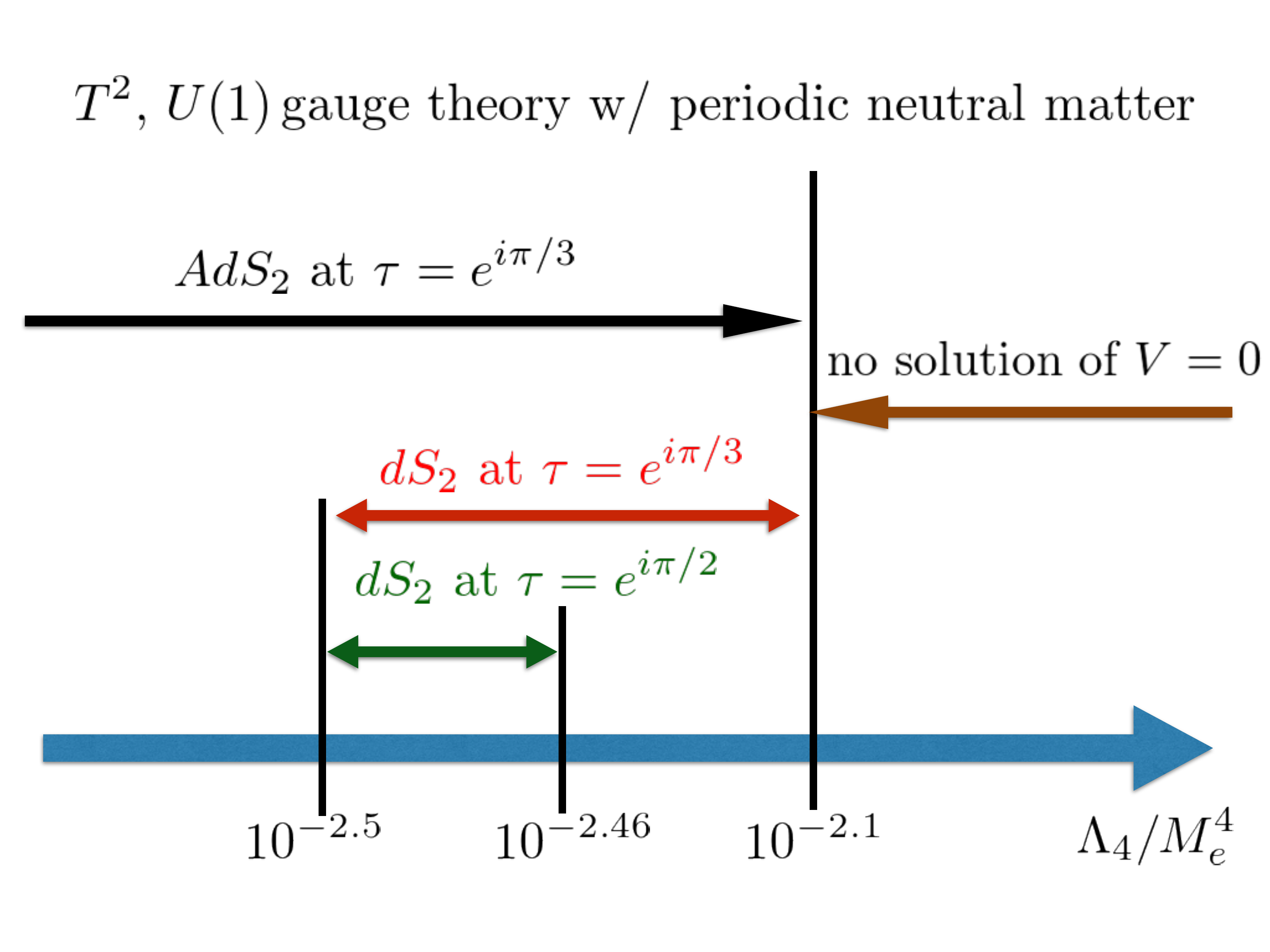}
\hfill\mbox{}
\end{center}
\caption{
A schematic picture of the perturbatively stable vacuum structure of $T^2$ compactification of $U(1)$ gauge theory with neutral matter.
}
\label{Fig:Result_T2U(1)}
\end{figure}

Next, we need to check the perturbative stability of the vacua, whose condition is summarized in Eq.~\eqref{Eq:stability1}.
In order to examine the vacuum stability, the mass-to-curvature ratio, $8m_{\tau_a}^2/|R^{(2)}|$ is plotted in Fig.~\ref{Fig:QED_T2_neutral_massratio}.
If this is smaller than $0$ (for $dS_2/M_2$) or  $-1$ (for $AdS_2$), the vacuum is perturbatively unstable.
It can be seen that only the $\tau=e^{i\pi/3}$ $AdS_2$ vacuum is stable for $\Lambda_4\lesssim10^{-2}M_e^4$.
Furthermore, if $\Lambda_4$ is close to $10^{-2}M_e^4$ both the $\tau=e^{i\pi/3}$ and the $e^{i\pi/2}$ $dS_2$ vacua can be stable.
These are the results for a periodic fermion.
The vacuum structure of this model is summarized in Fig.~\ref{Fig:Result_T2U(1)}.

If we slightly change the boundary condition from a periodic one, we still have extrema around $\tau=e^{i\pi/3}, e^{i\pi/2}$.
We found essentially the same result, i.e., the existence of $AdS_2$ and $dS_2$ vacua.
More comprehensive analysis with general boundary conditions will be presented elsewhere.


\subsection{SM on $T^2$}\label{Sec:T2 SM}
Now we move on to consider the vacuum structure of $T^2$ compactification of the SM.
Unfortunately, it is difficult to completely analyze the extrema of a multi-dimensional potential.
Nevertheless even though a general analysis is too complicated, we can argue that if the charged matter contribution dominates the potential, the global minimum in the $\tau$ plane disappears.
To see this, let us consider the $\tau_2\to\infty$ limit.
The potential is
\al{\label{Eq:large tau2 limit}
V^{(1)}_{T^2}(\rho, \,&\tau,\theta_1,\theta_2)
\to
-{\tau_2^2\over32\pi^6\rho^2}\br{\text{Li}_4(e^{2\pi i\theta_1})+c.c}.
}
For massless contribution, this is valid for $\tau_2\gg 1$.
A necessary condition for the existence of a global minimum is the positivity of Eq.~\eqref{Eq:large tau2 limit}, which we will check below.
More precisely, we check the positivity of Eq.~\eqref{Eq:large tau2 limit} for each $\rho$, where only the contribution from particles whose mass $\sqrt{\rho}M<1$ is considered. 
The results are summarized in Tables~\ref{Table:tau_stability}, \ref{Table:tau_stability2}, \ref{Table:tau_stability3}, \ref{Table:tau_stability4} and \ref{Table:tau_stability5}.
In the tables, the minimum of $-(\text{Li}_4(e^{2\pi i\theta_1})+c.c)$ as a function of the Wilson line moduli is evaluated for Majorana and Dirac neutrinos with various boundary conditions.
Then, we can see that the $\tau$ moduli do not have a global minimum for $\rho^{-1/2}\gtrsim \GeV$ and for boundary conditions other than the periodic one.
This situation does not change if we start instead from the high energy vacuum in $4$ dimensions.

\begin{figure}
\begin{center}
\hfill
\includegraphics[width=.6\textwidth]{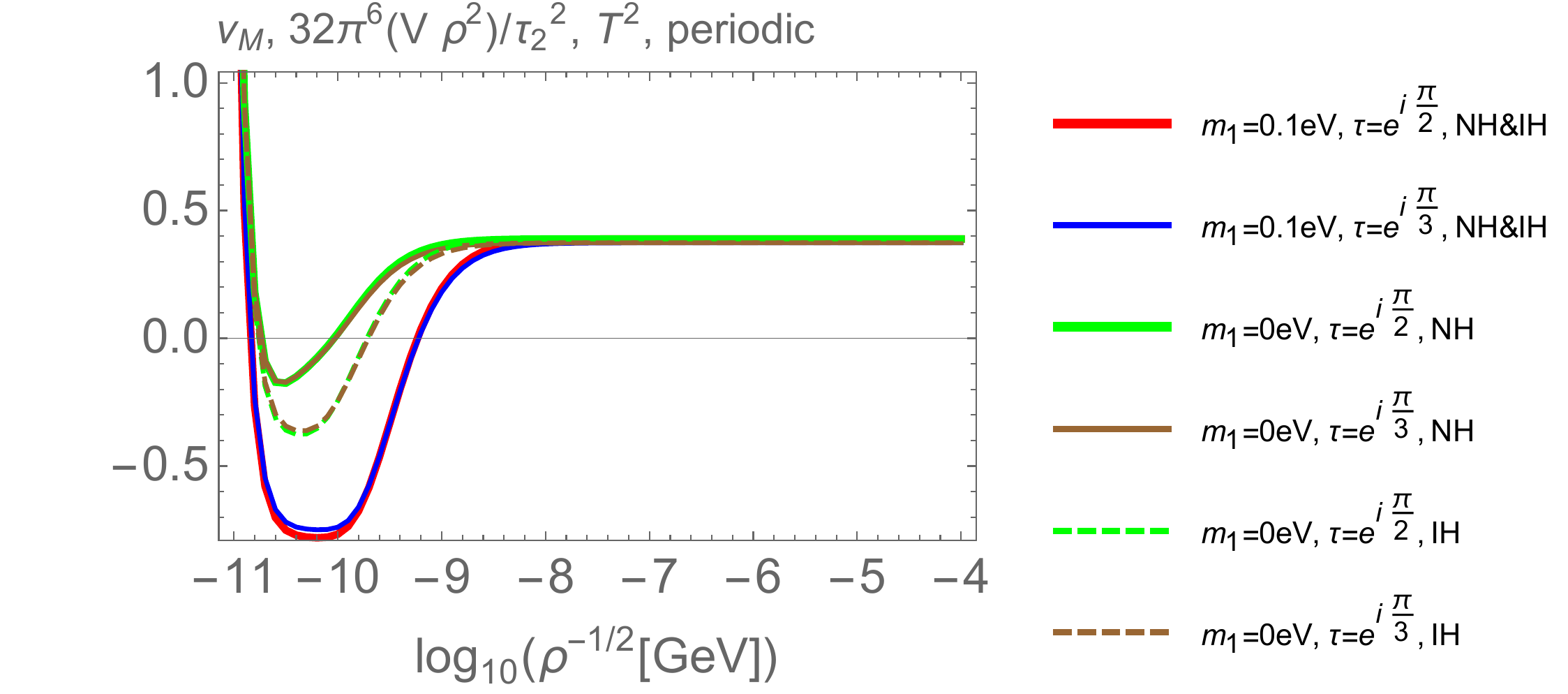}
\hfill\mbox{}
\end{center}
\caption{
Periodic Majorana neutrino.
The potential for the volume moduli $\rho$.
}
\label{Fig:T2PnuM}
\end{figure}

\begin{figure}
\begin{center}
\hfill
\includegraphics[width=.49\textwidth]{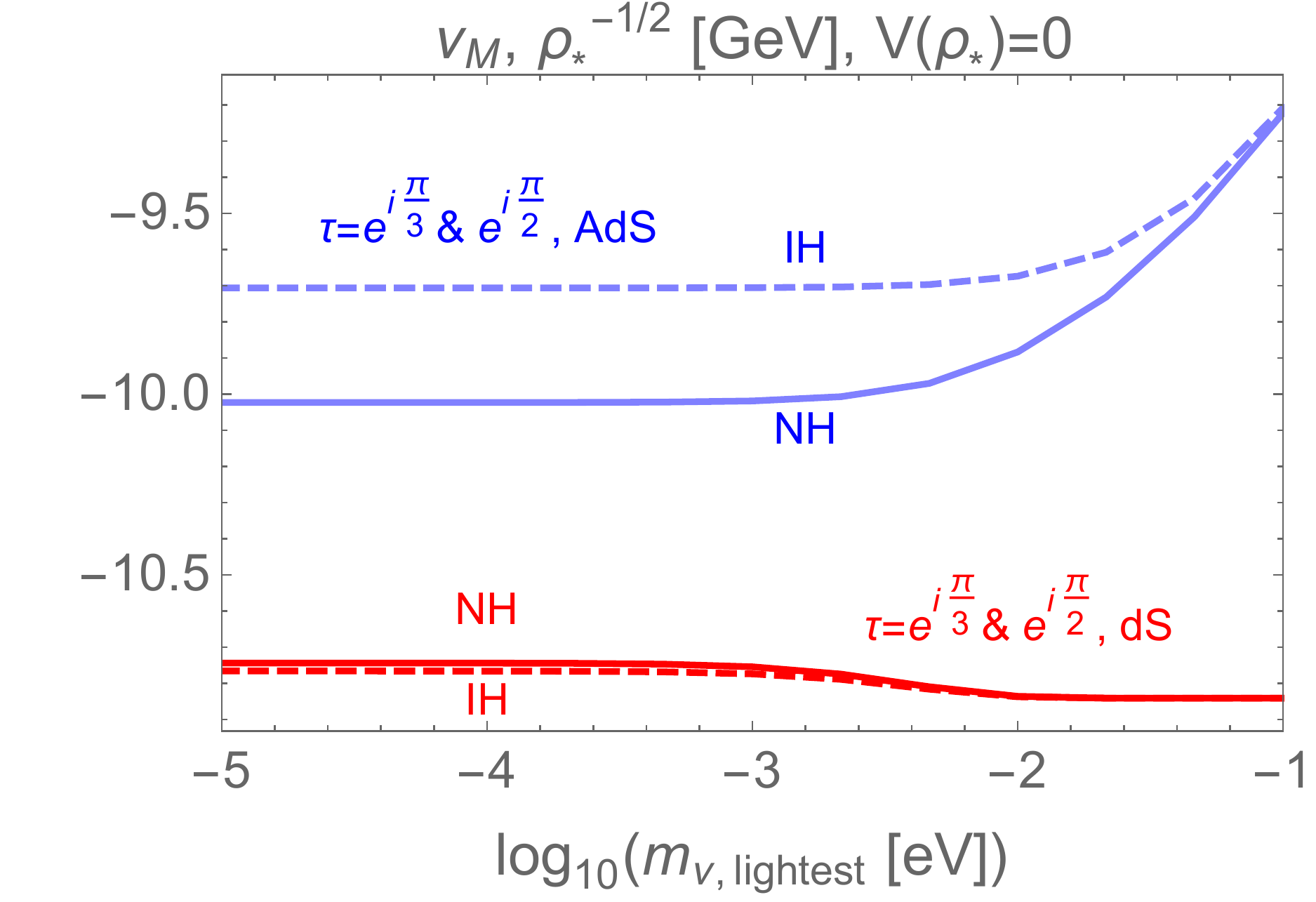}
\includegraphics[width=.49\textwidth]{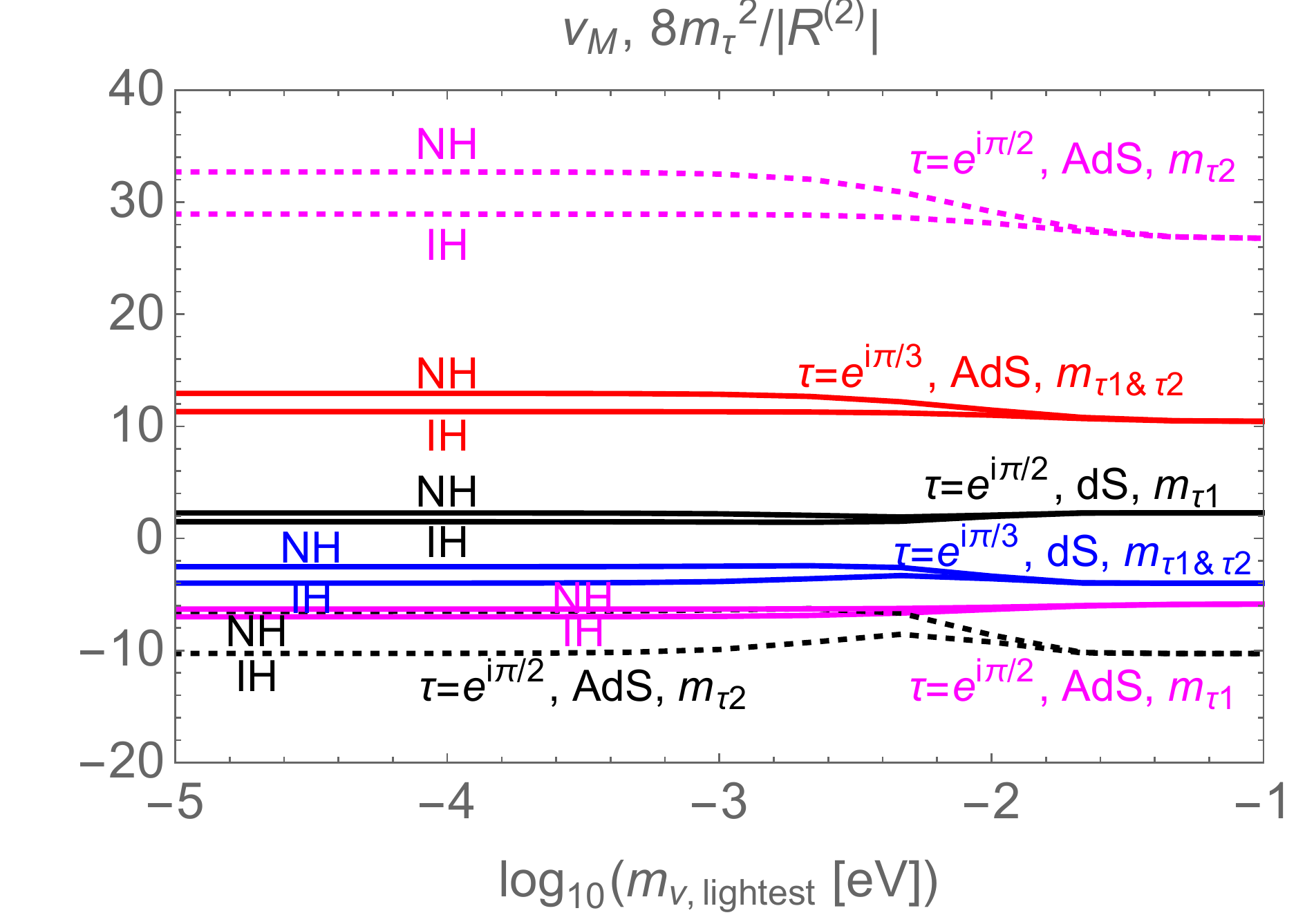}
\hfill\mbox{}
\end{center}
\caption{
Left: The value of $\rho_*$ which satisfies $V(\rho_*)=0$.
Right: The ratio between the mass of the $\tau$ moduli and the curvature of the $2$-dimensional spacetime. One can see that the only perturbatively stable vacuum is the $AdS$ minimum with $\tau=e^{i\pi/3}$.
}
\label{Fig:T2PnuM_massratio}
\end{figure}

\begin{figure}[t]
\begin{center}
\hfill
\includegraphics[width=.5\textwidth]{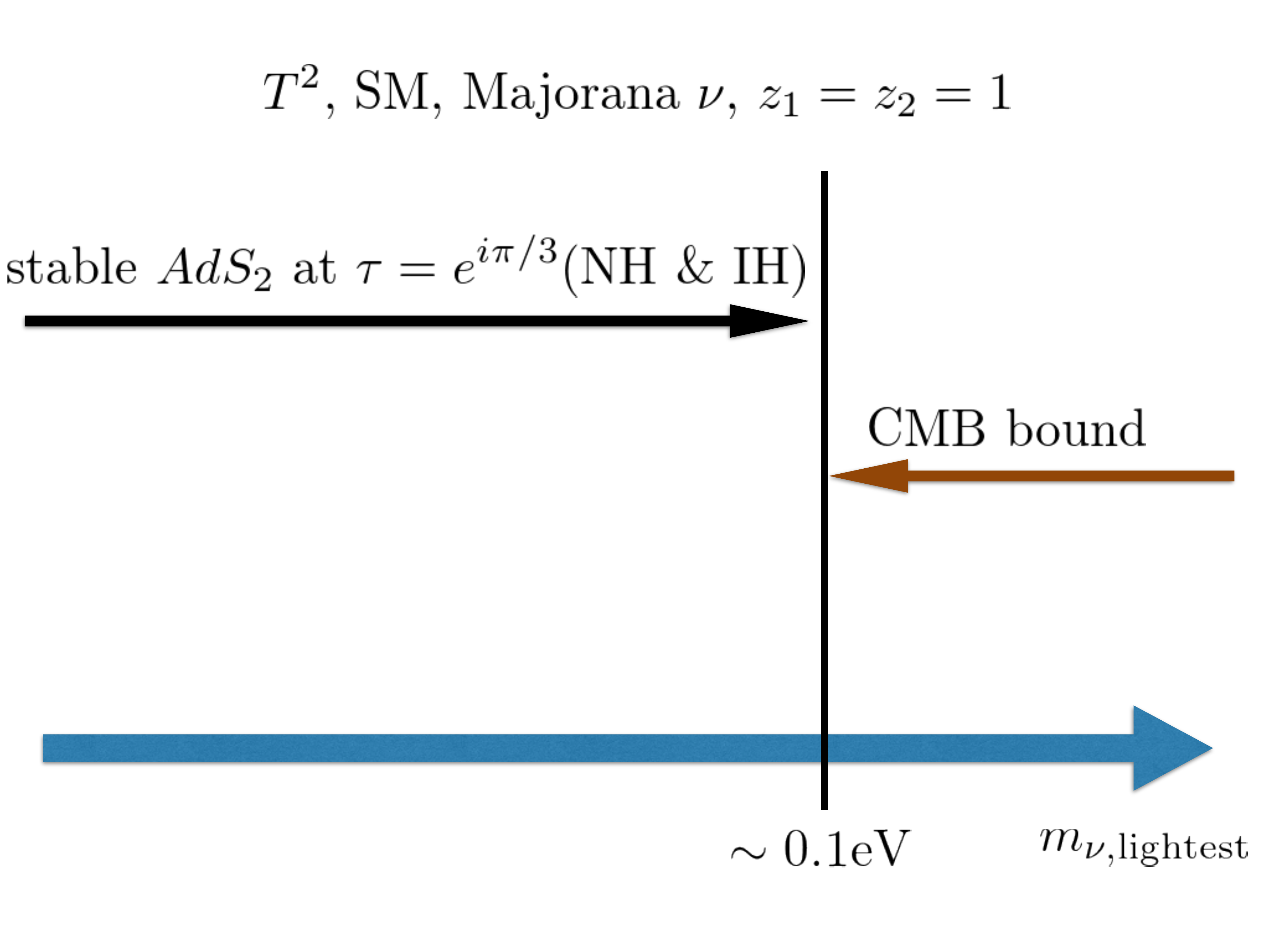}
\hfill\mbox{}
\end{center}
\caption{
Same as Fig.~\ref{Fig:Result_T2U(1)}, but for the SM with periodic Majorana neutrinos.
}
\label{Fig:Result_T2SMMajorana}
\end{figure}

\begin{figure}
\begin{center}
\hfill
\includegraphics[width=.49\textwidth]{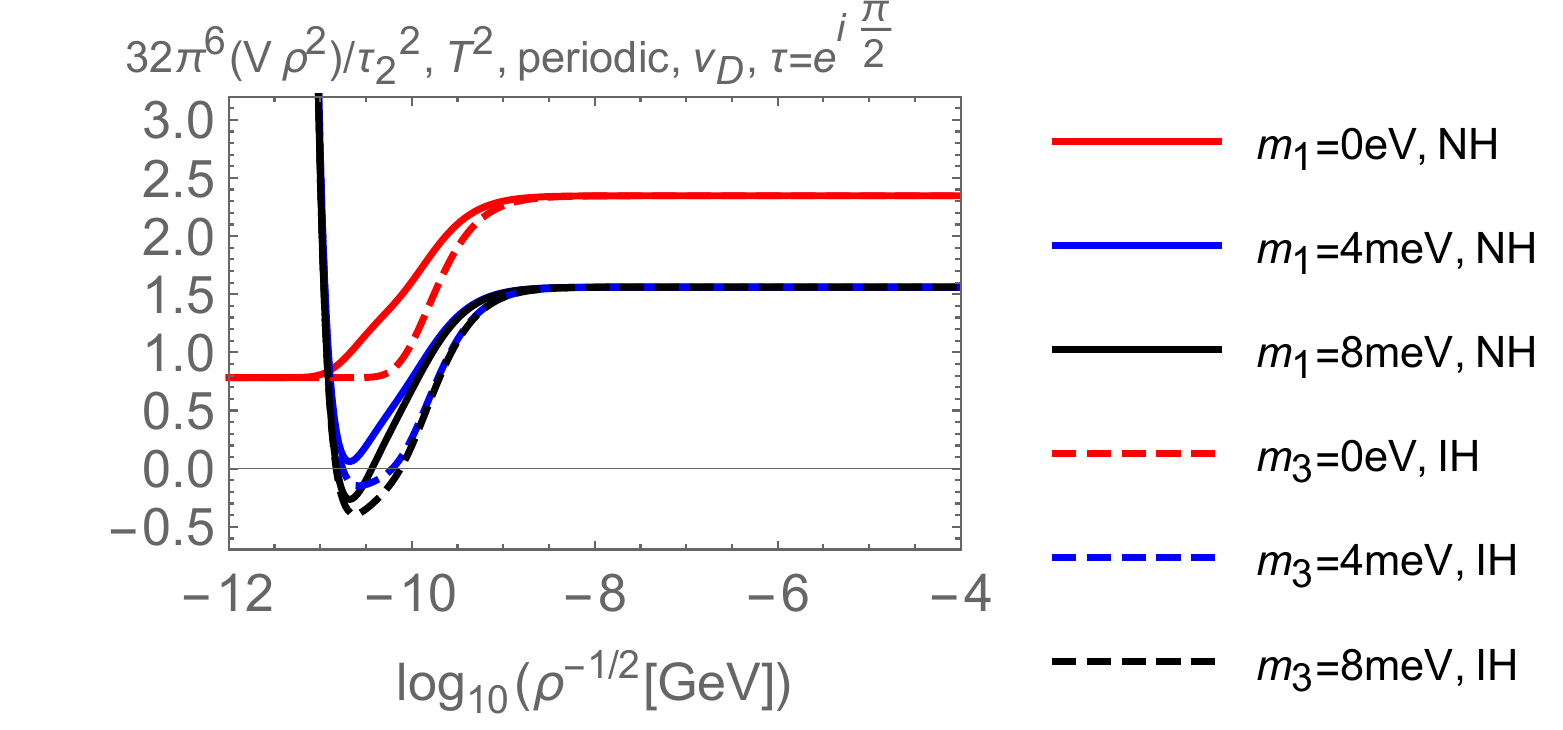}
\includegraphics[width=.49\textwidth]{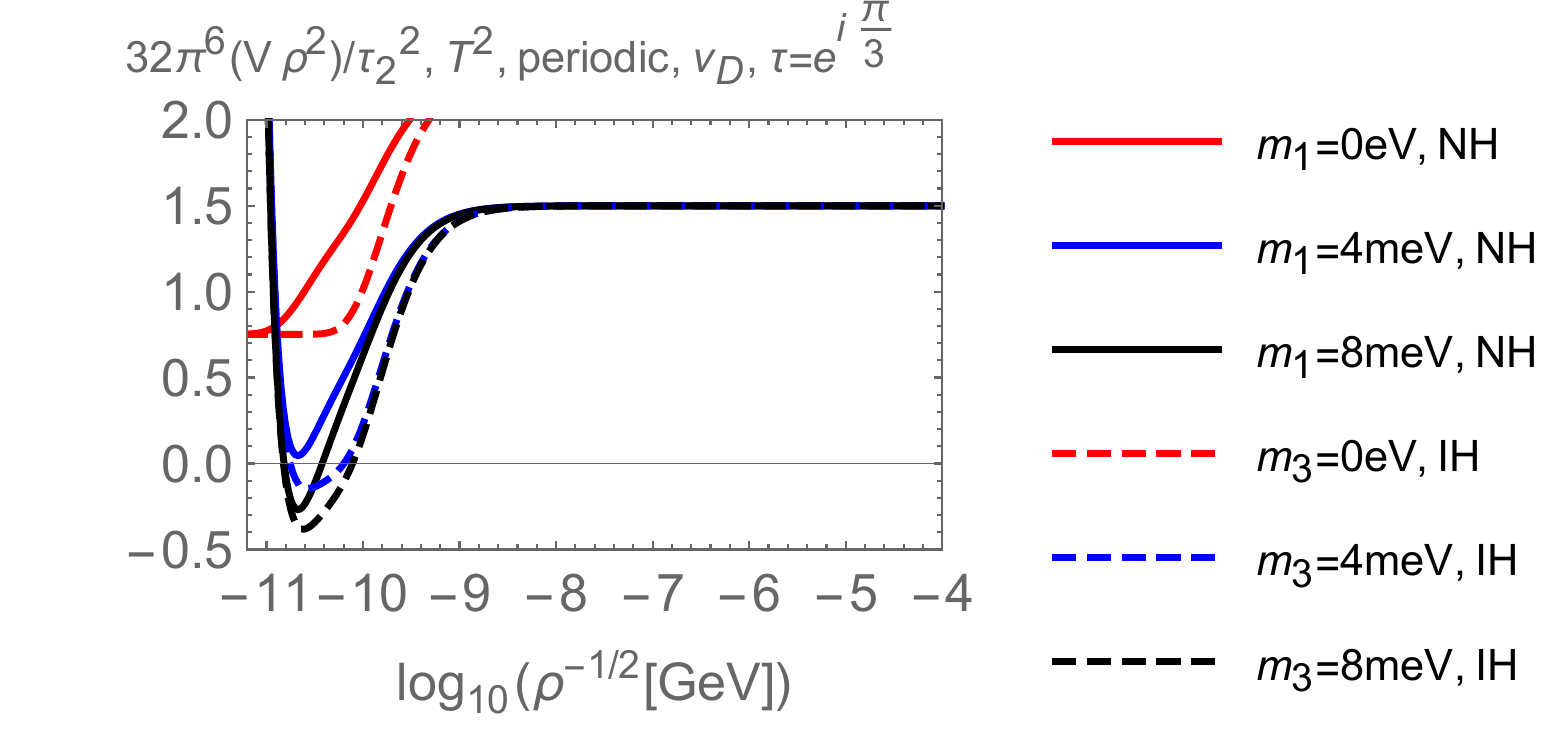}
\hfill\mbox{}
\end{center}
\caption{
Periodic Dirac neutrino with normal and inverted hierarchies.
}
\label{Fig:T2PnuD}
\end{figure}

When the neutrino term is dominant in the potential, we may have lower dimensional vacua according to the discussion of Sec.~\ref{Sec:T2 neutral}.
We show the potential $V=V_{T^2}^\text{all}+\Lambda_4$ below the MeV scale in Figs.~\ref{Fig:T2PnuM}, \ref{Fig:T2PnuD} in the case of periodic boundary condition.
Fig.~\ref{Fig:T2PnuM} corresponds to Majorana neutrino, and there are two solutions of $V=0$ for any value of $m_{\nu,\text{lightest}}$ allowed by the experiment, $m_{\nu,\text{lightest}}\lesssim0.1\eV$~\cite{Ade:2015xua,Bilenky:2012qi}.
In the left panel of Fig.~\ref{Fig:T2PnuM_massratio}, we plot the value of $\rho$ moduli, $\rho_*$, corresponding to the solution of $V=0$.
In the right panel of Fig.~\ref{Fig:T2PnuM_massratio}, the perturbative stability of each solution is investigated, from which it is concluded that only the $AdS_2$ vacuum corresponding to $\tau=e^{i\pi/3}$ is stable.
This result is summarized in Fig.~\ref{Fig:Result_T2SMMajorana}, and is consistent with Ref.~\cite{Arnold:2010qz}.

\begin{figure}
\begin{center}
\hfill
\includegraphics[width=.6\textwidth]{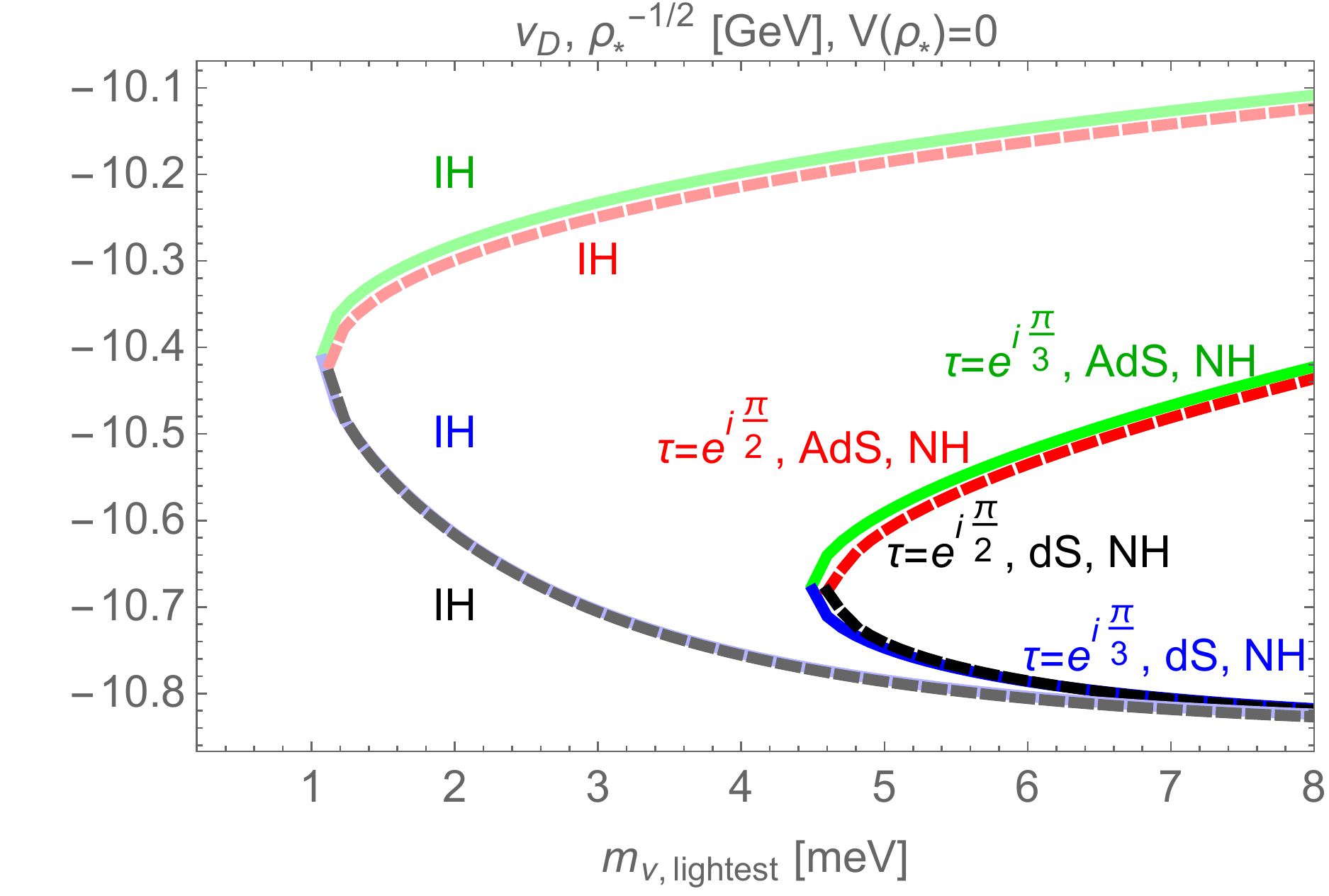}
\hfill\mbox{}
\end{center}
\caption{
The value of $\rho_*$ which gives $V(\rho_*)=0$ as a function of the lightest neutrino mass, $m_{\nu,\text{lightest}}$.
In order to have a solution, we need to have $4.5\meV\lesssim m_{\nu,\text{lightest}}$ for NH and $1.1\meV\lesssim m_{\nu,\text{lightest}}$ for IH, which is consistent with the result of Ref.~\cite{Arnold:2010qz}.
}
\label{Fig:T2PnuD15}
\end{figure}

\begin{figure}
\begin{center}
\hfill
\includegraphics[width=.49\textwidth]{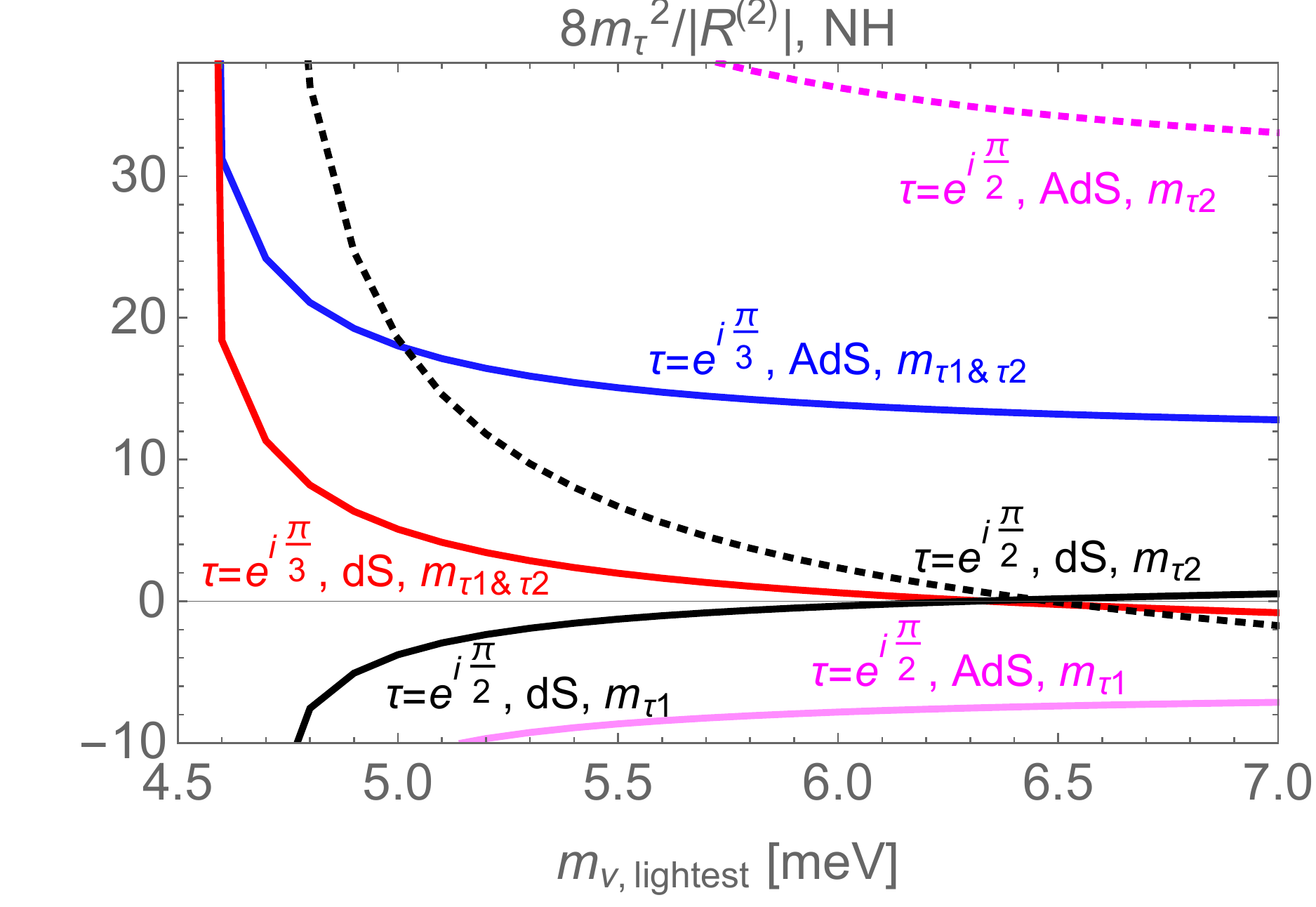}
\includegraphics[width=.49\textwidth]{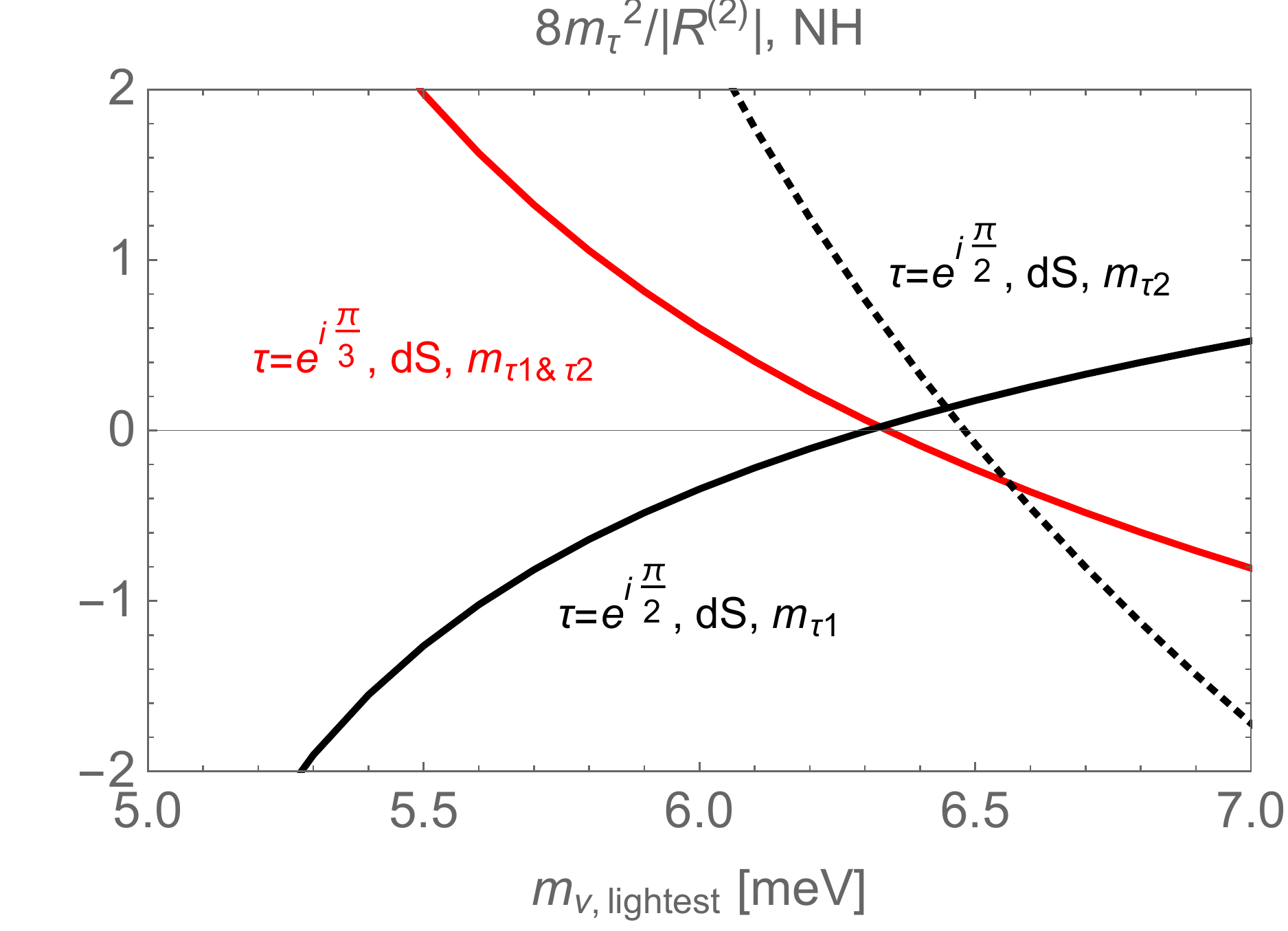}
\hfill\mbox{}
\\
\hfill
\includegraphics[width=.49\textwidth]{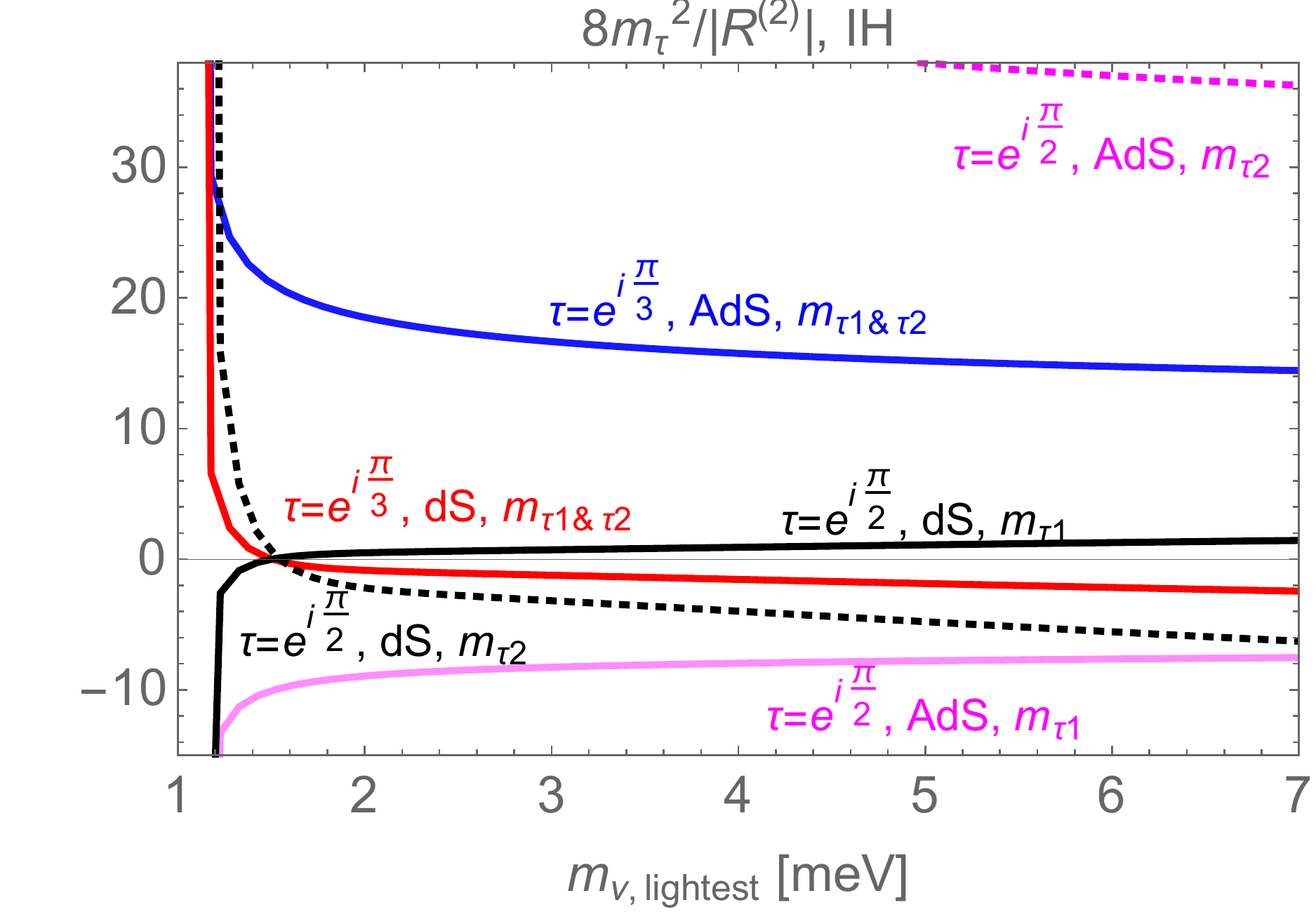}
\includegraphics[width=.49\textwidth]{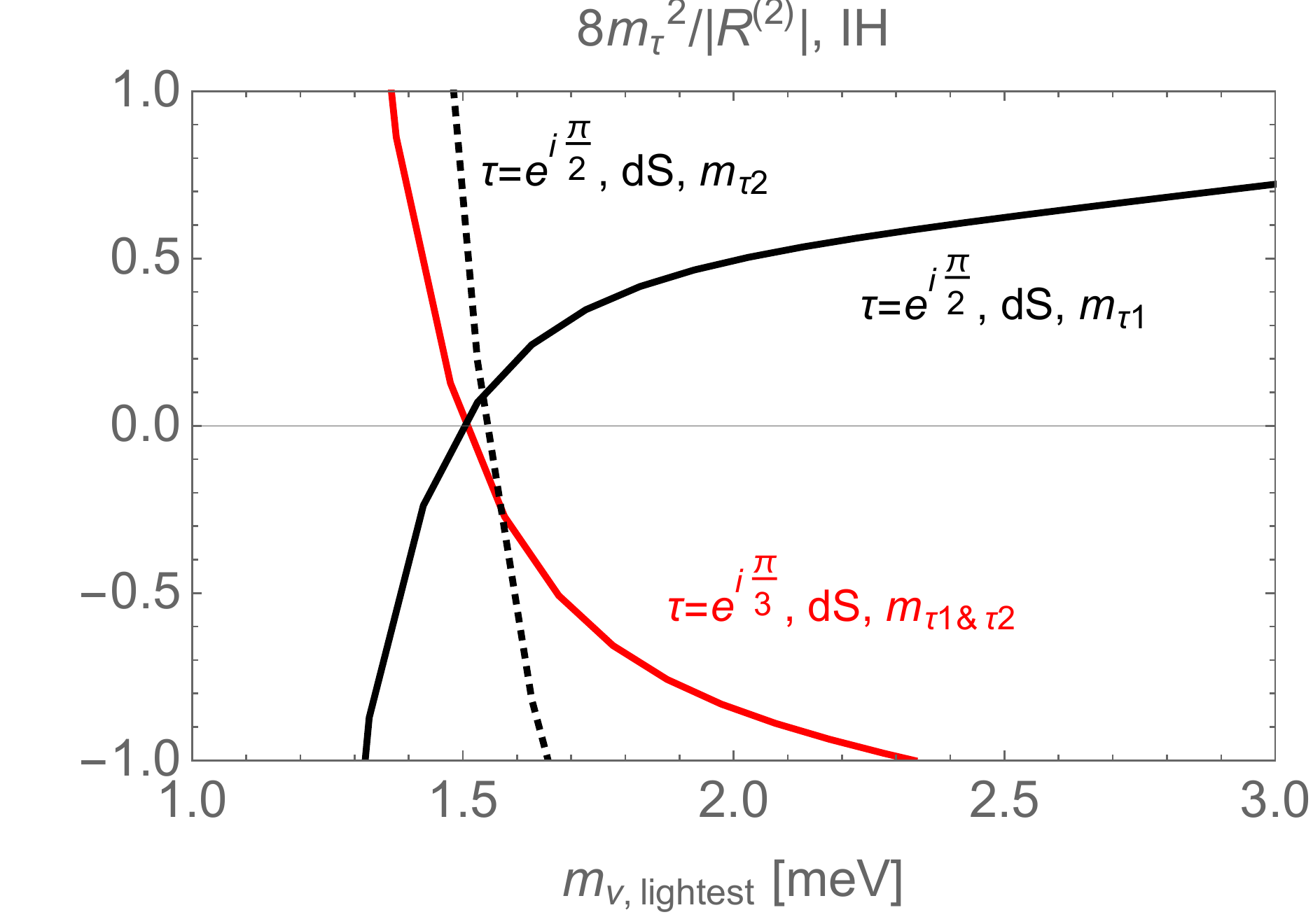}
\hfill\mbox{}
\end{center}
\caption{
Upper left: The mass and curvature ratio is plotted for the periodic Dirac neutrino with NH.
The $AdS_2$ minimum corresponding to $\tau=e^{i\pi/3}$ is always stable.
Upper right: The enlarged view of the upper left figure around $6\meV$.
One can see that the $dS_2$ minimum where $\tau=e^{i\pi/3}$ is stable for $4.5\meV\lesssim m_{\nu,\text{lightest}} \lesssim 6.3\meV$, while the $dS_2$ minimum where $\tau=e^{i\pi/2}$ is stable for $6.3\meV\lesssim m_{\nu,\text{lightest}} \lesssim 6.5\meV$. 
Lower left: Same as upper left figure, but for IH.
Lower right: The enlarged view of the lower left figure around $2\meV$.
If $1.1\meV\lesssim m_{\nu,\text{lightest}} \lesssim 1.5\meV$, the $dS_2$ minimum with $\tau=e^{i\pi/3}$ is stable. 
The $dS_2$ minimum where $\tau=e^{i\pi/2}$ is stable for $1.5\meV\lesssim m_{\nu,\text{lightest}} \lesssim 1.55\meV$.
The $dS_2$ solutions were overlooked in Ref.~\cite{Arnold:2010qz}.
}
\label{Fig:T2PnuD2}
\end{figure}

\begin{figure}[t]
\begin{center}
\hfill
\includegraphics[width=.47\textwidth]{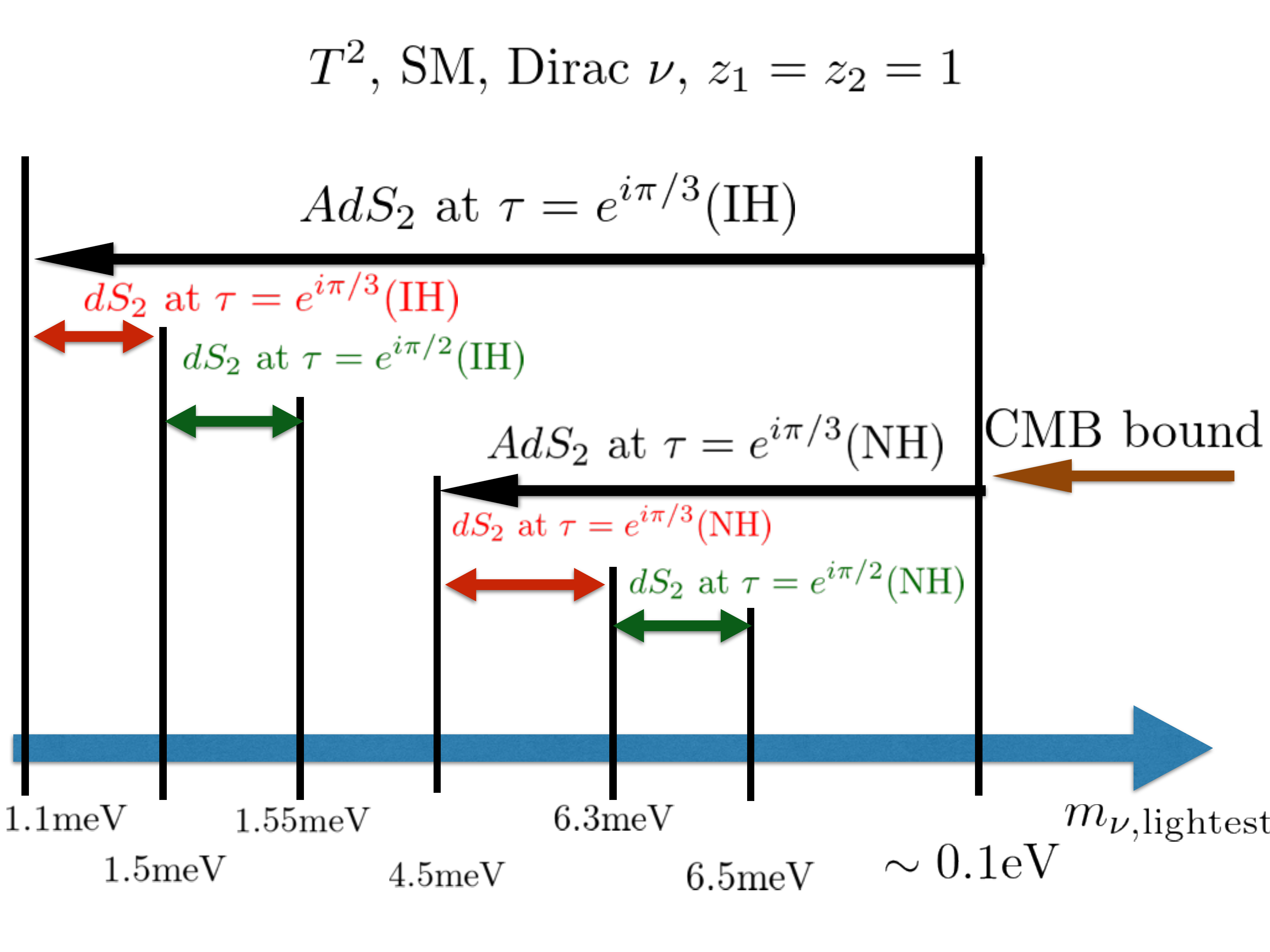}
\includegraphics[width=.47\textwidth]{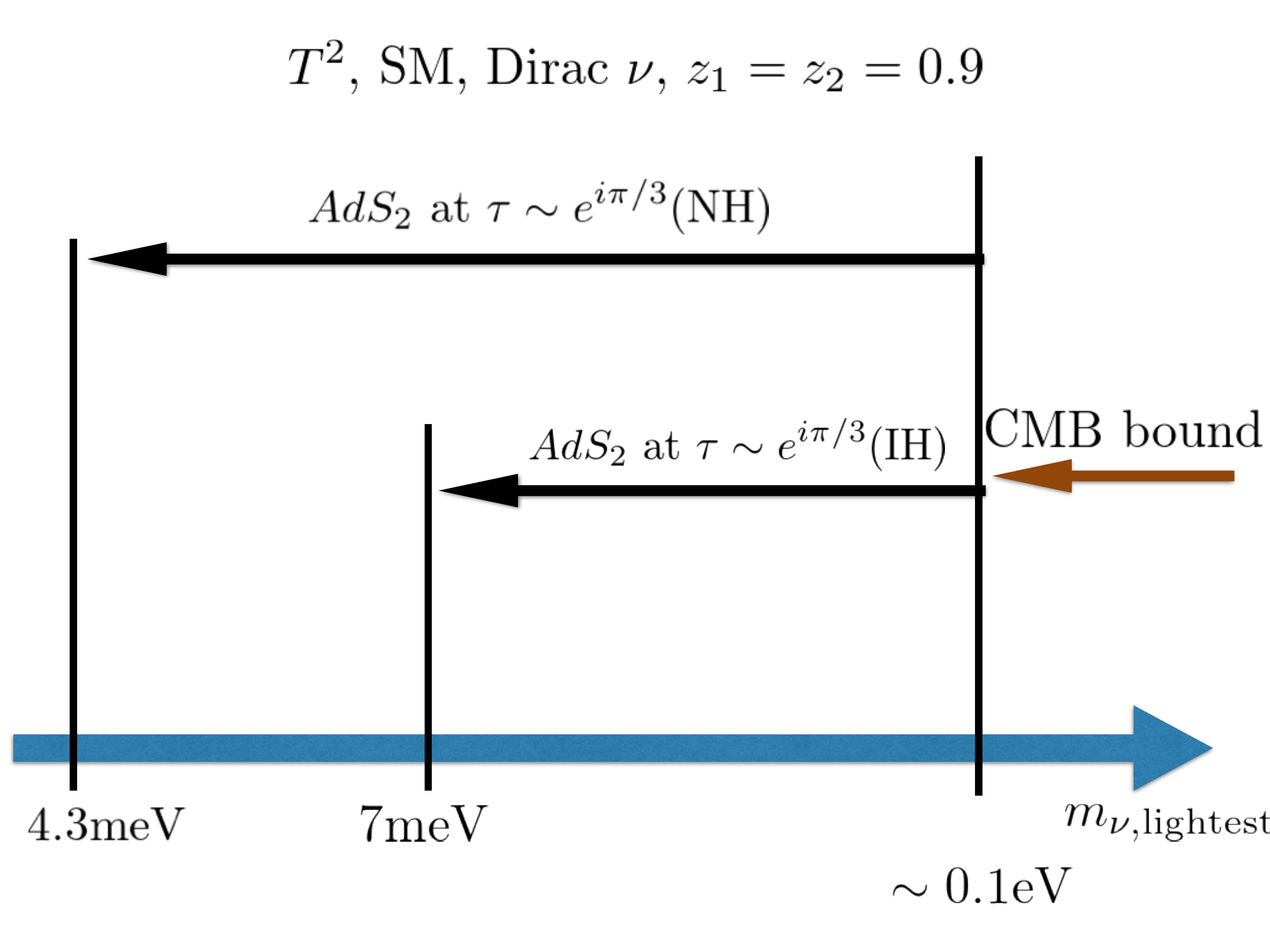}
\hfill\mbox{}
\end{center}
\caption{
Same as Fig.~\ref{Fig:Result_T2U(1)}, but for the SM with Dirac neutrino.
Regarding the boundary condition $z_1=z_2=0.9$, we concentrate on the $AdS_2$ vacuum where the value of $\tau$ is around $e^{i\pi/3}$.
}
\label{Fig:Result_T2SMDirac}
\end{figure}

In the Fig.~\ref{Fig:T2PnuD}, the potential corresponding to Dirac neutrino is plotted. If $m_1\lesssim 4.5\meV$ or $m_3\lesssim 1.1\meV$, no solution of $V=0$ exist.
The value of $\rho$ corresponding to the $V=0$ is plotted in Fig.~\ref{Fig:T2PnuD15}.
The stability of the solution is shown in Fig.~\ref{Fig:T2PnuD2}.
Upper and lower figures corresponds to the NH and IH neutrinos, respectively.
Regarding the case of NH, there always exists the $AdS_2$ vacuum where $\tau=e^{i\pi/3}$ for $m_1\gtrsim 4.5\meV$. 
The $dS_2$ vacua where $\tau=e^{i\pi/3}, e^{i\pi/2}$ are perturbatively stable for $4.5\meV\lesssim m_1\lesssim6.3\meV$ and $6.3\meV\lesssim m_1\lesssim6.5\meV$, respectively, as summarized in the left panel of Fig.~\ref{Fig:Result_T2SMDirac}.
On the other hand, in the case of IH, the stable $AdS_2$ vacuum with $\tau=e^{i\pi/3}$ can be obtained for $m_3\gtrsim 1.1\meV$.
The $dS_2$ vacua with $\tau=e^{i\pi/3}$ and $e^{i\pi/2}$ appear if $1.1\meV\lesssim m_3\lesssim1.5\meV$ and $1.5\meV\lesssim m_3\lesssim1.55\meV$, respectively.
The non-perturbative stability of these vacua is not clear. It  would be interesting to investigate this issue further.
If it turns out that these vacua are stable, we can constrain the neutrino parameters according to the conjecture~\cite{Ooguri:2016pdq,Freivogel:2016qwc}.
 
Interestingly, if we apply the multiple point criticality principle, as in the $S^1$ compactification, the lightest neutrino mass is predicted to be around $\mathcal{O}(1\text{--}10)\meV$ where the $T^2$ vacuum has a curvature close to the our four-dimensional vacuum.

\begin{figure}
\begin{center}
\hfill
\includegraphics[width=.49\textwidth]{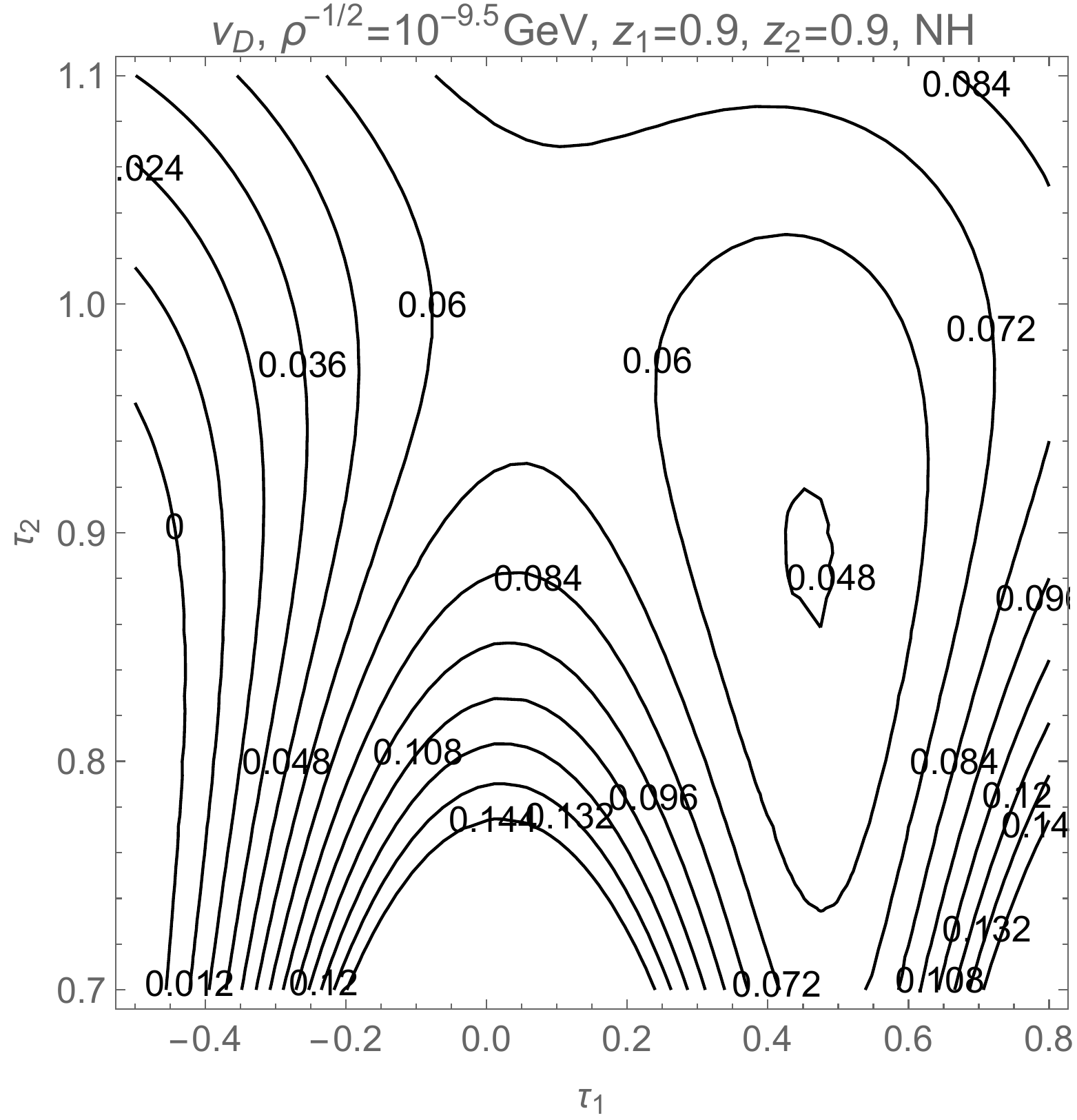}
\includegraphics[width=.49\textwidth]{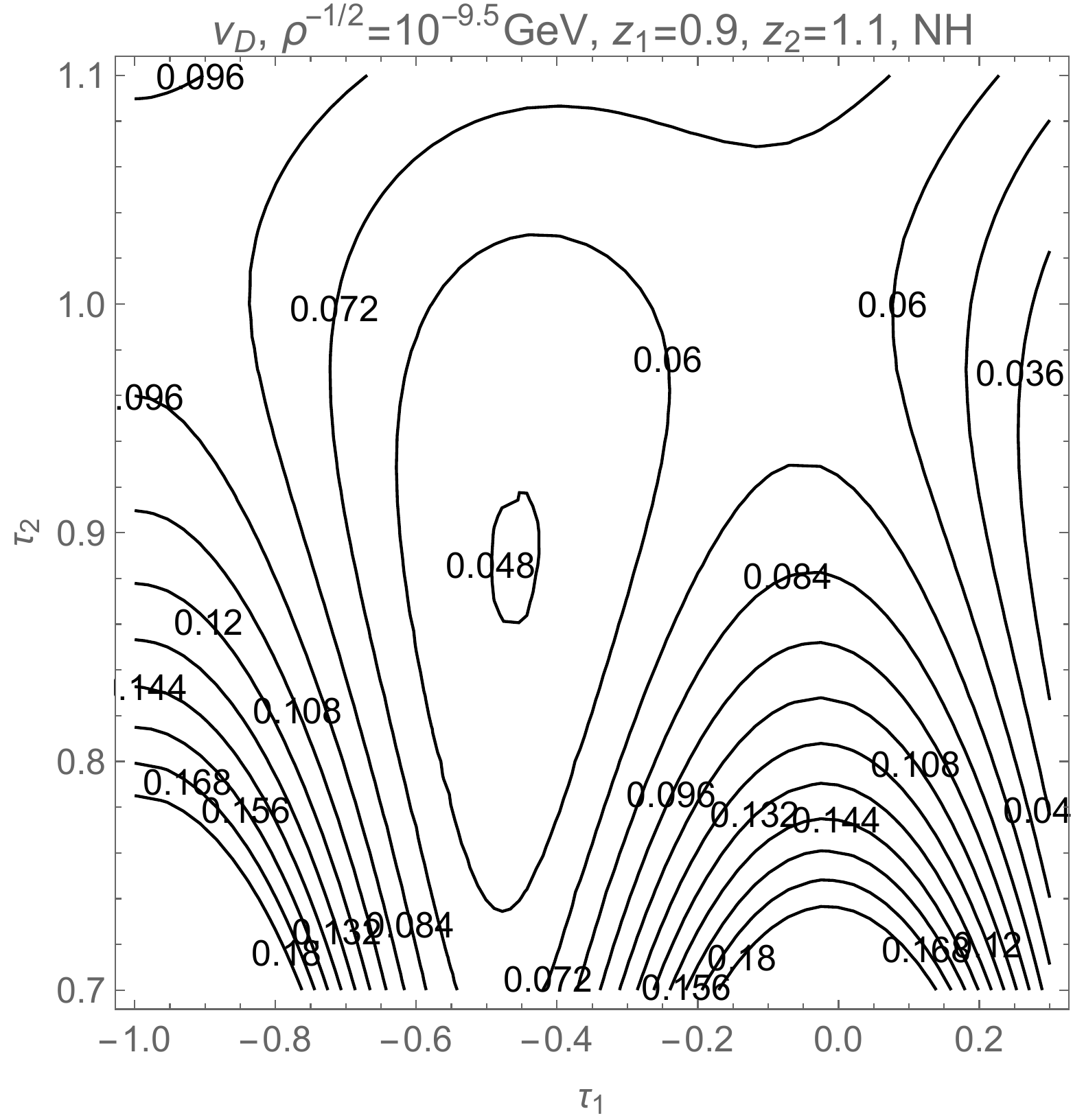}
\hfill\mbox{}
\\
\hfill
\includegraphics[width=.49\textwidth]{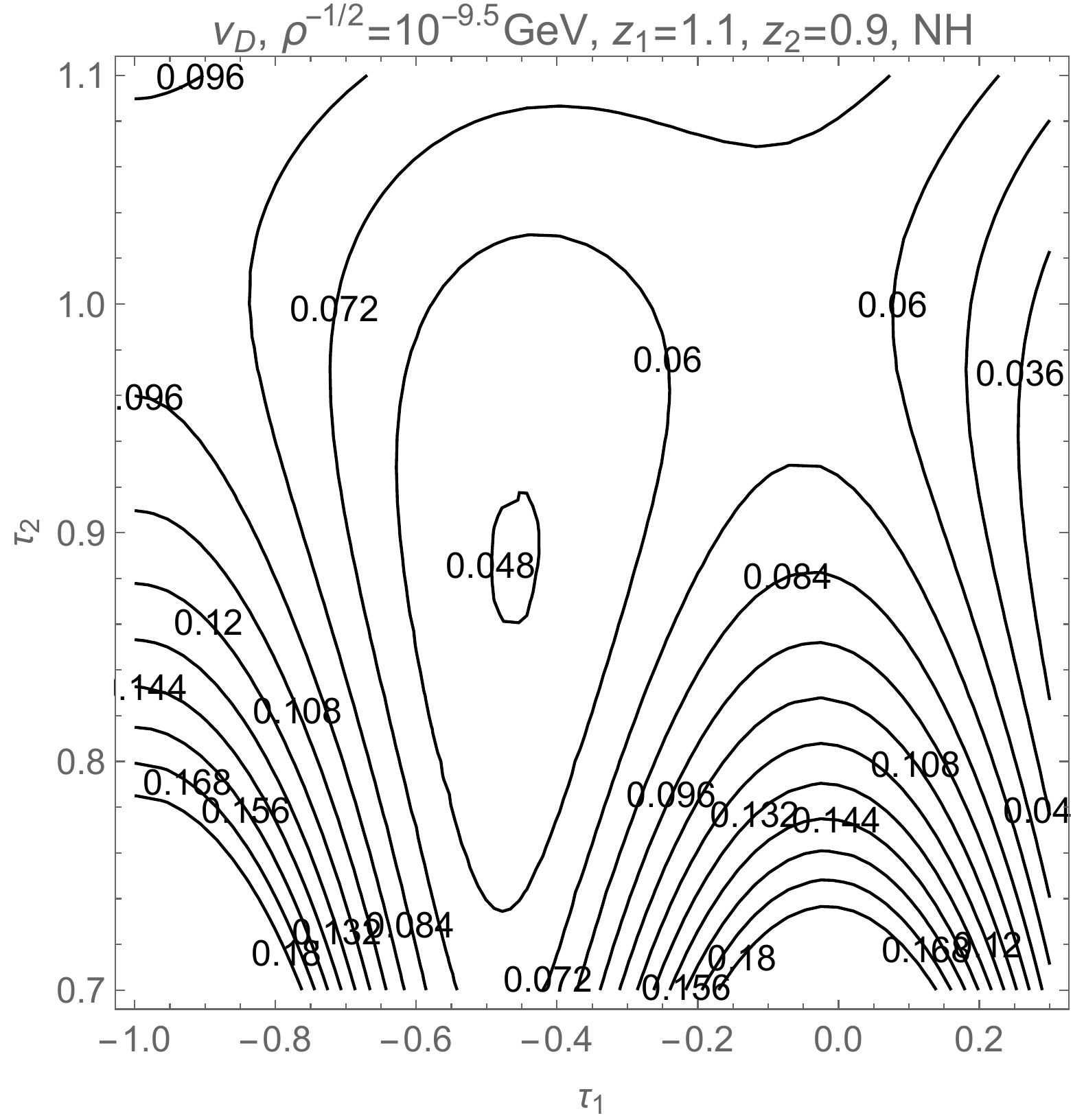}
\includegraphics[width=.49\textwidth]{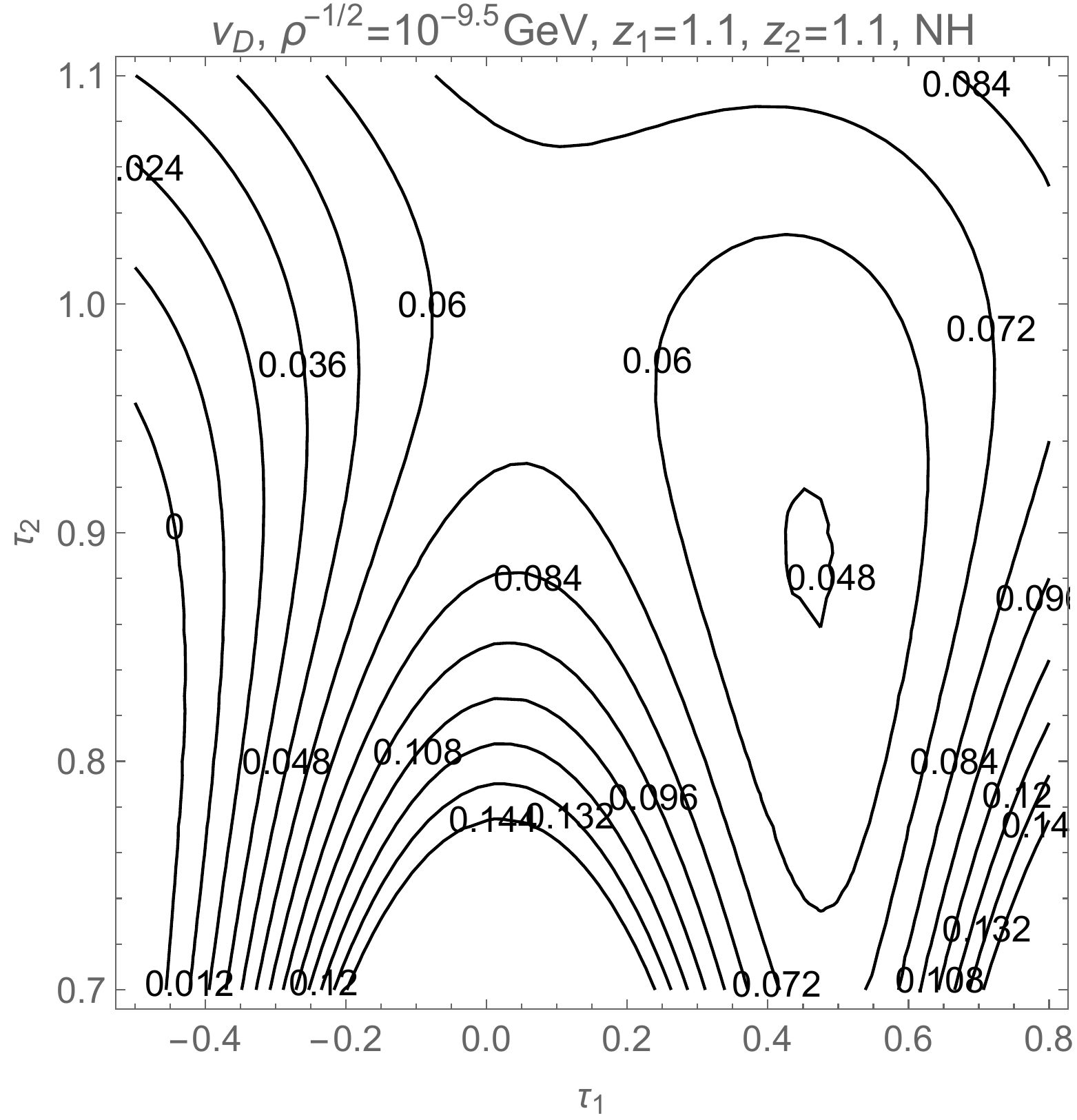}
\hfill\mbox{}
\end{center}
\caption{
The minimum around $\tau=e^{i\pi/3}$ survives as a local minimum if the boundary condition is close to periodic.
}
\label{Fig:tau_minimum}
\end{figure}

Even if the values of $z_{1,2}$ are away from $1$, as long as they are close to $1$, the minimum in the $\tau$ plane survives though it is no longer a global one.
Numerically, we have checked that the local minima exists for $0.9\lesssim z_{1,2}\lesssim1.1$, see Fig.~\ref{Fig:tau_minimum}.
Indeed, we can find perturbatively stable vacua for this range of boundary conditions.
In the right panel of Fig.~\ref{Fig:Result_T2SMDirac}, the condition for the neutrino mass to obtain perturbatively stable $AdS_2$ vacua with $\tau\sim e^{i\pi/3}$ is presented.
\begin{table}
  \begin{center}
    \begin{tabular}{|c||c|c|} \hline
particle &  (potential at minimum)$\times(32\pi^6\rho^2/\tau_2^2)$ & $(A_{\gamma1},A_{g11},A_{g21})$ \\ \hline \hline
graviton, $\gamma$    & $-8.7$ &$(\text{--},\text{--},\text{--})$ \\
+$\nu$ &  $4.3$ & $(\text{--},\text{--},\text{--})$ \\
+$e$ &   $-3.2$ & $(1/2,\text{--},\text{--})$ \\
+$\mu$ &  $-11$ & $(1/2,\text{--},\text{--})$\\
+$\pi$&  $-9.2$ &  $(1/2,\text{--},\text{--})$\\
+$K$&  $-9.7$ & $(1/2,\text{--},\text{--})$\\
+$\eta_8$&  $-11$ & $(1/2,\text{--},\text{--})$\\
SM$+$graviton wo/ $t,W,Z,H$   & $-86$ &$(1.3,0.1,0.7)$ \\
Full SM$+$graviton   & $-110$ &$(2.7,0.1,0.7)$ \\
\hline
    \end{tabular}
  \end{center}
\caption{
The minimum of Eq.~\eqref{Eq:large tau2 limit} as a function of the Wilson line is shown. Here the neutrino is periodic and Majorana.
The third column is the value of the Wilson line field corresponding to the minimum.
}
\label{Table:tau_stability}
\end{table}

\begin{table}
  \begin{center}
    \begin{tabular}{|c||c|c|} \hline
particle &  (potential at minimum)$\times(32\pi^6\rho^2/\tau_2^2)$ & $(A_{\gamma1},A_{g11},A_{g21})$ \\ \hline \hline
graviton, $\gamma$    & $-8.7$ &$(\text{--},\text{--},\text{--})$ \\
+$\nu$ &  $-20$ & $(\text{--},\text{--},\text{--})$ \\
+$e$ &   $-28$ & $(0,\text{--},\text{--})$ \\
+$\mu$ &  $-35$ & $(0,\text{--},\text{--})$\\
+$\pi$&  $-42$ &  $(0,\text{--},\text{--})$\\
+$K$&  $-50$ & $(0,\text{--},\text{--})$\\
+$\eta_8$&  $-44$ & $(0,\text{--},\text{--})$\\
SM$+$graviton wo/ $t,W,Z,H$   & $-180$ &$(1,1,0.3)$ \\
Full SM$+$graviton   & $-220$ &$(1,1,0.3)$ \\
\hline
    \end{tabular}
  \end{center}
\caption{
Same as Fig.~\ref{Table:tau_stability}, but for anti-periodic, Majorana neutrinos.
}
\label{Table:tau_stability2}
\end{table}

\begin{table}
  \begin{center}
    \begin{tabular}{|c||c|c|} \hline
particle &  (potential at minimum)$\times(32\pi^6\rho^2/\tau_2^2)$ & $(A_{\gamma1},A_{g11},A_{g21})$ \\ \hline \hline
graviton, $\gamma$    & $-8.7$ &$(\text{--},\text{--},\text{--})$ \\
+$\nu$ &  $17$ & $(\text{--},\text{--},\text{--})$ \\
+$e$ &   $9.7$ & $(1/2,\text{--},\text{--})$ \\
+$\mu$ &  $2.2$ & $(1/2,\text{--},\text{--})$\\
+$\pi$&  $3.8$ &  $(1/2,\text{--},\text{--})$\\
+$K$&  $3.2$ & $(1/2,\text{--},\text{--})$\\
+$\eta_8$&  $1.6$ & $(1/2,\text{--},\text{--})$\\
SM$+$graviton wo/ $t,W,Z,H$   & $-73$ &$(1.3,0.1,0.7)$ \\
Full SM$+$graviton   & $-94$ &$(2.7,0.1,0.7)$ \\
\hline
    \end{tabular}
  \end{center}
\caption{
Same as Fig.~\ref{Table:tau_stability}, but for periodic, Dirac neutrinos.
}
\label{Table:tau_stability3}
\end{table}

\begin{table}
  \begin{center}
    \begin{tabular}{|c||c|c|} \hline
particle &  (potential at minimum)$\times(32\pi^6\rho^2/\tau_2^2)$ & $(A_{\gamma1},A_{g11},A_{g21})$ \\ \hline \hline
graviton, $\gamma$    & $-8.7$ &$(\text{--},\text{--},\text{--})$ \\
+$\nu$ &  $-10$ & $(\text{--},\text{--},\text{--})$ \\
+$e$ &   $-18$ & $(3/4,\text{--},\text{--})$ \\
+$\mu$ &  $-25$ & $(3/4,\text{--},\text{--})$\\
+$\pi$&  $-28$ &  $(0.8,\text{--},\text{--})$\\
+$K$&  $-34$ & $(0.8,\text{--},\text{--})$\\
+$\eta_8$&  $-30$ & $(0.8,\text{--},\text{--})$\\
SM$+$graviton wo/ $t,W,Z,H$   & $-170$ &$(1.8,0,0)$ \\
Full SM$+$graviton   & $-182$ &$(0.9,1,0.7)$ \\
\hline
    \end{tabular}
  \end{center}
\caption{
Same as Fig.~\ref{Table:tau_stability}, but for $z=1/2$, Dirac neutrinos.
}
\label{Table:tau_stability4}
\end{table}

\begin{table}
  \begin{center}
    \begin{tabular}{|c||c|c|} \hline
particle &  (potential at minimum)$\times(32\pi^6\rho^2/\tau_2^2)$ & $(A_{\gamma1},A_{g11},A_{g21})$ \\ \hline \hline
graviton, $\gamma$    & $-8.7$ &$(\text{--},\text{--},\text{--})$ \\
+$\nu$ &  $-31$ & $(\text{--},\text{--},\text{--})$ \\
+$e$ &   $-39$ & $(0,\text{--},\text{--})$ \\
+$\mu$ &  $-47$ & $(0,\text{--},\text{--})$\\
+$\pi$&  $-53$ &  $(0,\text{--},\text{--})$\\
+$K$&  $-62$ & $(0,\text{--},\text{--})$\\
+$\eta_8$&  $-55$ & $(0,\text{--},\text{--})$\\
SM$+$graviton wo/ $t,W,Z,H$   & $-188$ &$(1,1,0.3)$ \\
Full SM$+$graviton   & $-230$ &$(1,1,0.3)$ \\
\hline
    \end{tabular}
  \end{center}
\caption{
Same as Fig.~\ref{Table:tau_stability}, but for anti-periodic, Dirac neutrinos.
}
\label{Table:tau_stability5}
\end{table}

\subsection{Flux vacua}\label{Sec:T2 flux vacua}
Similarly to the $S^1$ compactification, we can consider  flux vacua at the tree level.
One crucial difference from the $S^1$ compactification case is that we can introduce a magnetic field without violating the Lorentz symmetry in $2$ dimensions,
with a contribution: 
\al{
S_\text{flux}&=\int d^2x\sqrt{-g_{(2)}}
\bigg[
-{\rho\over2}F_{23}F^{23}
-\rho\Lambda_4
\bigg]
=
\int d^2x\sqrt{-g_{(2)}}
\bigg[
-{1\over2\rho}F_{23}F_{kl} \gamma^{2k}\gamma^{3l}
-\rho\Lambda_4
\bigg]
\nn&
\simeq
\int d^2x\sqrt{-g_{(2)}}
\bigg[
-{1\over2\rho}F_{23}^2-\rho\Lambda_4
\bigg].
}
In the last line, we have assumed a flat $2$d spacetime.
The magnetic flux in compactified space should be quantized: $F_{23}=2\pi m$ where $m\in \mathbb{Z}$.

Therefore, if we add a magnetic flux, the potential is modified to
\al{
V=\rho\Lambda_4+ {1\over2}{(2\pi m)^2\over \rho}.
}
Then, we can find a solution to the Hamiltonian constraint, $V=0$ for a negative cosmological constant.
Note that the $\tau$ moduli does not acquire a tree-level potential from the flux contribution, and we may need the Casimir energy to fix $\tau$.

Next, let us consider the effect of the axion flux.
If we add two axion-like particles, we can fix the $\rho$ and $\tau$ moduli at the tree level if $\Lambda_4<0$.
Explicitly, 
we have
\al{
S_\text{axion}&=
\int d^4x\sqrt{-g}
\bigg[
 \Phi_1^* \Delta \Phi_1
+\Phi_2^* \Delta \Phi_2
-\Lambda_4
\bigg]\nn
&\simeq\int d^2x\sqrt{-g_{(2)}}\,\rho
\sqbr{
-{|m_1-n_1\tau|^2\over\rho\tau_2}f_{a,1}^2
-{|m_2-n_2\tau|^2\over\rho\tau_2}f_{a,2}^2
-\Lambda_4
}
}
where $\Phi_i$ is the $U(1)_{PQ}$ breaking field, and we have put $\Phi_i=f_{a,i} \exp\sqbr{in_i y_1+i m_i y_2}$.
\al{&
V=\rho \Lambda_4 +{|m_1-n_1\tau|^2\over\tau_2}f_{a,1}^2+{|m_2-n_2\tau|^2\over\tau_2}f_{a,2}^2,
&&
m_i, n_i\in \mathbb{Z}.
}

\begin{figure}
\begin{center}
\hfill
\includegraphics[width=.35\textwidth]{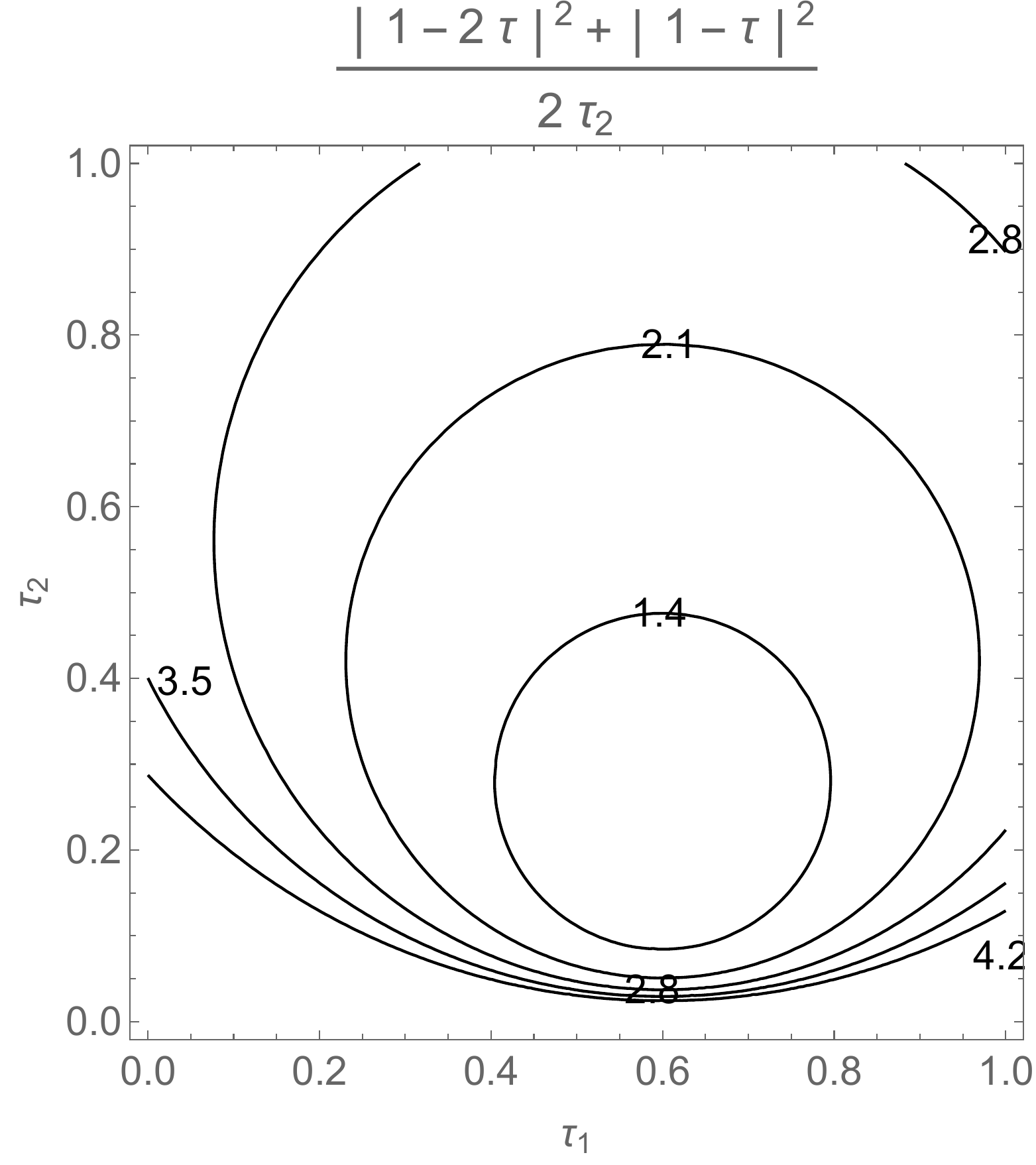}
\hfill\mbox{}
\end{center}
\caption{
The tree level stabilization of $\tau$ moduli.
}
\label{Fig:tau_potential}
\end{figure}

This shows that if $\Lambda_4<0$, we can fix all moduli at the tree level.
The $\tau$ moduli  can be fixed by an appropriate choice of $(m_{1,2},n_{1,2})$, and $V$ can become zero for $\rho\sim-f_a^2/\Lambda_4$.
The corresponding two-dimensional vacua is $AdS_2\times T^2$.
For example, if we take the parameter set $(m_1,n_1,m_2,n_2)=(1,1,1,2)$, we can fix the $\tau$ moduli as shown in Fig.~\ref{Fig:tau_potential}.
We can see that the minimum is around $(\tau_1,\tau_2)=(0.6,0.3)$.
As discussed in Sec.~\ref{Sec:S1 flux vacua}, this may imply that metastable electroweak vacuum with the addition of axions cannot be consistently embedded into a quantum theory of gravity.
There is also a possibility of moduli stabilization thanks to both the tree level potential and the one-loop Casimir energy. The tree level potential fixes the $\tau$ moduli at $\tau_1=m/n$, and the one-loop potential fixes the $\tau_2$ moduli.


\section{Summary and discussion}\label{Sec:summary}
In this paper, we have investigated the vacuum structure of the standard model upon compactification to lower dimensions.
Our work was motivated by the weak gravity conjecture and the multiple point principle, though our results
are of interest in their own right.
Understanding the myriad of vacua in the SM landscape may give us insights as to how  we evolve to our four-dimensional universe with the observed particle spectrum and interactions. 
Results from the LHC suggest the possibility that we can extrapolate the SM to rather high energies. 
Thus, studies along the lines of the present work may also elucidate what kinds of vacua  (albeit lower-dimensional ones) are permissible in an ultraviolet completable theory.

\begin{table}
  \begin{center}
  \scalebox{0.9}[0.9]{  
    \begin{tabular}{|c||c|c|c|c|} \hline
		& model 			                          & $AdS$ 									& flat 										& $dS$\\ \hline \hline
		& $U(1)$, neutral  		            	 & $\Lambda_4\lesssim10^{-2.8}M_e^4$			&$\Lambda_4\simeq10^{-2.8}M_e^4$					& $10^{-2.8}M_e^4\lesssim\Lambda_4\lesssim 10^{-2.6}M_e^4$\\ 
		& $U(1)$, charged   				 & -- 										& -- 											& -- \\ \cline{2-5}
		& SM, $\nu_M$ 				 & always 									& --											& -- \\ \cline{3-5}	
$S^1$	& SM, $\nu_D$, NH        			 & $8.4\meV\lesssim m_{\nu,\text{lightest}}$ 		&$m_{\nu,\text{lightest}}\simeq8.4\meV$				& $7.3\meV\lesssim m_{\nu,\text{lightest}}\lesssim8.4\meV$ \\ \cline{3-5}
		& SM, $\nu_D$, IH        			 & $3.1\meV\lesssim m_{\nu,\text{lightest}}$ 		&$m_{\nu,\text{lightest}}\simeq3.1\meV$				& $2.5\meV\lesssim m_{\nu,\text{lightest}}\lesssim3.1\meV$ \\ \cline{3-5}
		& SM, $\nu_M$, high scale	  	& --										& --											& -- \\ \cline{3-5}	
		& SM, $\nu_D$, high scale  		& $\Lambda_4\ll \text{(neutrino mass)}^4$ 		& --											& -- \\ \cline{2-5}						
		& axion  						& $\Lambda_4<0$ 							&-- 											&--\\ \cline{1-5}
		& $U(1)$, neutral 				& $\Lambda_4\lesssim10^{-2.1}M_e^4$ 			&$\Lambda_4\simeq10^{-2.1}M_e^4$				&$10^{-2.5}M_e^4\lesssim\Lambda_4\lesssim10^{-2.1}M_e^4$ \\
		& $U(1)$, charged 				& -- 										&--											&--\\ \cline{2-5}
$T^2$	& SM, $\nu_M$ 				& always 									&--											&-- \\ \cline{3-5}
		& SM, $\nu_D$, NH  				& $4.5\meV\lesssim m_{\nu,\text{lightest}}$ 		&$m_{\nu,\text{lightest}}\simeq4.5\meV$	& $4.5\meV\lesssim m_{\nu,\text{lightest}}\lesssim6.5\meV$\\ \cline{3-5}
		& SM, $\nu_D$, IH  				& $1.1\meV\lesssim m_{\nu,\text{lightest}}$ 		&$m_{\nu,\text{lightest}}\simeq1.1\meV$	& $1.1\meV\lesssim m_{\nu,\text{lightest}}\lesssim1.55\meV$ \\ \cline{2-5}		
		& axion 						& $\Lambda_4<0$ 				& --					 & -- \\ \hline
\end{tabular}
}
\end{center}
\caption{
A summary of the analysis in this paper.
Here the periodic boundary condition is taken.
We also impose the current upper bound on the neutrino mass, $m_{\nu,\text{lightest}}\lesssim0.1\eV$~\cite{Ade:2015xua,Bilenky:2012qi}.
}
\label{Table:Summary}
\end{table}

The vacuum structure of the SM (and the warmup $U(1)$ gauge theory example) compactified on $S^1$ and $T^2$ is summarized in Table \ref{Table:Summary}.
For an $S^1$ compactification, we found that there are no 3D vacua except for the neutrino one, and this neutrino vacuum is likely to be unstable under tunneling.
However, if the SM is supplemented with an additional axion, we found a lot of flux vacua from compactifications of the high scale vacuum with a negative 4D cosmological constant.
For  a $T^2$ compactification, we have calculated the Casimir energy for general boundary conditions of fields in the compact space. We have clarified the criteria for a perturbatively stable vacuum upon compactifiying on an $T^2$.
As a result, we found new $dS_2\times T^2$ vacua which were overlooked before. 
Moreover, previous studies have mostly been focussing on the compactifying the electroweak vacuum.
In this work, we have considered compactifications not only of the electroweak vacuum but also of the high scale Higgs vacuum.
The non-perturbative stability of $T^2$ vacuum is more subtle than the $S^1$ case. Following the discussion in the first part of Sec.~\ref{Sec:T2 SM} around Tables~\ref{Table:tau_stability},\ref{Table:tau_stability2},\ref{Table:tau_stability3},\ref{Table:tau_stability4},\ref{Table:tau_stability5}, we have found that for sufficiently small $\rho$, the potential for $\tau$ moduli is unbounded. However, $\rho$ is not dynamical field, and hence it is not clear that this unbounded potential implies the instability of the neutrino vacuum. This point needs further investigation.

In the case of $S^1$ compactification with Casimir energy, our results seem to be consistent with a recent conjecture that all non-supersymmetric $AdS$ solution are unstable~\cite{Ooguri:2016pdq,Freivogel:2016qwc}.
However, the fate of these $AdS$ solutions is not entirely clear, and we leave
the construction of the solution which describes the decay of the lower dimensional vacuum for a future publication.
On the contrary, if it is found that the lower dimensional $AdS$ vacuum
cannot decay (e.g., due to arguments alone the lines of Ref.~\cite{Banks:2010tj}), we can constrain the nature (Majorana vs Dirac) and the mass of the lightest neutrino according to the discussion of Refs.~\cite{Ooguri:2016pdq,Freivogel:2016qwc}.

The present work fits in the broader context of distinguishing the landscape from the swampland 
based on the requirement that quantum gravity should be well-behaved under compactification.
This consistency requirement has been tested against the weak gravity conjecture \cite{Brown:2015iha,Heidenreich:2015nta,Heidenreich:2016aqi,Montero:2017yja}. 
Recently, Ref.~\cite{Montero:2017yja} discussed the consistency of quantum gravity upon compactification to two dimensions.
In this work, we presented the criterion to obtain perturbatively stable vacuum in two dimensions, and thus we expect our findings to have applications in this and related contexts.

Furthermore, we speculated on the nature and value of the neutrino mass, based on the multiple point criticality principle.
By requiring the existence of lower dimensional vacua is close to the flat vacua.
we predict 
that the neutrinos have a Dirac mass, with the mass of the lightest neutrino $\sim \mathcal{O}(1\text{--}10)\meV$.
This prediction implies the absence of the neutrino-less double beta decay.
Current CMB measurements put a bound on the sum of the neutrino mass to be $\sim0.2\eV$~\cite{Ade:2015xua}.
Our prediction for the sum of the neutrino mass is $\sum_i m_{\nu i}\sim0.06\text{--}0.07\eV$ for NH and $0.10\text{--}0.11\eV$ for IH.
Future $21$cm observations such as SKA~\cite{2013ExA....36..235M}, CMB observation such POLARBEAR-2 and the Simons Array Experiment~\cite{Suzuki:2015zzg}, and baryon acoustic oscillation observations such as DESI~\cite{Levi:2013gra} further constrain the neutrino mass to a precision that our prediction could be tested~\cite{Liu:2015txa,Oyama:2015gma}.

It would be also interesting to study the cosmological consequences of the existence of anisotropic~\cite{Turner:1986vp} vacua such as those arising from compactifications on $S^1$ and $T^2$. We hope to return to this issue in the future.

\subsection*{Acknowledgement}
We thank Daniel Chung, Luis E. Ibanez, Victor Martin-Lozano, Pablo Soler, Wieland Staessens and Irene Valenzuela for useful comments.
This work is supported in part by the Grant-in-Aid for Japan Society for the JSPS Fellows No.16J06151 (YH),  the DOE grant DE-FG-02-95ER40896 (GS) and the Kellett Award of the University of Wisconsin (GS).

\appendix
\section{One-loop effective potential in curved spacetime}\label{App:curved space}
In this paper, we calculate the one-loop effective potential by the path-integral formalism.
Since we work in  curved spacetime, a careful definition of the measure of the path integral is needed.

The measure is determined once we fix the infinitesimal distance in the functional space.
Because we want to preserve general covariance, the following definition of the distance might be most appropriate:
\al{
||\delta \phi||^2:=\int d^4x \,\sqrt{-g}\delta \phi^*(x)\delta \phi(x).
}
Namely, by defining $\tilde{\phi}:=(-g)^{1/4}\phi, \tilde{\phi}^*:=(-g)^{1/4}\phi^*$, it is suitable to use $\mathcal{D}\tilde{\phi} \mathcal{D}\tilde{\phi}^*$ as a measure of the path integral.
This definition fixes our calculation of the one-loop effective potential.

\section{Calculation of the Casimir energy}\label{App:T2 calculation}
In this appendix, we present the calculation of the Casimir energy for $S^1$ and $T^2$ compactifications.
A related reference is Ref.~\cite{Elizalde:1997jv}.
The calculation of $S^1$ compactification is the review of known material. 
As for $T^2$ compactification, the new result is presented. 

\subsection{$S^1$ compactification}\label{App:S1 compactification}
The effective potential is
\al{\label{Eq:S1 calculation}
V_{S^1}^{(1)}&=(-1)^{2s_p+1}{n_p\over2}\sum_{n=-\infty}^\infty{1\over2\pi L} \int {d^3k\over\paren{2\pi}^3}\log\paren{k_0^2+k_1^2+k_2^2+M^2+{(n+\theta)^2\over L^2}}
\nn
&=-(-1)^{2s_p+1}{n_p\over2}{d\over ds}\sum_{n=-\infty}^\infty{1\over2\pi L}\int {d^3k\over\paren{2\pi}^3}\paren{k_0^2+k_1^2+k_2^2+M^2+{(n+\theta)^2\over L^2}}^{-s}\bigg|_{s=0}
\nn
&=(-1)^{2s_p+1}{n_p\over2}{1\over12\pi^2}{1\over L^4}F\paren{-{3\over2};\theta,ML}.
}
Here $s_p$ is the spin of the particle while $n_p$ is the real degrees of freedom, and $F$ is~\cite{Ponton:2001hq}:
\al{
F(s;a,c)&:=\sum_{n=-\infty}^\infty {1\over\sqbr{(n+a)^2+c^2}^s}\nn
&
={\sqrt{\pi}\over\Gamma(s)}|c|^{1-2s}\paren{\Gamma\paren{s-{1\over2}}+4\sum_{p=1}^\infty\paren{\pi p |c|}^{s-1/2}\cos\paren{2\pi pa}K_{s-1/2}\paren{2\pi p|c|}},
}
In general, this is divergent.
But if we consider sufficiently large $s$, this summation is convergent, and we can define the summation with $s=0$ by the analytic continuation.
Note that the factor $1/(2\pi L)$ in the above expression comes from the normalization of the wavefunction in $S^1$.
Namely if we use
\al{
\psi_n={1\over\sqrt{2\pi L}} e^{ik_3x_3},
\quad
\int_0^{2\pi} dx_3 \sqrt{g_{S^1}}\psi_m^*\psi_n=\delta_{nm}
}
as a normalized and orthogonal basis, the functional trace of the $S^1$ part is
\al{
\mathrm{tr}_{5D}(-\p^2_{5D})=\sum_{n} \psi_n^*\mathrm{tr}_{4D}(-\p^2_{4D}-L^{-2}\p_{x_3}^2)\psi_n
={1\over2\pi L}\sum_{n} \mathrm{tr}_{4D}(-\p^2_{4D}+L^{-2}k_3^2).
}
Finally, we obtain
\al{
\text{(Eq.~\eqref{Eq:S1 calculation})}=
-{1\over2}{M^4\over16\pi^2}(-1)^{2s_p}n_p\,\Gamma(-2)
-(-1)^{2s_p}{n_p}\sum_{n=1}^\infty {M^2\over8\pi^4 L^2}{K_2(2\pi n ML)\over n^2}\cos(2\pi n\theta),
}
where we have used $K_s(x)=K_{-s}(x)$.
The first divergent term is removed by the counterterm of the cosmological constant which is same as the flat space one.
Explicitly, in the case of the flat spacetime, we have\footnote{See e.g. (7.85) of Ref.~\cite{Peskin:1995ev} for the calculation here.}
\al{
 &\int {d^4k\over\paren{2\pi}^4}\log\paren{k_0^2+k_1^2+k_2^2+k_3^2+M^2}
 =-{d\over ds}\int {d^4k\over\paren{2\pi}^4}\paren{k_0^2+k_1^2+k_2^2+k_3^2+M^2}^{-s}\bigg|_{s=0}
 \nn
 &=-{d\over ds}\sqbr{{1\over(4\pi)^2}{\Gamma(s-2)\over\Gamma(s)}M^{4-2s}}\bigg|_{s=0}
 =-{1\over2}\paren{1\over4\pi}^2\Gamma(-2)M^4.
}
Therefore, only the second contribution should be taken into account.
%
%
As a result, we have
\al{
V_{S^1}^{(1)}&=
-(-1)^{2s_p}n_p{M^2\over8\pi^4 L^2}\sum_{n=1}^\infty {\cos(2\pi n\theta)\over n^2}K_2(2\pi n M L).
}
If $M=0$, this reduces to 
\al{\label{Eq:S1 massless}
V_{S^1,M=0}^{(1)}&=
-(-1)^{2s_p}n_p{1\over32\pi^6 L^4}\br{\text{Li}_4(e^{-2\pi i\theta})+c.c.}
\nn
&\simeq 
-(-1)^{2s_p}n_p{1\over16\pi^6 L^4}\cos\paren{2\pi\theta}.
}
In the second line, we have made the approximation that only the leading term of the polylogarithm is taken.


\subsection{Generalized Chowla-Selberg formula for $T^2$ compactification}\label{App:Generalized CS formula}
Our purpose is to calculate 
\al{
f(s;A,B,C,D,E,Q):=\sum_{n,m=-\infty}^\infty 
{1\over \paren{Am^2+Bmn+Cn^2+Dm+En+Q}^s},
}
under the zeta functional regularization. This summation naturally appears in the calculation of the Casimir energy on $T^2$, as we will see in the next subsection.
First we divide the summation as
\al{
f&=\sum_{n,m=-\infty,n\neq0}^\infty 
{1\over \paren{Am^2+Bmn+Cn^2+Dm+En+Q}^s}
+
\sum_{m=-\infty}^\infty 
{1\over \paren{Am^2+Dm+Q}^s}
\nn
&=:
f_1(s;A,B,C,D,E,Q)+f_2(s;A,D,Q)
.
}
The second term can be calculated as
{\footnotesize
\al{\label{Eq:second term}
&
f_2(s;A,D,Q)
=
{1\over\Gamma(s)}\sum_{m=-\infty}^\infty 
\int^\infty_0 dt \,t^{s-1}\exp\sqbr{-\br{A\paren{m+{D\over2A}}^2+\paren{Q-{D^2\over4A}}}t}
\nn
&=
{1\over\Gamma(s)}\sqrt{\pi\over A}\int^\infty_0 dt \,t^{s-3/2}\exp\sqbr{-\paren{Q-{D^2\over4A}}t}
\sqbr{1+2\sum_{p=1}^\infty\cos\paren{\pi pD\over A}e^{-\pi^2p^2\over At}}
\nn
&={1\over\Gamma(s)}\sqrt{\pi\over A}\bigg[\paren{Q-{D^2\over4A}}^{1/2-s}\Gamma(s-1/2)
\nn
&+4\sum_{p=1}^\infty\cos\paren{\pi pD\over A}\paren{\pi p\over\sqrt{A}\sqrt{Q-{D^2\over 4A}}}^{s-1/2}K_{s-1/2}\paren{{2\pi p\over\sqrt{A}}\sqrt{Q-{D^2\over 4A}}}
\bigg],
}
}
where we have used the formula
\al{
\sum_{n=-\infty}^\infty e^{-(n+z)^2w}
=\sqrt{\pi\over w}\sqbr{1+2\sum_{n=1}^\infty e^{-\pi^2 n^2\over w^2}\cos(2\pi n z)}
,
}
in the second line.
This formula can be easily derived by using the Poisson summation formula,
\al{&
\sum_{k=-\infty}^\infty \hat{f}(k)=\sum_{n=-\infty}^\infty f(n),
&&
\hat{f}(k)=\int^\infty_{-\infty}f(x) e^{-2\pi i k x}dx,
}
with $f(x)=e^{-(x+z)^2w}$.
We have also used the property of the modified Bessel function of the second kind,
\al{
\int^\infty_0 dt \,t^{s-1}\exp\sqbr{-\alpha^2t-{\beta^2\over t}}=2\paren{\beta\over\alpha}^s K_s(2\alpha\beta),
}
in the third line of Eq.~\eqref{Eq:second term}.

Similarly, the first term becomes
{\footnotesize
\al{
&
f_1(s;A,B,C,D,E,Q)
\nn
&=
\sum_{n\neq0}^\infty {1\over\Gamma(s)}\int dt\, t^{s-1}\exp\sqbr{-\br{A\paren{m+{Bn+D\over2A}}^2+\paren{C-{B^2\over4A}}n^2+\paren{E-{BD\over2A}}n+\paren{Q-{D^2\over4A}}}t}
\nn
&=
\sum_{n\neq0} {1\over\Gamma(s)}\int^\infty_0 dt \,t^{s-1}\sqrt{\pi\over At}
\br{1+2\sum_{m=1}^\infty e^{-{\pi^2m^2\over At}}\cos\paren{2\pi m{Bn+D\over2A}}}
\exp\sqbr{-\paren{{\Delta\over4A}\paren{n-n_0}^2+q}t}
\nn
&=
{1\over\Gamma(s)}\sqrt{\pi\over A}\sum_{n\neq0}{\Gamma(s-1/2)\over\br{{\Delta\over4A}\paren{n-n_0}^2+q}^{s-1/2}}
\nn
&+{4\over\Gamma(s)}\sqrt{\pi\over A}\sum_{m=1}^{n\neq0}\cos\paren{2\pi m{Bn+D\over2A}}\paren{\pi m/\sqrt{A}\over{\sqrt{{\Delta\over4A}\paren{n-n_0}^2+q}}}^{s-1/2}K_{s-1/2}\paren{{2\pi m\over\sqrt{A}}\sqrt{{\Delta\over4A}\paren{n-n_0}^2+q}}
\nn
&=
{\Gamma(s-1/2)\over\Gamma(s)}\sqrt{\pi\over A}
\br{
f_2\paren{s-{1\over2};{\Delta\over4A},-{\Delta \over2A}n_0, {\Delta \over4A}n_0^2+q}
-{1\over\paren{{\Delta \over4A}n_0^2+q}^{s-1/2}}
}
\nn
&+{4\over\Gamma(s)}\sqrt{\pi\over A}\sum_{m=1}^{n\neq0}\cos\paren{2\pi m{Bn+D\over2A}}\paren{\pi m/\sqrt{A}\over{\sqrt{{\Delta\over4A}\paren{n-n_0}^2+q}}}^{s-1/2}K_{s-1/2}\paren{{2\pi m\over\sqrt{A}}\sqrt{{\Delta\over4A}\paren{n-n_0}^2+q}}
,
}
}
where we define
\al{&
n_0=-{1\over2}{E-{BD\over2A}\over C-{B^2\over4A}},
&&
q=Q-{D^2\over4A}-{1\over4}{\paren{E-{BD\over2A}}^2\over C-{B^2\over4A}},
&&
\Delta=4AC-B^2.
}

For the application to $T^2$, we calculate the following quantities,
\al{
df_2(A,D,Q):=&-{d\over ds}{1\over4\pi}{1\over1-s}f_2(s-1;A,D,Q)\bigg|_{s=0}
\nn
&=
\frac{1}{8 \pi ^3 A}
\bigg[{\pi^3\over3A}\paren{4AQ-D^2}^{3/2}+A \bigg\{\pi  \sqrt{4 A Q-D^2}
   \left(\text{Li}_2\left(e^{\eta_+}\right)+\text{Li}_2\left(e^{\eta_-}\right)\right)
   \nn
   &+A
   \br{\text{Li}_3\left(e^{\eta_+}\right)+\text{Li}_3\left(e^{\eta_-}\right)}\bigg\}\bigg],
}
\al{
&df_1(A,B,C,D,E,Q):=-{d\over ds}{1\over4\pi}{1\over1-s}f_1(s-1;A,D,Q)\bigg|_{s=0}
\nn
&=
{1\over3\sqrt{A}}f_2\paren{-{3\over2};{\Delta\over4A},-{\Delta\over2A}n_0,{\Delta\over4A}n_0^2+q}
\nn
&
-{1\over3\sqrt{A}}\paren{{\Delta\over4A}n_0^2+q}^{3/2}
\nn
&
+\frac{1}{8\pi ^3}\sum_{n\neq0}\bigg[\pi  \sqrt{(n-n_0)^2\Delta+4 A q}
   \br{\text{Li}_2\paren{e^{\sigma_+}
   }
   +\text{Li}_2\left(e^{\sigma_-}\right)}
   +A
   \br{\text{Li}_3\left(e^{\sigma_+}
   \right)
   +\text{Li}_3\left(e^{\sigma_-}\right)}\bigg],
}
where
\al{
&
\eta_\pm:={\pm i\pi D-\pi\sqrt{4AQ-D^2}\over A}
,
&&
\sigma_\pm:=
{\pm i\pi(Bn+D)-\pi\sqrt{(n-n_0)^2\Delta+4Aq}\over A}.
}

\subsection{$T^2$ compactification}
As discussed in Eq.~\eqref{Eq:T2 eigenvalue}, the eigenvalue corresponding to the quadratic term in the action is
\al{
&M^2+{|(m+\theta_2)-(n+\theta_1)\tau|^2\over \rho\,\tau_2}
=
{1\over \rho\,\tau_2}m^2-{2\tau_1\over \rho\,\tau_2}nm+{|\tau|^2\over \rho\,\tau_2}n^2\nn
&+{2\theta_2-2\tau_1\theta_1\over \rho\,\tau_2}m+{-2\tau_1\theta_2+2\theta_1|\tau|^2\over \rho\,\tau_2}n+\paren{M^2+{\theta_2^2-2\tau_1\theta_1\theta_2+|\tau|^2\theta_1^2\over \rho\,\tau_2}}.
}
Hence, the Casimir energy consists of
\al{
&{1\over2}{1\over(2\pi)^2\rho}\sum_{m,n=-\infty}^{\infty}\int {d^2k\over(2\pi)^2}\ln\paren{k_0^2+k_1^2+M^2+{1\over \rho}{1\over\tau_2}|(m+\theta_2)-(n+\theta_1)\tau|^2}\nn
&=
{1\over8\pi^2\rho}{d\over ds}\sum_{m,n=-\infty}^{\infty}\int {d^2k\over(2\pi)^2}\paren{k_0^2+k_1^2+M^2+{1\over \rho}{1\over\tau_2}|(m+\theta_2)-(n+\theta_1)\tau|^2}^{-s}\bigg|_{s=0}\nn
&=
{1\over8\pi^2\rho}{d\over ds}\sum_{m,n=-\infty}^{\infty}{1\over4\pi}{1\over1-s}{1\over\br{M^2+|(m+\theta_2)-(n+\theta_1)\tau|^2/(\rho\tau_2)}^{s-1}}\bigg|_{s=0}.
}
Using the functions defined in App.~\ref{App:Generalized CS formula}, we find that the one field contribution is
\al{
&V^{(1)}_{T^2}(\rho,\tau,\theta_1,\theta_2)
\nn&={1\over8\pi^2\rho}{d\over ds}{1\over4\pi}{1\over1-s}\nn&\times f\paren{s-1;{1\over \rho\tau_2},-{2\tau_1\over \rho\tau_2},{|\tau|^2\over \rho\tau_2},{2\theta_2-2\tau_1\theta_1\over \rho\tau_2},{-2\tau_1\theta_2+2\theta_1|\tau|^2\over \rho\tau_2},M^2+{\theta_2^2-2\tau_1\theta_1\theta_2+|\tau|^2\theta_1^2\over \rho\tau_2}}\Bigg|_{s=0}
\nn
&=
-{1\over8\pi^2\rho}\bigg[df_1\paren{{1\over \rho\tau_2}, -{2\tau_1\over \rho\tau_2}, {|\tau|^2\over \rho\tau_2}, {2\theta_2-2\tau_1\theta_1\over \rho\tau_2}, {-2\tau_1\theta_2+2\theta_1|\tau|^2\over \rho\tau_2}, M^2+{\theta_2^2-2\tau_1\theta_1\theta_2+|\tau|^2\theta_1^2\over \rho\tau_2}}
\nn
&+df_2\paren{{1\over \rho\tau_2},  {2\theta_2-2\tau_1\theta_1\over \rho\tau_2}, M^2+{\theta_2^2-2\tau_1\theta_1\theta_2+|\tau|^2\theta_1^2\over \rho\tau_2}}\bigg]
.
}
Note that above parametrization gives
\al{&
\Delta={4\over \rho^2},
\quad
n_0=-\theta_1,
\quad
q=M^2,
\quad
\eta_\pm=2\pi\paren{\pm i(\theta_2-\tau_1\theta_1)-\sqrt{\tau_2^2\theta_1^2+\rho\tau_2 M^2}},
\nn
&
4AQ-D^2=4\paren{{M^2\over A^2\tau_2}+{\theta_1^2\over A^4}},
\quad
\sigma_\pm=2\pi\paren{\pm i\br{-(n+\theta_1)\tau_1+\theta_2}-\sqrt{(n+\theta_1)^2\tau_2^2+M^2 \rho\tau_2}}.
}
The divergent cosmological constant term is contained in the $df_2$ term
\al{{1\over3\sqrt{A}}f_2\paren{-{3\over2};{\Delta\over4A},-{\Delta\over2A}n_0,{\Delta\over4A}n_0^2+q}
&={\rho\over4}M^{4}\Gamma(-2)+\text{(finite term)},
}
which should be subtracted by the counterterm of the cosmological constant in flat spacetime.
The total Casimir energy is
\al{
&V^{(1)}_{T^2}(\rho,\tau,\theta_1,\theta_2)
= 
-{1\over8\pi^2\rho}\Bigg[{\rho\over4}\br{M^4\Gamma(-2)+4\sum_{p=1}^\infty\paren{\sqrt{\tau_2}M\over\pi p \sqrt{\rho}}^2\cos\paren{2\pi p\theta_1}K_2\paren{2\pi p \sqrt{\rho} M\over\sqrt{\tau_2}}}
\nn&
+\frac{1}{8\pi ^3 \rho\tau_2}\sum_{n\neq0}\bigg[2\pi  \sqrt{(n+\theta_1)^2\tau_2^2+M^2\rho\tau_2}
   \br{\text{Li}_2\paren{e^{\sigma_+}
   }
   +\text{Li}_2\left(e^{\sigma_-}\right)}
   +
   \br{\text{Li}_3\left(e^{\sigma_+}
   \right)
   +\text{Li}_3\left(e^{\sigma_-}\right)}\bigg]
\nn&   
+{1\over8\pi^3} \bigg\{2\pi  \sqrt{{M^2\over \rho\tau_2}+{\theta_1^2\over \rho^2}}
   \left(\text{Li}_2\left(e^{\eta_+}\right)+\text{Li}_2\left(e^{\eta_-}\right)\right)
+{1\over \rho\tau_2}
   \br{\text{Li}_3\left(e^{\eta_+}\right)+\text{Li}_3\left(e^{\eta_-}\right)}\bigg\}\Bigg].
}
The finite Casimir energy after the subtraction is
\al{\label{Eq:finite Casimir T2}
&V^{(1)}_{T^2}(\rho,\tau,\theta_1,\theta_2)
=
-{1\over8\pi^2\rho}\Bigg[\sum_{l=1}^\infty{\tau_2 M^2\over\pi^2 l^2}\cos\paren{2\pi l\theta_1}K_2\paren{2\pi l \sqrt{\rho} M\over\sqrt{\tau_2}}
\nn
&
+\frac{1}{8\pi ^3 \rho\tau_2}\sum_{n=-\infty}^\infty\br{2\pi  \sqrt{(n+\theta_1)^2\tau_2^2+M^2 \rho \tau_2}
   \br{\text{Li}_2\paren{e^{\sigma_+}}
   +\text{Li}_2\left(e^{\sigma_-}\right)}
   +\br{\text{Li}_3\left(e^{\sigma_+}\right)
   +\text{Li}_3\left(e^{\sigma_-}\right)}}\Bigg].
}

The Casimir energy including all fields in the theory can be written as
\al{
V^{\text{all}}_{T^2}(\rho,\tau,\theta_1,\theta_2)
=\sum_\text{particle} (-1)^{2s_p}n_p V^{(1)}_{T^2}\paren{\rho,\tau,q_p A_1+{1-z_{1p}\over2},q_p A_2+{1-z_{2p}\over2}},
}
where $s_p$ is the spin, $n_p$ is the degrees of freedom, $z_p=0(1)$ corresponds to anti-periodic(periodic) boundary condition.

\subsubsection{Consistency with Ref.~\cite{Arnold:2010qz}}
For the periodic particles ($\theta_1=\theta_2=0$), Eq.~\eqref{Eq:finite Casimir T2} becomes
\al{
V^{(1)}_{T^2}(\rho,\tau,0,0)=
&-{1\over8\pi^2\rho}\Bigg[\sum_{l=1}^\infty{\tau_2M^2\over\pi^2 l^2}K_2\paren{2\pi l \sqrt{\rho} M\over\sqrt{\tau_2}}
\nn&
+\frac{1}{8\pi ^3 \rho\tau_2}\sum_{n\neq0}\bigg[2\pi  \sqrt{n^2\tau_2^2+M^2 \rho\tau_2}
   \br{\mathrm{Li}_2\paren{e^{\sigma_+}
   }
   +\mathrm{Li}_2\left(e^{\sigma_-}\right)}
   +\br{\mathrm{Li}_3\left(e^{\sigma_+}
   \right)
   +\mathrm{Li}_3\left(e^{\sigma_-}\right)}\bigg]
\nn&   
+{1\over8\pi^3}
\bigg[2\pi  {M\over \sqrt{\rho\tau_2}}
   \br{\mathrm{Li}_2\left(e^{\eta_+}\right)+\mathrm{Li}_2\left(e^{\eta_-}\right)}
+{1\over \rho\tau_2}
   \br{\mathrm{Li}_3\left(e^{\eta_+}\right)+\mathrm{Li}_3\left(e^{\eta_-}\right)}\bigg]   \Bigg],
}
where we have used
\al{
\eta_\pm\to-2\pi M\sqrt{\rho\tau_2},
\quad
\sigma_\pm\to-2\pi\paren{\pm i n\tau_1 +\sqrt{n^2\tau_2^2+M^2 \rho\tau_2}}.
}
Moreover, in the massless limit, we have
\al{
&V^{(1)}_{T^2}(\rho,\tau,0,0)\nn
&\to
-{1\over8\pi^2\rho}\Bigg[{\tau_2^2\over180\rho}+{\zeta(3)\over4\pi^3}{1\over \rho\tau_2}
\nn&\phantom{++}+\frac{1}{8\pi ^3 \rho\tau_2}\sum_{n\neq0,-\infty}^\infty\bigg[2\pi  |n|\tau_2
   \br{\mathrm{Li}_2\paren{e^{\sigma_+}
   }
   +\mathrm{Li}_2\left(e^{\sigma_-}\right)}
   +\br{\mathrm{Li}_3\left(e^{\sigma_+}
   \right)
   +\mathrm{Li}_3\left(e^{\sigma_-}\right)}\bigg]\Bigg]
\nn&=-{1\over16\pi^4\rho^2}
\bigg[
{\pi^2\tau_2^2\over90}+{\zeta(3)\over2\pi}{1\over\tau_2}
\nn&\phantom{++++}+{1\over2\pi\tau_2}\sum_{n=1}^\infty\bigg\{2\pi  n\tau_2
   \br{\mathrm{Li}_2\paren{e^{-2\pi i n\bar{\tau}}
   }
   +\mathrm{Li}_2\left(e^{2\pi i n\tau}\right)}
   +\br{\mathrm{Li}_3\left(e^{-2\pi i n\bar{\tau}}
   \right)
   +\mathrm{Li}_3\left(e^{2\pi i n\tau}\right)}\bigg\}
\bigg],
}
where
\al{&
\eta_\pm\to0,
&&
\sigma_\pm\to-2\pi\paren{\pm in\tau_1+|n|\tau_2}.
}
Recalling the identity, 
\al{
&4\sqrt{\tau_2}\sum_{n,p=1}^\infty \paren{n\over p}^{3/2}\cos\paren{2\pi n p\tau_1}K_{3/2}\paren{2\pi n p\tau_2}
\nn
&=
{1\over2\pi\tau_2}\sqbr{2\pi n\tau_2\br{\mathrm{Li}_2(e^{2\pi i n\tau})+\mathrm{Li}_2(e^{-2\pi i n\bar{\tau}})}
+\br{\mathrm{Li}_3(e^{2\pi i n\tau})+\mathrm{Li}_3(e^{-2\pi i n\bar{\tau}})}
}
,
}
our result in the massless and periodic case is consistent with Eq.~$(17)$ in Ref.~\cite{Arnold:2010qz}.

Next let us compare our massive periodic expression  with Eq.~$(16)$ in Ref.~\cite{Arnold:2010qz}.
Our expression is
{\footnotesize
\al{
V^{(1)}_{T^2}(\rho,\tau,0,0)=
&-{1\over8\pi^2\rho}\Bigg[\sum_{l=1}^\infty\paren{\tau_2M^2\over\pi^2 l^2}K_2\paren{2\pi l \sqrt{\rho} M\over\sqrt{\tau_2}}
\nn&
+\frac{1}{4\pi ^3 \rho\tau_2}\sum_{n=1}^\infty\bigg[2\pi  \sqrt{n^2\tau_2^2+M^2\rho\tau_2}
   \br{\mathrm{Li}_2\paren{e^{\sigma_+}}
   +\mathrm{Li}_2\left(e^{\sigma_-}\right)}
   +\br{\mathrm{Li}_3\left(e^{\sigma_+}\right)
   +\mathrm{Li}_3\left(e^{\sigma_-}\right)}\bigg]
\nn&   
+{1\over8\pi^3}
\bigg[2\pi  {M\over \sqrt{\rho\tau_2}}
   \br{\mathrm{Li}_2\left(e^{\eta_+}\right)+\mathrm{Li}_2\left(e^{\eta_-}\right)}
+{1\over \rho\tau_2}
   \br{\mathrm{Li}_3\left(e^{\eta_+}\right)+\mathrm{Li}_3\left(e^{\eta_-}\right)}\bigg]   \Bigg]
\nn
&=
-{1\over16\pi^4\rho^2}
\bigg[
\sum_{p=1}^\infty {2\rho\tau_2M^2\over p^2}K_2\paren{2\pi p \sqrt{\rho} M\over\sqrt{\tau_2}}
\nn&
+\frac{1}{2\pi\tau_2}\sum_{n=1}^\infty\bigg\{2\pi  \sqrt{n^2\tau_2^2+M^2\rho\tau_2}
   \br{\mathrm{Li}_2\paren{e^{\sigma_+}}
   +\mathrm{Li}_2\left(e^{\sigma_-}\right)}
   +\br{\mathrm{Li}_3\left(e^{\sigma_+}
   \right)
   +\mathrm{Li}_3\left(e^{\sigma_-}\right)}\bigg\}
\nn&
+{\sqrt{\rho} M\over\sqrt{\tau_2}}
   \mathrm{Li}_2\left(e^{-2\pi M\sqrt{\rho\tau_2}}\right)
+{1\over2\pi\tau_2}
   \mathrm{Li}_3\left(e^{-2\pi M\sqrt{\rho\tau_2}}\right)   \bigg\}
\bigg].
}
}
$\!\!$We confirmed that this is consistent with Eq.~$(16)$ in Ref.~\cite{Arnold:2010qz}.
We note that the identities:
{\footnotesize
\al{
&{2(\sqrt{\rho} M)^{3/2}\over\tau_2^{1/4}}\sum_{p=1}^\infty K_{3/2}(2\pi p \sqrt{\rho} M\sqrt{\tau_2})
=
\frac{2 \pi  \sqrt{\rho} M
   \sqrt{\tau_2}\,
   \text{Li}_2\left(e^{-2 \sqrt{\rho} M \pi 
   \sqrt{\tau_2}}\right)+\text{Li}_3\left(e^{-2 \sqrt{\rho} M \pi  \sqrt{\tau_2}}\right)}{2 \pi \tau_2 },
   \nn&
4\sqrt{\tau_2}\sum_{n,p=1}^\infty \paren{1\over p}^{3/2}\paren{n^2+{\rho M^2\over\tau_2}}^{3/4}\cos\paren{2\pi n p\tau_1}K_{3/2}\paren{2\pi p\tau_2\sqrt{\rho M^2\over\tau_2}}
\nn
&=
\frac{1}{2 \pi  \tau_2}
\bigg[\text{Li}_3\left(e^{-2 i n\pi  \tau_1-2 \pi \tau_2
   \sqrt{\frac{\rho M^2}{\tau_2}+n^2} }\right)
   +\text{Li}_3\left(e^{2 i n \pi  \tau_1-2\pi \tau_2 \sqrt{\frac{\rho M^2}{\tau_2}+n^2}}\right)
   \nn&+2 \pi \tau _2 \sqrt{\frac{\rho M^2}{\tau _2}+n^2}
   \br{\text{Li}_2\left(e^{-2 in \pi \tau_1-2 \pi \tau_2\sqrt{\frac{\rho M^2}{\tau_2}+n^2} }\right)+\text{Li}_2\left(e^{2 i n \pi  \tau_1-2\pi \tau_2 \sqrt{\frac{\rho M^2}{\tau_2}+n^2} }\right)}
\bigg],   
}
}
are needed for comparison. 
More explicitly, the correspondence is
$V_{T^2}^{(1)}=\rho_{(4)}^\text{obs}$,
where $\rho_{(4)}^\text{obs}$ is the quantity which appears in Ref.~\cite{Arnold:2010qz}, and we identify $\rho=a^2$.

We recall the well-known equivalence between
\al{
F:=\int{d^Nk\over(2\pi)^N}\ln(k^2+M^2-i\epsilon)\paren{=i\int{d^Nk_E\over(2\pi)^N}\ln(k_E^2+M^2)},
}
and
\al{
G:=\int{d^{N-1}k\over(2\pi)^{N-1}}\sqrt{|\vec{k}|^2+M^2},
}
up to a $M^2$-independent constant.
The calculation in Ref.~\cite{Arnold:2010qz} employs  $G$ for deriving the Casimir energy. 
The equivalence can be easily seen by taking the derivative respect with $M^2$:
\al{
{\p\over\p M^2}F&=\int {d^Nk\over(2\pi)^N}{1\over k^2+M^2-i\epsilon}
\nn&=\int {d^Nk\over(2\pi)^N}{-1\over \br{k^0-(E_k-i\epsilon)}\br{k^0+(E_k-i\epsilon)}}
\nn&=-i\int {d^{N-1}k\over(2\pi)^{N-1}}{-1\over2E_k}=i\p_M^2G,
}
where $E_k=\sqrt{|\vec{k}|^2+M^2}$. 
Hence, one can see $F=iG+\text{const.}$.


\subsection{Consistency between $S^1$ and $T^2$ compactifications}
If we take the limit where the one of the radius of $T^2$ becomes infinite, the potential should becomes that of the $S^1$ case.
Note that
\al{&
\rho=L_1 L_2 {\tau_2\over|\tau|},
&&
L_2=L_1|\tau|
}
where $L_1$ and $L_2$ are the radii of the two 1-cycles.
If one take $L_1$ to be fixed and $L_2\to \infty$, then
\al{&
|\tau|\to\infty, 
&&
\rho\to\infty, 
&&
L_1^2={\rho\over\tau_2}=\text{fixed}.
}
We can see that the first term dominates Eq.~\eqref{Eq:T2 result}, and we have
\al{
V_{T^2}^{(1)}\to
-\sum_{l=1}^\infty{\tau_2\over\rho}{M^2\over8\pi^4 l^2}\cos\paren{2\pi l\theta_1}K_2\paren{2\pi l M\sqrt{\tau_2\over\rho}}
}
which is the same as Eq.~\eqref{Eq:S1 result} under the identification $\rho/\tau_2\sim L^2$.

\section{Vacuum condition}\label{Sec:vacuum condition}
In this appendix, we examine the conditions for the stable vacuum against localized perturbations in $S^1$ and $T^2$ compactifications.
These conditions are different from that for the conventional $4$d flat spacetime because there can be negative curvature, and the dynamical degrees of freedom is different in lower dimensions.
The material in this appendix is new except for App.~\ref{App:BF bound}.

\subsection{Breitenlohner-Freedman bound in $AdS_{d+1}$ spacetime}\label{App:BF bound}
It is well-known that $AdS_{d+1}$ spacetime allows some amount of 
tachyonic mass.
In order to see this quickly~\cite{Moroz:2009kv}, we start from the equation of motion (EoM) of the scalar
\al{&
\paren{{1\over\sqrt{-g}}\p_\mu\paren{\sqrt{-g}g^{\mu\nu}\p_\nu}-M^2}\phi=0,
&&
ds^2={1\over z^2}\paren{dz^2+\eta_{\mu\nu}dx^\mu dx^\nu}.
}
More explicitly, we have
\al{
\paren{\p_z^2-{d-1\over z}\p_z+\eta^{\mu\nu}\p_\mu\p_\nu-{M^2\over z^2}}\phi=0.
}
By performing the Fourier transformation except for the $z$ direction, this becomes
\al{&
\paren{\p_z^2-{d-1\over z}\p_z-k_\mu k^\mu-{M^2\over z^2}}\varphi=0,
&&
\phi(z,x)=\varphi(z) e^{ik\cdot x},
}
which leads to
\al{&
\paren{-\p_z^2+V(z)}\varphi(z)=\omega^2\varphi(z),
&&
V=\vec{k}^2+{1\over z^2}\paren{M^2+{d^2-1\over4}},
}
by $\varphi(z)\to z^{-d+1\over2}\varphi(z)$.
Combining the fact that a time dependent Schrodinger equation only admits a stable solution if $V>-1/4$, we get
\al{
M^2 L_{AdS}^2>-{d^2\over4},
}
where $L_{AdS}^2$ is the radius of the $AdS$.


\subsection{Vacuum condition of the $S^1$ compactification}\label{App:vacuum condition S1}
In the case of the $AdS_3$ vacuum, the BF bound reads
\al{\label{Eq:AdS3 tachyon}
-{1\over L_{AdS}^2}={R_{(3)}\over6}\leq M^2.
}
Let us derive the stability condition for the Wilson line moduli.
The action is
\al{
S&=\int d^4x \sqrt{-g}\paren{-{1\over4}F_{\mu\nu}F^{\mu\nu}-V_{S^1}^\text{all}+...}
\nn&=\int_{x_{3d}} (L_0)\sqbr{{1\over2}M_P^2R^{E(3)}-{1\over2 L^2}(\p_i A_3)^2-{\Lambda_4 L_0^2\over(2\pi L)^2}-{V_{S^1}^\text{all} L_0^2\over(2\pi L)^2}+...}
}
$L_0$ is the arbitrary length parameter which is introduced just for convenience.
As we will see below, the physical condition Eq.~\eqref{Eq:AdS3 tachyon} does not depend on $L_0$, as it should be.
The Einstein equation says
\al{\label{Eq:3d Einstein eq}
(L_0 M_P^2)\paren{-R_{\mu\nu}^{E(3)}+{1\over2}g_{\mu\nu}R^{E(3)}}={1\over2}V_{S^1}^\text{all}L_0^3 g_{\mu\nu},
}
whose trace leads to
\al{
R^{E(3)}={3\over(2\pi)^2}{V_{S^1}^\text{all}\over M_P^2}{L_0^2\over L^2}.
}
On the other hand, the mass of the Wilson line $a=A_4$ is
\al{
m_a^2={L_0^2\over(2\pi)^2}{\p^2V_{S^1}^\text{all}\over\p a^2}\bigg|_{a=a_*},
}
where $L_*, a_*$ are the spacetime independent solutions.
Now Eq.~\eqref{Eq:AdS3 tachyon} is
\al{
{m_a^2\over R^{E(3)}}={1\over3}{L_*^2M_P^2\over V_{S^1}^\text{all}}{\p^2V_{S^1}^\text{all}\over\p a^2}\bigg|_{a=a_*}\leq -{1\over6}.
}

Practically, the value in the left hand side is too large to save the tachyonic Wilson line field, because the $M_P$ factor in the numerator is much larger than the compactification scale.

\subsection{Vacuum condition of the $T^2$ compactification}\label{App:vacuum condition}
Unlike in other dimensions, we cannot go to the Einstein frame because of Weyl invariance of the Einstein Hilbert action.
The condition for perturbatively stable vacuum is briefly analyzed in Ref.~\cite{ArkaniHamed:2007gg} assuming a flat $2$d spacetime.
Here we present the extension to $dS_2$ and $AdS_2$ spacetime.

First, we derive the equation of motion starting from 
\al{
S=\int d^2x\sqrt{-g_{(2)}}
\sqbr{
{1\over2}M_P^2\paren{\rho R_{(2)}-{\rho\over2\tau_2^2}\paren{(\p_\alpha\tau_1)^2+(\p_\alpha\tau_2)^2}}-V(\rho,\tau)
}.
}
The variations by $\rho, g_{\alpha\beta}, \tau_a$ are given by
{\footnotesize
\al{
\delta_{\rho}S&=\int_x 
\sqbr{
{1\over2}M_P^2\paren{R_{(2)}-{1\over2\tau_2^2}\paren{(\p_\alpha\tau_1)^2+(\p_\alpha\tau_2)^2}}-\p_\rho V(\rho,\tau)
-{1\over4}F_{\alpha\beta}F^{\alpha\beta}
}\delta \rho,
\nn
\delta_{g_{\alpha\beta}}S&=\int_x 
\bigg[
{1\over2}M_P^2 \rho\paren{-g^{\alpha\beta}\nabla^2+\nabla^\alpha\nabla^\beta-R^{\alpha\beta}}\delta g_{\alpha\beta}
+{\rho\over2}F^{\alpha\gamma}F^{\beta\delta}g_{\gamma\delta}\delta g_{\alpha\beta}
\nn&+\br{{1\over4}M_P^2\rho R_{(2)}-{\rho\over4\tau_2^2}\paren{(\p_\alpha\tau_1)^2+(\p_\alpha\tau_2)^2}-{1\over2}V-{\rho\over8}F_{\alpha\beta}F^{\alpha\beta}}g^{\alpha\beta}\delta g_{\alpha\beta}
\bigg]
\nn&
=
\int_x 
\delta g_{\alpha\beta}\bigg[
{1\over2}M_P^2 \paren{-g^{\alpha\beta}\nabla^2+\nabla^\alpha\nabla^\beta}\rho
+{\rho\over2}\paren{F^{\alpha\gamma}{F^{\beta}}_{\gamma}-{1\over2}F_{\gamma\delta}F^{\gamma\delta}}
-\br{{\rho\over4\tau_2^2}\paren{(\p_\alpha\tau_1)^2+(\p_\alpha\tau_2)^2}+{1\over2}V}g^{\alpha\beta}
\bigg],
\nn
\delta_{\tau_a}S&=\int_x \bigg[
{M_P^2\over2}{\rho\over\tau_2^3}(\p_\alpha\tau_a)^2\delta\tau_2
-{M_P^2\over2}{\rho\over\tau_2^2}(\p_\alpha\tau_a)(\p^\alpha\delta\tau_a)-(\p_{\tau_a}V)\delta\tau_a
\bigg]
\nn&
=\int_x \bigg[
\delta\tau_1\paren{{M_P^2\over2}{\rho\over\tau_2^2}\nabla^2\tau_1-M_P^2{\rho\over\tau_2^3}(\p^\alpha\tau_2)(\p_\alpha\tau_1)-\p_{\tau_1}V}
\nn&+\delta\tau_2\paren{{M_P^2\over2}{\rho\over\tau_2^3}(\p_\alpha\tau_a)^2+M_P^2{\rho\over\tau_2^3}(\p^\alpha\tau_2)(\p_\alpha\tau_2)+{M_P^2\over2}{\rho\over\tau_2^2}\nabla^2\tau_2-\p_{\tau_2}V}
\bigg],
}
}
where $\int_x:=\int d^2x\sqrt{-g_{(2)}}$.
Therefore, EoM is
{\footnotesize
\al{&
{1\over2}M_P^2\paren{R_{(2)}-{1\over2\tau_2^2}\paren{(\p_\alpha\tau_1)^2+(\p_\alpha\tau_2)^2}}-\p_\rho V(\rho,\tau)-{1\over4}F_{\alpha\beta}F^{\alpha\beta}=0,
\nn&
{1\over2}M_P^2 \paren{-g^{\alpha\beta}\nabla^2+\nabla^\alpha\nabla^\beta}\rho
+{\rho\over2}\paren{F^{\alpha\gamma}{F^\beta}_\gamma-{1\over2}F_{\gamma\delta}F^{\gamma\delta}g^{\alpha\beta}}
-\br{{\rho\over4\tau_2^2}\paren{(\p_\alpha\tau_1)^2+(\p_\alpha\tau_2)^2}+{1\over2}V}g^{\alpha\beta}=0,
\nn&
{M_P^2\over2}{\rho\over\tau_2^2}\nabla^2\tau_1-M_P^2{\rho\over\tau_2^3}(\p^\alpha\tau_2)(\p_\alpha\tau_1)-\p_{\tau_1}V=0,
\nn&
{M_P^2\over2}{\rho\over\tau_2^3}(\p_\alpha\tau_a)^2+{M_P^2\rho\over\tau_2^3}(\p^\alpha\tau_2)(\p_\alpha\tau_2)+{M_P^2\over2}{\rho\over\tau_2^2}\nabla^2\tau_2-\p_{\tau_2}V=0.
}
}

If we focus on the solution which does not have spacetime dependence, the EoM becomes
\al{&
{1\over2}M_P^2R_{(2)}-\p_\rho V=0,
&&
V=0,
&&
\p_{\tau_a}V=0.
}
The curvature of the two-dimensional spacetime is determined by $\p_\rho V$.
Next, we consider localized fluctuations of the solution, $\delta g_{\alpha\beta}, \delta\rho, \delta\tau_a$, around the spacetime independent background in order to discuss the stability of the solution.
The equations for the localized fluctuations are written as
\al{\label{Eq:stability}
&
{1\over2}M_P^2\delta R_{(2)}-(\p_\rho^2 V)\delta\rho-(\p_\rho\p_{\tau_a}V)\delta\tau^a=0,
\nn&
{1\over2}M_P^2 \paren{-g_{\alpha\beta}\nabla^2+\nabla_\alpha\nabla_\beta}\delta\rho
-{1\over2}(\p_\rho V)g_{\alpha\beta}\delta\rho =0,
\nn&
{M_P^2\over2}{\rho\over\tau_2^2}\nabla^2\delta\tau_a-(\p_{\tau_a}\p_\rho V)\delta\rho-(\p_{\tau_a}\p_{\tau_b} V)\delta\tau_b=0.
}
The first equation is just the constraint. 
The perturbation of gravity, $\delta R_{(2)}$ is fixed by this equation.
The second equation is also not dynamical. 
By taking the trace,  we have
\al{
M_P^2\nabla^2\delta\rho+2(\p_\rho V)\delta\rho=0.
}
By substituting this again, the second equation becomes 
\al{
{1\over2}M_P^2 \nabla_\alpha\nabla_\beta\delta\rho
+{1\over2}(\p_\rho V)g_{\alpha\beta}\delta\rho 
=
{1\over4}M_P^2 \paren{-g_{\alpha\beta}\nabla^2+2\nabla_\alpha\nabla_\beta}\delta\rho
=0.
}
By employing the conformally flat gauge, $g_{\alpha\beta}=e^{2\omega}\eta_{\alpha\beta}$, this leads to
\al{&
\br{\p_0\p_1-(\p_1\omega)\p_0-(\p_0\omega)\p_1}\delta\rho=0,
\nn&
\br{\p_0^2-(\p_0\omega)\p_0-(\p_1\omega)\p_1-{\p_\rho V\over M_P^2}e^{2\omega}}\delta\rho=0,
\nn&
\br{\p_1^2-(\p_0\omega)\p_0-(\p_1\omega)\p_1+{\p_\rho V\over M_P^2}e^{2\omega}}\delta\rho=0.
}
The $e^{2\omega}=1/x_0^2, 1$ and $1/x_1^2$ correspond to $dS_2$, $M_2$ and $AdS_2$, respectively.
In these background, we can show $\delta\rho=0$ from
\al{&
\br{\p_0^2+\p_1^2-2(\p_0\omega)\p_0-2(\p_1\omega)\p_1}\delta\rho=0,
&&
\br{\p_0\p_1-(\p_1\omega)\p_0-(\p_0\omega)\p_1}\delta\rho=0.
}
More concretely, for $dS_2$, we have
\al{&
\br{\p_0^2+\p_1^2+{2\over x_0}\p_0}\delta\rho
=\br{\paren{\p_0+{1\over x_0}}^2+\p_1^2}\delta\rho
=0,
&&
\paren{\p_0+{1\over x_0}}\p_1 \delta\rho=0.
}
By introducing $\delta\rho=\delta\tilde{\rho}_{dS}/x_0$, this becomes
\al{
&
{1\over x_0}\paren{\p_0^2+\p_1^2}\delta\tilde{\rho}_{dS}=0,
&&
{1\over x_0}\p_0\p_1\delta\tilde{\rho}_{dS}=0.
}
Therefore, we cannot take the localized $\delta \rho$ as an initial condition.
Similarly, we get following equations in the case of $M_2$ and $AdS_2$,
\al{&
\paren{\p_0^2+\p_1^2}\delta{\rho}=0, 
&&
\p_0\p_1\delta\rho=0,
&&
\text{for $M_2$},
\nn
&
\paren{\p_0^2+\p_1^2}\delta{\tilde{\rho}_{AdS}}=0,
&&
\p_0\p_1\delta\tilde{\rho}_{AdS}=0, 
&&
\text{for $AdS_2$},
}
where  $\delta\rho=\delta\tilde{\rho}_{AdS}/x_1$.
Hence, we can safely put $\delta\rho=0$ to study the stability against localized perturbation.

Finally, let us move on the third equation of Eq.~\eqref{Eq:stability}.
Now, it is
\al{
\nabla^2\delta\tau_a-{2\tau_2^2\over\rho M_P^2}(\p_{\tau_a}\p_{\tau_b} V)\delta\tau_b=0.
}
If the $2$-dimensional space is flat or $dS$, this says that the matrix $\p_{\tau_a}\p_{\tau_b} V$ should be positive definite.
In the case of $AdS_2$, the stability condition is given by the BF bound:
\al{
{2\tau_2^2\over\rho M_P^2}(\p_{\tau_a}\p_{\tau_b} V)\geq -{1\over4}{1\over L_{AdS}^2}={R_{(2)}\over8},
}
where $R_{(2)}=-2/L_{AdS}^2$.\footnote{In general case, $R_{(n)}=-n(n-1)/L_{AdS}^2$ for $AdS_n$.} 

Note that, unlike the discussion in Ref.~\cite{Arnold:2010qz}, we conclude that $dS_2$ and $M_2$ are possible.
The point is that the discussion in Ref.~\cite{Arnold:2010qz} is applicable only to $2+\epsilon$ dimensions.
The limit $\epsilon\to0$ is not smooth because the Einstein Hilbert action becomes Weyl invariant.

Indeed, our argument matches the number of physical degrees of freedom.
The $4$-dimensional graviton has $2$ physical degrees of freedom.
In term of $T^2$ compactification, this corresponds to the fluctuation of the $\tau$ moduli.
So we only need to consider the stability of the $\tau$ fluctuation, and the other fluctuations are determined by the constraint equations.

To summarize, we need to solve
\al{\label{Eq:vacuum condition1}&
V=0,
&&
\p_{\tau_a}V=0,
}
in order to obtain the $2$d spacetime independent solution of $T^2$ compactification.
The curvature of $2$d is determined by $R_{(2)}M_P^2/2-\p_\rho V=0$, namely, $\p_\rho V>0, \p_\rho V=0$ and $\p_\rho V<0$ correspond to $dS_2$, $M_2$ and $AdS_2$, respectively.
To guarantee the perturbative stability of the vacuum,  it is required that 
\al{\label{Eq:vacuum condition2}&
\p_{\tau_a}\p_{\tau_b}V\geq 0, \quad\text{for $dS_2$ and $M_2$},
&&
{(4\tau_2)^2\over\rho M_P^2}\p_{\tau_a}\p_{\tau_b}V\geq R_{(2)}, \quad\text{for $AdS_2$},
}

The dynamics of the Wilson line is similar to that of the $\tau$ moduli.
The EoM requires that the Wilson line sits at the extremum of the potential.
There is a lower bound on the mass depending on whether the extremum is $dS_2, M_2$ or $AdS_2$.

\bibliographystyle{TitleAndArxiv}
\bibliography{Bibliography}

\end{document}